\documentclass[twocolumn]{aastex63}

\usepackage{amsmath}
\usepackage{gensymb}
\graphicspath{{./}{figures/}}
\usepackage{graphicx}
\usepackage{multirow}

\begin{document}

\title{Kilonova Emission from Black Hole-Neutron Star Mergers. II. Luminosity Function and Implications for Target-of-opportunity Observations of Gravitational-wave Triggers and Blind Searches}

\author[0000-0002-9195-4904]{Jin-Ping Zhu}
\affil{Department of Astronomy, School of Physics, Peking University, Beijing 100871, China}

\author[0000-0002-9188-5435]{Shichao Wu}
\affiliation{Department of Astronomy, Beijing Normal University, Beijing 100875, China}

\author[0000-0001-6374-8313]{Yuan-Pei Yang}
\affiliation{South-Western Institute for Astronomy Research, Yunnan University, Kunming, Yunnan, People’s Republic of China}

\author[0000-0002-9725-2524]{Bing Zhang}
\affiliation{Department of Physics and Astronomy, University of Nevada, Las Vegas, NV 89154, USA}

\author[0000-0002-3100-6558]{He Gao}
\affiliation{Department of Astronomy, Beijing Normal University, Beijing 100875, China}

\author[0000-0002-1067-1911]{Yun-Wei Yu}
\affiliation{Institute of Astrophysics, Central China Normal University, Wuhan 430079, China}

\author{Zhuo Li}
\affiliation{Department of Astronomy, School of Physics, Peking University, Beijing 100871, China}
\affiliation{Kavli Institute for Astronomy and Astrophysics, Peking University, Beijing 100871, China}

\author[0000-0002-1932-7295]{Zhoujian Cao}
\affiliation{Department of Astronomy, Beijing Normal University, Beijing 100875, China}

\author[0000-0002-8708-0597]{Liang-Duan Liu}
\affiliation{Department of Astronomy, Beijing Normal University, Beijing 100875, China}

\author{Yan Huang}
\affiliation{School of Physics and Materials Science, Anhui University, Hefei 230601, China}

\author{Xing-Han Zhang}
\affiliation{Department of Physics, Tsinghua University, Beijing 100084, China}

\correspondingauthor{Jin-Ping Zhu; Bing Zhang; Shichao Wu; Zhoujian Cao} 
\email{zhujp@pku.edu.cn; zhang@physics.unlv.edu; wushichao@mail.bnu.edu.cn; zjcao@amt.ac.cn}

\begin{abstract}

We present detailed simulations of black hole-neutron star (BH--NS) mergers  kilonova and gamma-ray burst (GRB) afterglow and kilonova luminosity function, and discuss the detectability of electromagnetic (EM) counterpart in connection with gravitational wave (GW) detections, GW-triggered target-of-opportunity observations, and time-domain blind searches. The predicted absolute magnitude of the BH--NS kilonovae at $0.5\,{\rm days}$ after the merger falls in $[-10,-15.5]$. The simulated luminosity function contains the potential viewing-angle distribution information of the anisotropic kilonova emission. We simulate the GW detection rates, detectable distances and signal duration, for the future networks of 2nd/2.5th/3rd-generation GW detectors. BH--NSs tend to produce brighter kilonovae and afterglows if the BH has a higher aligned-spin, and a less massive NS with a stiffer EoS. The detectability of kilonova is especially sensitive to the BH spin. If BHs typically have low spins, the BH--NS EM counterparts are hard to discover. For the 2nd generation GW detector networks, a limiting magnitude of $m_{\rm limit}\sim23-24\,{\rm mag}$ is required to detect the kilonovae even if BH high spin is assumed. Thus, a plausible explanation for the lack of BH--NS associated kilonova detection during LIGO/Virgo O3 is that either there is no EM counterpart (plunging events), or the current follow-ups are too shallow. These observations still have the chance to detect the on-axis jet afterglow associated with a sGRB or an orphan afterglow. Follow-up observations can detect possible associated sGRB afterglows, from which kilonova signatures may be studied. For time-domain observations, a high-cadence search in redder filters is recommended to detect more BH--NS associated kilonovae and afterglows.

\end{abstract}


\section{Introduction}

{Neutron-rich materials released from mergers of binary neutron star (BNS) and black hole--neutron star (BH--NS) binaries undergo the rapid neutron capture process ($r$-process), which has long been proposed to account for about half of the elements heavier than iron in the universe \citep{lattimer1974,lattimer1976,symbalisty1982}}. \cite{li1998} first predicted that the radioactive decays of these $r$-process nuclei can provide a heating source for powering a new type of transient { days to weeks following a BNS or BH--NS merger.} The detailed research by \cite{metzger2010} suggested that the luminosity of {such a radioactivity-powered transient is around several times $10^{41}\,{\rm erg}\,{\rm s}^{-1}$}, which is a factor of $\sim10^3$ higher than that of a typical nova, so that such a transient is called a ``kilonova''. The characteristics of the lightcurves, peak luminosities, timescales, and spectra of the kilonova emission have been widely investigated {theoretically \citep[e.g.,][]{kulkarni2005,rosswog2007,roberts2011,barnes2013,kasen2013,kasen2015,kasen2017,tanaka2013,yu2013,grossman2014,metzger2014b,perego2014,fernandez2016,metzger2017,kawaguchi2020}.}

Despite the clear predictions of kilonova emission from BNS and BH--NS mergers, its low luminosity and fast evolution compared with supernova emission makes difficult to detect. Traditional optical time-domain survey projects have been designed to discover ordinary supernovae or tidal disruption events whose rise and decay timescales in brightness are long, e.g., weeks or even months. Without target-of-opportunity observations, only some kilonovae candidates could be recorded in the such time-domain survey projects. Recently, thanks to improved cadence of supernova searches and the operation of some new large survey projects, some fast-evolving transients, which had lightcurves lasting days to weeks and brightness in the range of theoretically predicted kilonova brightness, have been discovered \citep[e.g.,][]{drout2014,prentice2018,rest2018,mcbrien2020}. However, none of these events have been firmly identified as kilonovae.

If short gamma-ray bursts (sGRBs) also originate from BNS or BH--NS mergers \citep{paczynski1991,paczynski1993,eichler1989,narayan1992}, then another plausible way to constrain kilonova models is via follow-up observations of nearby sGRBs on timescales of hours to a week. In the past few years, there have been several kilonova candidates detected in superposition with the decaying sGRB afterglows \citep{berger2013,tanvir2013,fan2013,gao2015,gao2017,jin2015,jin2016,jin2020,yang2015,compertz2018,ascenzi2019,rossi2020}. However, due to the scarce and ambiguous observational data, these kilonova candidates also could not be firmly confirmed. 

There are some limitations of searching kilonovae using time-domain survey observations or sGRBs follow-up observations. For time-domain survey observations, it is easy to miss the optimal observational time window for searching kilonovae due to their faint nature and short emission timescale. The lack of spectral observations or color evolution for potential kilonovae candidates make them hard to identify. For sGRBs follow-up observations, on the other hand, gamma-ray detectors usually have poor spatial positioning capabilities \citep[e.g.,][]{connaughton2015}. Current gamma-ray detectors also do not achieve an all-sky coverage. The beaming effect of sGRBs \citep{rhoads1999,sari1999} makes only a fraction of events being triggered. Also the putative kilonova emission is likely outshone by the luminous on-axis jet afterglow. In view that BNS and BH--NS mergers are targeted gravitational wave (GW) events for ground-based interferometer detectors, the strategy of optical follow-up observations for GW triggers to search kilonovae has been studied in detail \citep[e.g.,][]{metzger2012,cowperthwaite2015,gehrels2016,cowperthwaite2019}. Compared with sGRBs, GW radiation is essentially isotropic from the source \citep{creighton2011}. Also GW detectors such as the Advanced Laser Interferometer Gravitational Wave Observatory (LIGO) have all-sky coverage and can detect sources out to distances of $\sim300$ and $\sim650\,{\rm Mpc}$ for BNS and BH--NS mergers, respectively \citep{cutler2002}. GW sources will be localized to $\sim 10\,{\rm deg}^2$ by the network of the the Advanced LIGO, Advanced Virgo and KAGRA GW detectors \citep{abbott2018prospects}. Therefore, an optimal searching strategy for kilonovae would be to take advantage of target-of-opportunity follow-up observations of GW triggers. 

On 2017 August 17, a BNS merger event (GW170817) for the first time was observed in GWs by the Advanced LIGO and the Advanced Virgo detectors \citep{abbott2017gw170817}. Independent gamma-ray observations discovered a sGRB \citep[GRB\,170817A;][]{abbott2017gravitational,goldstein2017,savchenko2017,zhangbb2018}, and follow-up observations discovered an ultraviolet-optical-infrared transient \citep[AT\,2017gfo;][]{abbott2017multimessenger,andreoni2017,arcavi2017,chornock2017,coulter2017,covino2017,cowperthwaite2017,diaz2017,drout2017,evans2017,hu2017,kasliwal2017,kilpatrick2017,lipunov2017,mccully2017,nicholl2017,pian2017,shappee2017,smartt2017N,soaressantos2017,tanvir2017,troja2017,utsumi2017,valenti2017} and a broadband afterglow from radio to X-rays consistent with off-axis jet emission \citep{alexander2017,haggard2017,hallinan2017,margutti2017,troja2017,troja2020,davanzo2018,dobie2018,lazzati2018,lyman2018,piro2019,ghirlanda2019}. The ultraviolet-optical-infrared transient was well consistent with the theoretical predictions of kilonova emission \citep{cowperthwaite2017,perego2017,tanaka2017,tanaka2018,villar2017,gao2017b,ai2018,kawaguchi2018,li2018,wanajo2018,yu2018,ren2019,wu2019}. The multi-messenger observations of this BNS merger officially opened the era of GW-led astronomy, and provided a smoking-gun evidence for the long-hypothesized origin of sGRBs and kilonovae. 

Over the course of the third observing run (O3) of the LIGO Scientific Collaboration and Virgo Collaboration (LVC), several compact binary merger candidates have been announced. Even though many efforts of follow-up observations for GW triggers during O3 \citep{anand2020,antier2020a,antier2020b,coughlin2020a,coughlin2020b,compertz2020,kasliwal2020,page2020} have been carried out, no confirmed EM counterpart, especially kilonova, has been reported since the multi-messenger observations of GW170817\footnote{Some controversial  GW-associated electromagnetic (EM) counterparts have been reported, e.g. a sub-threshold GRB candidate GBM\,190816 occurring at $1.57\,{\rm s}$ after a sub-threshold GW event candidate \citep{goldstein2019,Yang2020}, and a plausible EM flare associated with a candidate binary black hole (BBH) merger S190521g \citep{graham2020P}.}.  {Observationally, due to the poor sky localization and the large distances of these GW events,} the incomplete coverage of the error boxes  {for the follow-up searches} could partially be the reason of non-detection. But it is also likely that previous and present follow-up observations are not deep enough to search for EM counterparts \citep{saguescarracedo2020,coughlin2020a}.  {Moreover, such candidates could occur in the accretion disks of active galactic nucleus \citep[e.g.,][]{mckernan2020,cantiello2020}. GRB jets could be choked by the disk materials and potential shock breakout signals could be fast-evolving and partially outshone by the disk emission \citep{perna2021,zhu2021b,zhu2021a}.}

Kilonovae from BH--NS mergers are another type of multi-messenger sources that astronomers expect to discover. In\defcitealias{zhu2020}{Paper I} \citetalias{zhu2020} \citep{zhu2020}, we have modeled the viewing-angle-dependent lightcurves of BH--NS merger kilonova in detail. In this second paper in the series, we  {attempt to explain the reason for the the lack of EM identification during O3 in theory and} study how likely the BH--NS merger kilonova can be discovered in the future and what would be the best strategy to discover them. In addition, we also take into account the contribution of jet afterglow to study the optical emission detectability. The paper is organized as follows. The physical models are presented in Section \ref{Sec:Model}. Section \ref{Sec:LF} shows our simulated luminosity functions for BH--NS merger kilonovae. The dependence on the BH--NS system parameters and the effect of viewing angle for the kilonova luminosity function are studied in detail. In Section \ref{Sec:ToO}, we simulate the GW detection and subsequent EM follow-up detectability for the networks of 2nd, 2.5th, and 3rd generation GW detectors. Some implications for past and future GW triggered target-of-opportunity observations are also given. The EM detectability for time-domain survey observations is studied in Section \ref{Sec:Time_domain}. We also make calculations of EM detection rates for specific survey projects. Finally, we summarize our conclusions and give discussions in Section \ref{Sec:Conclusions}. A cosmology of $H_0 = 67.8\,{\rm km}\,{\rm s}^{-1}\,{\rm Mpc}^{-1}$, $\Omega_{\rm m} = 0.308$, $\Omega_\Lambda = 0.692$ \citep{planck2016} is applied in this paper.

\section{Modelling}\label{Sec:Model}

In order to simulate the BH--NS merger kilonova luminosity distribution and the detectability of optical EM counterparts from BH--NS mergers, one needs to know (1) the event rate density and redshift distribution of BH--NS mergers in the Universe (Section \ref{Sec:EventRate}); (2) the radiation characteristics of optical EM signals produced after the mergers including both kilonovae and sGRB jet afterglows (Section \ref{Sec:KilonovaandAfterglow}); and (3) the parameter distributions of the BH--NS merger systems and their effects on the EM counterparts (Section \ref{Sec:BHNSdistribution}). We note that the more massive BH--NS systems have stronger GW radiation intensities and also would merge sooner than less massive BH--NS systems. Therefore the redshift and system parameter distributions for the observed BH--NS merger events {should in principle be the result of the convolution} of both system parameters and the intrinsic redshift distribution. In this paper, we ignore the possible redshift evolution of BH--NS system parameters to allow us to separately discuss BH--NS systems parameter distributions and the redshift distibution $f(z)$. We describe our concrete models in the following.

\subsection{Event Rate and Redshift Distribution\label{Sec:EventRate}}

The number density per unit time for BH--NS mergers at a given redshift $z$ can be estimated as
\begin{equation}
\label{Eq. EventRates}
\frac{d\dot{N}}{dz} = \frac{\dot{\rho}_{0} f(z)}{1 + z}\frac{dV(z)}{dz},
\end{equation}
where $\dot{\rho}_{0}$ is the local BH--NS merger rate, and $f(z)$ is the dimensionless redshift distribution factor. The comoving volume element $dV(z)/dz$ in Equation (\ref{Eq. EventRates}) is
\begin{equation}
\frac{dV(z)}{dz} = \frac{c}{H_0}\frac{4\pi D^2_{\rm L}}{(1 + z)^2\sqrt{\Omega_\Lambda + \Omega_{\rm m}(1 + z)^ 3}},
\end{equation}
where $c$ is the speed of light, and $D_{\rm L}$ is the luminosity distance which is expressed as
\begin{equation}
    D_{\rm L} = (1 + z)\frac{c}{H_0}\int_0^{z}\frac{dz}{\sqrt{\Omega_\Lambda + \Omega_{\rm m}(1 + z)^ 3}}.
\end{equation}

There are many previous studies on the local BH--NS merger rate density $\dot{\rho}_0$. For recent binary population synthesis simulations, by considering the constraints of observed compact star binary samples and supernova rates, \cite{oshaughnessy2008} and \cite{abadie2010} predicted that the BH--NS merger event rate density is $\dot{\rho}_0 = 34.8^{+1125}_{-34.2}\,{\rm Gpc}^{-3}\,{\rm yr}^{-1}$ in the local universe. By sampling from compact binary merger catalogs derived from $N$-body and population-synthesis simulations, \cite{santoliquido2020} \citep[see also][]{rastello2020} presented the local event rate density as $\dot{\rho}_0 = 38^{+32}_{-24}\,{\rm Gpc}^{-3}\,{\rm yr}^{-1}$ for BH--NS mergers in young star clusters, and $\dot{\rho}_0 = 45^{+45}_{-32}\,{\rm Gpc}^{-3}\,{\rm yr}^{-1}$ for BH--NS mergers in the field.  {Observationally, the lack of BH--NS merger events detected during the first and second observing runs by LIGO and LVC placed an upper limit of the local BH--NS event rate density of $\dot{\rho}_0 < 610\,{\rm Gpc}^{-3}\,{\rm yr}^{-1}$ \citep{abbott2019gwtc}}. \cite{antier2020b} systematically analyzed the GW candidates discovered during third run by LVC and estimated that the event rates of BH--NS and MassGap mergers are $\sim 27\,{\rm Gpc}^{-3}\,{\rm yr}^{-1}$ and $\sim3.3\,{\rm Gpc}^{-3}\,{\rm yr}^{-1}$, respectively. Based on these theoretical simulation results and observational constraints, we set the local event rate density of BH--NS mergers as $\dot{\rho}_0 \approx 35\,{\rm Gpc}^{-3}\,{\rm yr}^{-1}$ hereafter. 

Compact binary mergers occur with a delay time scale with respect to star formation history. The delay time distribution models include the Gaussian delay model \citep{virgili2011}, log-normal delay model \citep{wanderman2015}, and power-law delay model \citep{virgili2011}.  {The observations of sGRBs support either a Gaussian delay model \citep{virgili2011} or a log-normal delay model \citep{wanderman2015}. For completeness, in Appendix \ref{Sec:EmpiricalFormula} we derive analytical fitting expressions of $f(z)$ to all three delay time distribution models (cf. \citealt{sun2015}). } Hereafter, we adopt log-normal delay model as our merger delay model. 
 {We randomly generate a group of BH--NS events based on Equation (\ref{Eq. EventRates})  following the $f(z)$ function Equation (\ref{Eq: LogNomalDistribution}) for the log-normal delay model.}

\subsection{Properties of Kilonova and Afterglow Emission from BH--NS Mergers\label{Sec:KilonovaandAfterglow}}

\subsubsection{Kilonova Model\label{Sec:KilonovaModel}}

The  energy  that  powers  a  kilonova  comes  from  {the radioactive  decays  of $r$-process nuclei} \citep[e.g.,][]{metzger2010,korobkin2012,tanaka2013,wanajo2014} for different ejecta components. The asymmetric geometry of the ejecta provides viewing-angle-dependent properties of kilonova lightcurves and polarization properties \citep{kasen2015,bulla2019,bulla2020,kawaguchi2020,korobkin2020,zhu2020,darbha2020,kobori2020}. In \citetalias{zhu2020}, we present a detail semi-analytical model of kilonova emission from BH--NS mergers which has dependence on the latitudinal viewing angle $\theta_{\rm view}$ {measured from the $z$-axis} and longitudinal viewing angle $\varphi_{\rm view}$\footnote{{See Figure 7 in \citetalias{zhu2020} for visual definitions of $\theta_{\rm view}$ and $\varphi_{\rm view}$.}}. Three different ejecta components, i.e. tidal dynamical ejecta \citep[e.g.,][]{foucart2014,foucart2017,foucart2019,kyutoku2013,kyutoku2015,kyutoku2018,kawaguchi2015,kawaguchi2016,brege2018}, neutrino-driven wind ejecta \citep[e.g.,][]{fernandez2013,just2015,martin2015,wollaeger2018,perego2017}, and viscosity-driven wind ejecta \citep[e.g.,][]{fernandez2013,fernandez2015,just2015,wu2016,lippuner2017,siegel2017,metzger2019,fujibayashi2020}, are considered. We summary the properties of three ejecta components we assumed in Table \ref{table:ejecta} (For more detail, see Section 3 in \citetalias{zhu2020}).

\begin{deluxetable}{cccc}[tphb]
\tablecaption{Properties of ejecta components\label{table:ejecta}}
\tablecolumns{4}
\tablewidth{0pt}
\tablehead{
\colhead{Ejecta Component} &
\colhead{Mass\tablenotemark{a}} &
\colhead{$Y_e$\tablenotemark{b}} &
\colhead{Direction} 
}
\startdata
Tidal dynamical ejecta & $M_{\rm dyn}$ & $\lesssim0.2$ & Equator\\
Viscosity-driven ejecta & $\xi_{\rm v}M_{\rm disk}$ & $\sim0.15-0.30$ & Equator\\
Neutrino-driven ejecta & $\xi_{\rm n}M_{\rm disk}$ & $\gtrsim0.4$ & Polar
\enddata
\tablenotetext{a}{Calculations for the masses of dynamical ejecta ($M_{\rm dyn}$) and remnant disk ($M_{\rm disk}$) are introduced in Section \ref{Sec:EjectaMassAndBinarySystem}. Neutrino-driven ejecta and viscosity-driven ejecta are assumed to be a certain fraction of the disk mass, i.e., $\xi_{\rm n}\approx0.01$ and $\xi_{\rm v}\approx0.2$. }
\tablenotetext{b}{$Y_e$ is the eletron fraction. }
\end{deluxetable}

For ejecta with different origin, we assign the opacity according to the value of $Y_e$. {Including a large range of $r$-process nuclei in radiative transfer simulations, \cite{tanaka2020} presented that the average effective gray opacities for the mixture of $r$-process elements are nearly a constant for different $Y_e$ values}, i.e., $\kappa\sim20-30\,{\rm cm}\,{\rm g}^{-1}$ for electron fraction $Y_e\leq 0.20$, $\kappa\sim3-5\,{\rm cm}^2\,{\rm g}^{-1}$ for $Y_e = 0.25 - 0.35$, and $\kappa\sim 1\,{\rm cm}^2\,{\rm g}^{-1}$ for $Y_e = 0.4$. We thus simply set the gray opacity of dynamical ejecta, neutrino-driven ejecta, and viscosity-driven ejecta as $\kappa_{\rm d} = 20\,{\rm cm}^2\,{\rm g}^{-1}$, $\kappa_{\rm n} = 1\,{\rm cm}^2\,{\rm g}^{-1}$, and $\kappa_{\rm v} = 5\,{\rm cm}\,{\rm g}^{-1}$, respectively.

\subsubsection{sGRB Afterglow Model \label{Sec:sGRBAfterglowModel}}

 {The remnant BH formed after a BH--NS merger would accrete from the surrounding remnant disk if the NS undergoes tidal disruption and  launches a pair of collimated relativistic jets, possibly along the polar directions via the Blandford$-$Znajek mechanism \citep{blandford1977}}. The relativistic jets would collide with the surrounding interstellar medium to generate a luminous afterglow with emission ranging from radio to X-rays \citep{rees1992,meszaros1993,paczynski1993,meszaros1997,sari1998}. \cite{barbieri2019} noted that the kinetic energy of the jet generated after the BH--NS merger could be estimated as 
\begin{equation}
    E_{\rm K,jet} = \epsilon(1 - \xi_{\rm v} - \xi_{\rm n})M_{\rm disk}c^2\Omega^2_{\rm H}f(\Omega_{\rm H}),
\end{equation}
where $\epsilon = 0.015$, $\Omega_{\rm H}$ is the dimensionless angular frequency at the horizon, which is determined by the final spin of the remnant BH $\chi_{\rm BH,f}$, i.e., $\Omega_{\rm H} = {\chi_{\rm BH,f}}/{2\left(1 + \sqrt{1 - \chi_{\rm BH,f}^2}\right)} $
and $f(\Omega_{\rm H}) = 1 + 1.38\Omega_{\rm H}^2 - 9.2\Omega_{\rm H}^4$ is a correction factor for high-spin values \citep{tchekhovskoy2010}. The final spin of BH $\chi_{\rm BH,f}$, which depends on the initial BH--NS system parameter, is calculated using Equation (11) from \cite{pannarale2013}\citep[see also][]{Deng2020}.

The observations of late-time multiband afterglow associated with GW170817 revealed that the sGRB jet plausibly carries an angular structure \citep{alexander2017,davanzo2018,gill2018,ghirlanda2019,lamb2017,lazzati2018,lyman2018,margutti2018GW170817,mooley2018,ren2020,troja2018,tan2020,xie2018}. Here, we consider a power-law jet model \citep{rossi2002,zhang2002}, where the angular distributions of the kinetic energy and Lorentz factor are adopted as \citep{ghirlanda2019,salafia2019}:
\begin{equation}
\begin{split}
    \frac{dE}{d\Omega}(\theta) &= \frac{E_{\rm c}/4\pi}{1 + (\theta / \theta_{\rm c})^{s_1}} , \\
    \Gamma(0 , \theta) &= 1 + \frac{\Gamma_{\rm c} - 1}{1 + (\theta / \theta_{\rm c})^{s_2}},
\end{split}    
\end{equation}
where we set the core Lorentz factor $\Gamma_{\rm c} = 250$, core half-opening angle $\theta_{\rm c} = 3.7^\circ$, energy power law slope $s_1 = 5.5$ and Lorentz factor slope $s_2 = 3.5$ based on the observational constraints of GW170817/GRB\,170817A, and the core isotropic equivalent kinetic energy $E_{\rm c} = E_{\rm K,jet}/\pi\theta_{\rm c}^2$.

The standard synchrotron emission spectra from the relativistic electrons are employed following \cite{sari1998}. For more details of the afterglow model we used to calculate the sGRB lightcurves along the line of sight, one can see Appendix C in \citetalias{zhu2020} \citep[for reviews of afterglow models see][]{gao2013,zhang2018}.

\subsection{Parameter Distributions\label{Sec:BHNSdistribution}}

\subsubsection{Ejecta Mass and BH--NS System Parameters \label{Sec:EjectaMassAndBinarySystem}}

The lightcurve of a BH--NS kilonova mainly depends on the amount of the masses of the three ejecta components. In Section \ref{Sec:KilonovaModel}, we mentioned that the masses of the neutrino-driven wind ejecta and the viscosity-driven wind ejecta could simply be considered as a constant fraction of the remnant disk mass. We model the dynamical ejecta mass $M_{\rm dyn}$ based on Equation (7) in \citetalias{zhu2020} \citep[see also][for other fitting model]{kawaguchi2016,kruger2020} and the disk mass $M_{\rm disk}$ given by Equation (4) from \cite{foucart2018} \citep[see also][]{foucart2012} which are associated with the location of the radius at which the tidal disruption occurs $R_{\rm tidal}$ and the radius of the innermost stable circular orbit (ISCO) $R_{\rm ISCO}$ \citep{kyutoku2011,shibata2011} (see more details in Section 2 of \citetalias{zhu2020}). The mass of three components are {thus a function of} the BH mass $M_{\rm BH}$, the BH dimensionless spin parameter $\chi_{\rm BH}$, the NS mass $M_{\rm NS}$, and the NS EoS. The parameter distributions of the BH--NS binary systems would determine detection results of the kilonovae. Besides, the kinetic energy of the jet we modeled, affected on the afterglow lightcurve, is dependent on the disk mass $M_{\rm disk}$ and the final spin of the BH $\chi_{\rm BH,f}$. Therefore, the afterglow lightcurve is also determined by BH--NS system parameters. Furthermore, one should also consider the distribution of the viewing angles since both afterglow and kilonova emissions are highly viewing-angle-dependent. 

 {So far, LVC only reported a candidate BH--NS GW event during O3a \citep{Abbott2020gwtc2}. Also no BH--NS system has been discovered in our galaxy. We therefore cannot consider BH--NS system parameters based on observations.} Here, we assume the following independent distributions of several properties for BH--NS systems based on some models and simulation results as described below.

\begin{figure}
    \centering
    \includegraphics[width = 0.99\linewidth , trim = 70 30 95 60, clip]{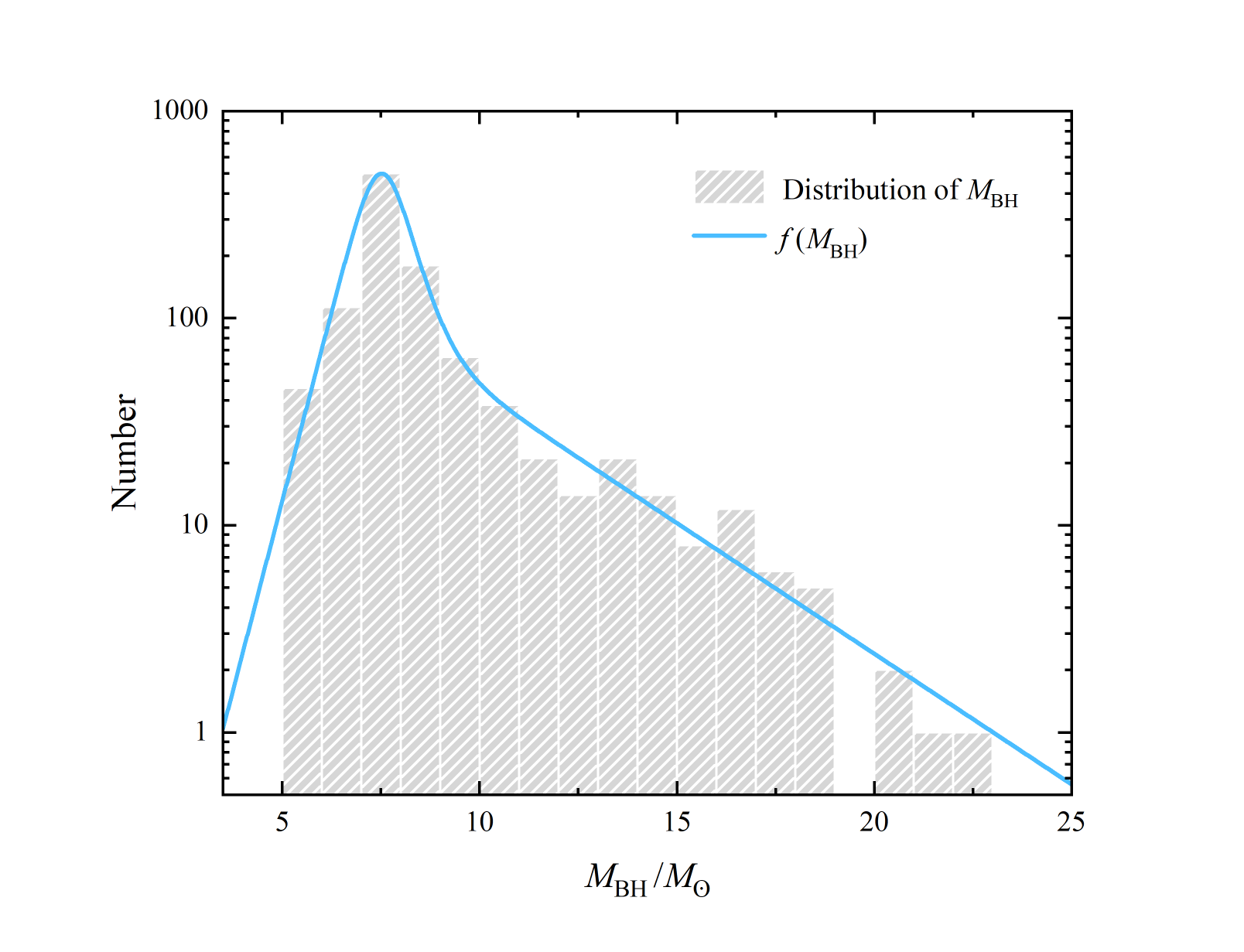}
    \caption{Distribution of the BH mass from BH--NS systems. The gray bar chart represents the BH mass distribution of population-synthesis simulations $\alpha3$ with metallicity $0.008\leq Z \leq 0.02$ presented in \cite{giacobbo2018}. The blue line is our best fit probability density function of the BH mass distribution.}
    \label{fig:BHMassDistribution}
\end{figure}

1. {\em BH mass} --- The probability density function of BH mass distribution depends on the progenitor’s metallicity $Z$, the common-envelope parameter $\alpha$ and the natal kick. However, population-synthesis simulations from \cite{giacobbo2018,mapelli2018} showed that most BH masses in merging BH--NS systems are $<10\,M_\odot$ and are mainly distributed in the range from $\sim 5\,M_\odot$ to $\sim 8\,M_\odot$ for different conditions of the BH--NS binaries progenitors. So we can define a specific probability density function of BH mass distribution. Observationally, the metallicity of sGRB host galaxy spans in the range of $0.6\,Z_\odot- 1.6\,Z_\odot$ where $Z_\odot = 0.012$ \citep{berger2009,berger2014,davanzo2009}. Here, in order to obtain our probability density function of BH mass distribution, we assume a formula with an exponential rise and two broken exponential decays \citep[see][{for a theoretical exponential decay}]{fryer2001,ozel2010,steiner2012} to fit the $\alpha3$ model with metallicity $0.008\leq Z \leq 0.02$ presented in \cite{giacobbo2018} as our BH mass distribution. The best fitting results of the probability density function showed in Figure \ref{fig:BHMassDistribution} is
\begin{equation}
\begin{split}
    f(M_{\rm BH}) &\propto \left(
    \frac{1}{a_1\exp(-b_1M_{\rm BH})+a_2\exp(-b_2M_{\rm BH})}\right. \\
    &\left.+\frac{1}{a_3\exp(b_3 M_{\rm BH})} \right)^{-1},\, {\rm for}\,M_{\rm BH} > M_{\rm TOV}, 
\end{split}
\end{equation}
where $M_{\rm TOV}$ is the NS TOV mass (non-rotating NS maximum mass) and the best fit values of each parameter are $a_1 = 1.04\times10^{11}$, $b_1 = 2.1489$, $a_2 = 799.1$, $b_2 = 0.2904$, $a_3 = 0.002845$, and $b_3 = 1.686$. According to the rapid core-collapse SN model by \cite{fryer2012}, \cite{giacobbo2018} set a minimum possible mass of a BH as $\sim5\,M_\odot$, which is the reason of a sharp cut-off in Figure \ref{fig:BHMassDistribution}. However, the existence of low-mass BHs cannot be theoretically ruled out \citep[e.g.,][]{Drozda2020}. We extend the minimum mass limit of our model to $M_{\rm TOV}$, although the region of $M_{\rm TOV} < M_{\rm BH} < 5M_\odot$ only occupies a very small portion.

2. {\em BH spin} --- It could be plausibly presumed that the orientation of the BH spin is nearly aligned with the orbit due to the direction of the BH spin could retain some memory of its birth orientation and be affected by torquing of the accretion disk \citep{miller2015}. So we assume that the spin of the primary BH is almost aligned with the orbit. The measurement of GWs from binary black holes (BBHs) by LVC showed that the majority of the observational BBH systems have a low effective spin \citep{abbott2019gwtc}, i.e., $\chi_{\rm eff} \sim 0$, which indicated that the spin of each BH in BBH systems could also be low. In consideration of possible similar formation paths between BH--NS and BBH systems \citep[e.g.,][]{colpi2017}, the spin of BH in BH--NS systems could have similar properties. In addition, \cite{qin2018,fuller2019} predicted that most BHs born from single stars would rotate very slowly, indicating that the BH in the BH--NS system could have a low spin distribution. On the other hand, the BH spin can be indirectly measured by fitting the Fe $K\alpha$ line profiles or the continuum spectra from the accretion disks in black hole X-ray binary systems. The spin of the BH in these systems is usually high which is concentrated in the range of $\chi_{\rm BH} \in [0.7, 0.98]$ \citep{qin2019,mcclintock2011,miller2011}. Since part of X-ray binaries would evolve into BH--NS binaries, the BHs in BH--NS systems could also have high-spin distributions. The uncertainties in theory encourage us to define two simple cases for the BH spin distribution, i.e., $\chi_{\rm BH}\sim\mathcal{N}(0 , 0.15^2)$ and $\chi_{\rm BH}\sim\mathcal{N}(0.8 , 0.15^2)$, for the low-spin case and the high-spin case, respectively,  {where $\mathcal{N}$ represents normal distribution}. We assume that $\chi_{\rm BH}$ is in the range of $\chi_{\rm BH}\in[0 , 0.99]$.

3. {\em NS EoS} --- As listed in Table \ref{table:1}, we select three representative EoSs. From soft to stiff, they are AP4 \citep{akmal1997}, DD2 \citep{typel2010}, and Ms1 \citep{muller1996}. Among them, AP4 is supported as the most possible EoS while DD2 is the stiffer EoS allowed by present constraints \citep[e.g.,][]{gao2016,abbott2019properties}. \cite{gao2020} presented a relation between NS mass $M_{\rm NS}$ and baryonic mass $M_{\rm NS}^{\rm b}$ which reads
\begin{equation} 
\label{Eq: CalculateBaryonicMass}
    M_{\rm NS}^{b} = M_{\rm NS} + A_1\times M_{\rm NS}^2 + A_2\times M_{\rm NS}^3,
    \end{equation}
where $M_{\rm NS}^{\rm b}$ and $M_{\rm NS}$ in this formula are in units of $M_\odot$. The best fit values of $A_1$ and $A_2$ for each EoS are collected in Table \ref{table:1}. For each EoS, the compactness of the NS can be expressed as \cite{coughlin2017}
\begin{equation}
    C_{\rm NS} =  1.1056\times(M_{\rm NS}^{\rm b} / M_{\rm NS} - 1)^{0.8277}.
\end{equation}
During the inspiral phase of a BH--NS merger, the matter effect coming from the tidal deformation could enhance GW emission and accelerate the coalescence \citep{gao2016,flanagan2008,damour2012measurability}. Therefore, the amount of tidal deformation, which is described by the dimensionless tidal deformalibity of the NS, i.e., $\Lambda = (2/3)k_2C_{\rm NS}^{-5}$ with $k_2$ is the Love number, depends on the NS mass. The EoS could thus affect the GW signal  {and $\Lambda$ will be an input parameter to calculate the GW signals used in Section \ref{sec:GWDetectionMethod}}.

\begin{deluxetable}{ccccc}[t!]
\tablecaption{Characteristic Parameters for Various EoSs\label{table:1}}
\tablecolumns{5}
\tablewidth{0pt}
\tablehead{
\colhead{EoS} &
\colhead{$M_{\rm TOV}/M_\odot$} &
\colhead{$R_{\rm 1.4} / {\rm km}$} &
\colhead{$A_1$} &
\colhead{$A_2$}
}
\startdata
AP4 & 2.22 & 11.36 & 0.045 & 0.023 \\
DD2 & 2.42 & 13.12 & 0.046 & 0.014 \\
Ms1 & 2.77 & 14.70 & 0.042 & 0.010
\enddata
\tablecomments{The columns from left to right represent the NS EoSs we selected, the NS TOV mass $M_{\rm TOV}$, the NS radius for $1.4M_\odot$, and the best fit values of $A_1$ and $A_2$ for each EoS in Equation (\ref{Eq: CalculateBaryonicMass}). All of the values of each parameter are cited from \cite{gao2020}}
\end{deluxetable}

4. {\em NS mass} --- We assume that the distribution of the NS mass in BH--NS systems is approximately the observationally derived normal distribution of galactic BNS systems, i.e., $M_{\rm NS}/M_\odot\sim \mathcal{N}(1.33, 0.11^2)$ \citep{lattimer2012,kiziltan2013}. In addition, the population-synthesis simulation results from \cite{mapelli2018} and \cite{giacobbo2018} predicted that the masses of most of NSs in BH--NS systems are possibly larger than the NSs in BNS systems, with the median value as $\sim 1.6\,M_\odot$. In order to consider the effect of different NS mass distribution, we assume another normal distribution $M_{\rm NS}/M_\odot\sim \mathcal{N}(1.6, 0.11^2)$ for specific DD2 NS EoS. The range of NS mass adopted here is $M_{\rm NS}/M_\odot \in [1.1,M_{\rm TOV}]$, where the lower boundary $1.1\,M_\odot$ is the minimum mass of a NS formed from a core-collapse supernova based on the rapid supernova model of \cite{fryer2012}.

5. {\em Viewing angles} --- The distributions of latitudinal viewing angle $\sin\theta_{\rm view}$\footnote{{Latitudinal viewing angle is randomly distributed, so it is proportional to $N(\theta_{\rm view})\propto d\Omega \propto\sin\theta d\theta$ \citep[e.g.,][]{zhang2002}}} and longitudinal viewing angle $\varphi_{\rm view}$ are adopted to be uniform, so we set $\sin\theta_{\rm view} \sim \mathcal{U}(0 , 1)$ and $\varphi_{\rm view} \sim \mathcal{U}(0 , \pi)$,  {where $\mathcal{U}$ represents the uniform distribution}..

\subsubsection{Afterglow}

Except the redshift distribution, the properties of the observed sGRB afterglow emission are mainly affected by the the isotropic equivalent kinetic energy $E_{\rm c}$, the power-law index of the electron distribution $p$, the density of circumburst environments $n$, the shock microphysics parameters $\varepsilon_e$ and $\varepsilon_B$, and the latitudinal viewing angle $\theta_{\rm view}$ whose distribution has been obtained. For $E_{\rm c}$, we assume that it would be dependent on the BH--NS binary system parameters, which has been introduced in Section \ref{Sec:sGRBAfterglowModel}. We fix $p = 2.15$ and $\varepsilon_e = 0.1$, since these parameters have relatively narrow distributions from the GRB afterglow modeling \citep[e.g.,][]{santana2014,wang2015}. Therefore, only $n$ and $\varepsilon_B$ remain free parameters that can significantly effect the brightness of the observed afterglow emission.

The ranges of $n$ and $\varepsilon_B$ distributions are much wider in the literature. By systematically analysing a sample of $38$ well-observed broadband sGRB afterglow lightcurves, \cite{fong2015} found that sGRBs usually occur in low-density environments, with a median density of $n\approx(3-15)\times 10^{-3}\,{\rm cm}^{-3}$, and $\lesssim16\%$ of sGRBs in their sample took place at densities $\lesssim1\,{\rm cm}^{-3}$. Moreover, \cite{oconnor2020} found that the medium densities of $\lesssim 16 \%$ sGRBs have $n\lesssim 10^{-4}\,{\rm cm}^{-3}$ by utilizing upper limits on the deceleration time and lower limits on the peak X-ray flux with a sample of $52$ sGRB X-ray afterglow observations. For the limit of $\varepsilon_B$, \cite{santana2014} performed a systematic study of 60 sGRBs X-ray afterglows by constraining the observed flux at the end of the steep decay phase. Setting a constant density $n = 1\,{\rm cm}^{-3}$, they showed that $\varepsilon_B$ has a wide range of $\sim 10^{-8}-10^{-3}$ and is centered at $\sim {\rm few}\times10^{-5}$. Similar results were obtained by \cite{wang2015}. Despite the constraints of $\varepsilon_B$ are usually given in the case of a given $n$, $n$ and $\varepsilon_B$ are actually independent physical parameters. So, based on these observational constraints, we simply assume that they have independent log-normal distributions as follow:

1. {\em Density of circumburst environments} --- We assume its distribution as $\log(n/{\rm cm}^{-3})\sim \mathcal{N}(-2.0 , 1.1 ^ 2)$.

2. {\em Fraction of shock energy carried by magnetic field } --- We assume its distribution as $\log\varepsilon_B\sim\mathcal{N}(-3.0 , 1.0^2)$.

\section{Luminosity Function \label{Sec:LF}}

{Kilonovae are thought to be fast-evolving with bolometric lightcurve peak times distributed from $0.2\,{\rm days}$ to $1\,{\rm day}$ \citep[e.g.,][]{korobkin2020}}. Early fast evolution of kilonovae makes it difficult to observationally catch the lightcurve peak times and reconstruct the bolometric lightcurve near the peak. With the consideration of actual kilonova searches, we adopt the absolute magnitude at $0.5\,{\rm days}$ post-merger, e.g. the first detection times of GW170817/AT\,2017gfo \citep{abbott2017multimessenger,coulter2017} and the possible kilonova associated with GRB\,070809 \citep{jin2020}, rather than the peak absolute magnitude, to define the kilonova luminosity function $\Phi(M)$.  {Here, $M$ represents the absolute magnitude at $0.5\,{\rm days}$ after the merger.} Traditionally, the luminosity function $\Phi(M)$ of BH--NS merger kilonovae is defined as $\Phi(M)dM \propto \dot{N}(M)dM$, with the integration of $\Phi(M)$ normalized to unity, i.e., 
\begin{equation}
\label{Eq: LFIntegration}
    \int_{M_{\rm max}}^{M_{\rm min}}\Phi(M)dM = 1.
\end{equation}
However, the luminosity function $\Phi(M)$ cannot show the relative fraction of the case where the NS directly plunges into the BH without a kilonova out of all BH--NS mergers by considering the parameters distributions of different BH--NS system, so we look for other alternative parameters to display $\Phi(M)$.

We define the local event rate density per unit absolute magnitude for kilonova emission from BH--NS mergers as 
\begin{equation}
    \dot{\rho}_{0,M}=\frac{d\dot{\rho}_0}{dM}.
\end{equation}
Therefore, the number density in the absolute magnitude interval from $M$ to $(M+dM)$ can be obtained by Equation (\ref{Eq. EventRates}), i.e.,
\begin{equation}
    \frac{d\dot{N}}{dzdM} = \frac{\dot{\rho}_{0,M} f(z)}{1 + z}\frac{dV(z)}{dz}.
\end{equation}
 {Suppose that $\Delta\dot{N}$ events are detected in a finite absolute magnitude bin from $M$ to $(M + \Delta M)$, we then have}
\begin{equation}
\label{Eq: LFNumber}
    \dot{\rho}_{0,M} \simeq \frac{\Delta \dot{N}}{\Delta M}\left[\int_0^\infty\frac{f(z)}{1 + z}\frac{dV(z)}{dz}dz\right]^{-1}.
\end{equation}
The local event rate density is therefore 
\begin{equation}
\begin{split}
\label{Eq: LFEventRates}
    \dot{\rho}_0 &= \frac{1}{\xi}\int^{M_{\rm min}}_{M_{\rm max}}\dot{\rho}_{0,M}dM  \\
    &\simeq \frac{1}{\xi} \sum_{M_{\rm max}}^{M_{\rm min}} \frac{\Delta \dot{N}}{\Delta M}\left[\int_0^\infty\frac{f(z)}{1 + z}\frac{dV(z)}{dz}dz\right]^{-1},
\end{split}
\end{equation}
{where $\xi$ is introduced to consider {the fraction of BH--NS mergers for which tidal disruption can occur} and, hence, kilonovae can be produced}, $M_{\rm min}$ and $M_{\rm max}$ are the minimum and maximum values of the absolute magnitude, respectively. 

With the definitions in Equation (\ref{Eq: LFIntegration}) and (\ref{Eq: LFEventRates}), the kilonova luminosity function can be derived by displaying the local event rate density per unit absolute magnitude $\dot{\rho}_{0,M}/\dot{\rho}_0\propto\Phi(M)$ as a function of $M$. In view of the relation between absolute magnitude $M$ and bolometric luminosity $L$, i.e., $\Delta M = 2.5\Delta \log L$, one can also map $\dot{\rho}_{0,\log L}/\dot{\rho}_0$ as a function of $L$ to replace $\dot{\rho}_{0,M}/\dot{\rho}_0$. Similar to the concerned absolute magnitude, $L$ should be the kilonova bolometric luminosity at $0.5\,{\rm days}$ after a BH--NS merger.  {Moreover, we also show the $griJ$-band luminosity function $\dot{\rho}_{0,M_{\rm X}}/\dot{\rho}_0$ of BH--NS kilonovae, where $X = g,r,i,J$, since $griJ$ bands are frequently used survey bands for the present and future survey telescopes. Here, we assume that the kilonovae peaking in multi-band can be detected, since the peak times of multi-band lightcurves are always reached after $0.5\,{\rm days}$ post mergers, and hence $M_{\rm X}$ are defined as the absolute magnitude at the peak time in the $X$ band.}

\begin{figure}
    \centering
    \includegraphics[width = 0.99\linewidth , trim = 60 130 40 20, clip]{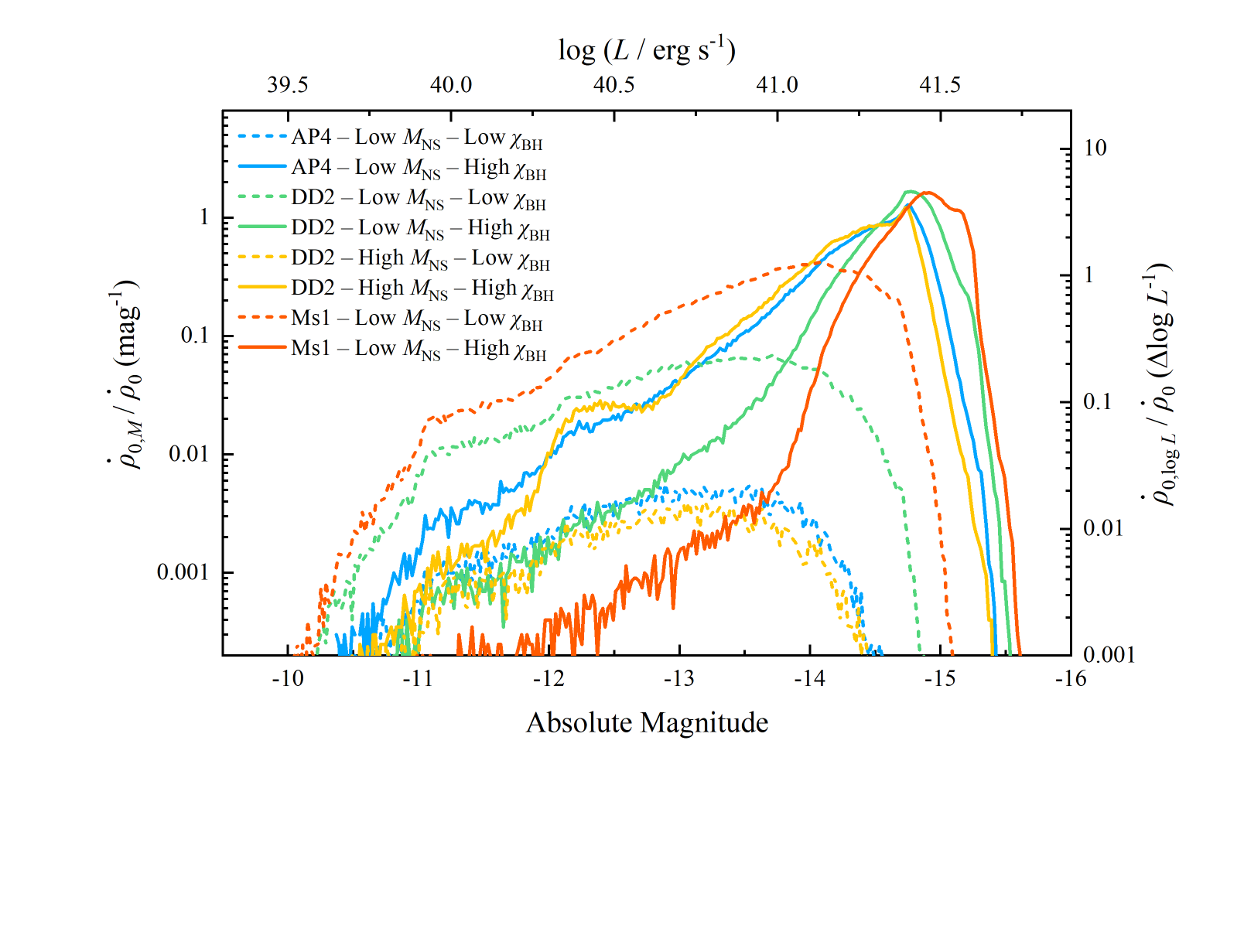}
    \caption{Kilonova luminosity functions for different parameter distributions of BH--NS system. The colored dashed lines and solid lines denote BHs in BH--NS systems having a low-spin $\chi_{\rm BH}$ distribution and a high-spin $\chi_{\rm BH}$ distribution, respectively. The blue, yellow and orange lines represent three selected EoSs, i.e., AP4, DD2, and Ms1, with the Galactic NS mass (low $M_{\rm NS}$) distribution, while the green lines represent DD2 with high NS mass $M_{\rm NS}$ distribution predicted by \cite{mapelli2018} and \cite{giacobbo2018}.} 
    \label{fig:LF}
\end{figure}

\begin{figure}
    \centering
    \includegraphics[width = 0.99\linewidth , trim = 65 45 100 65, clip]{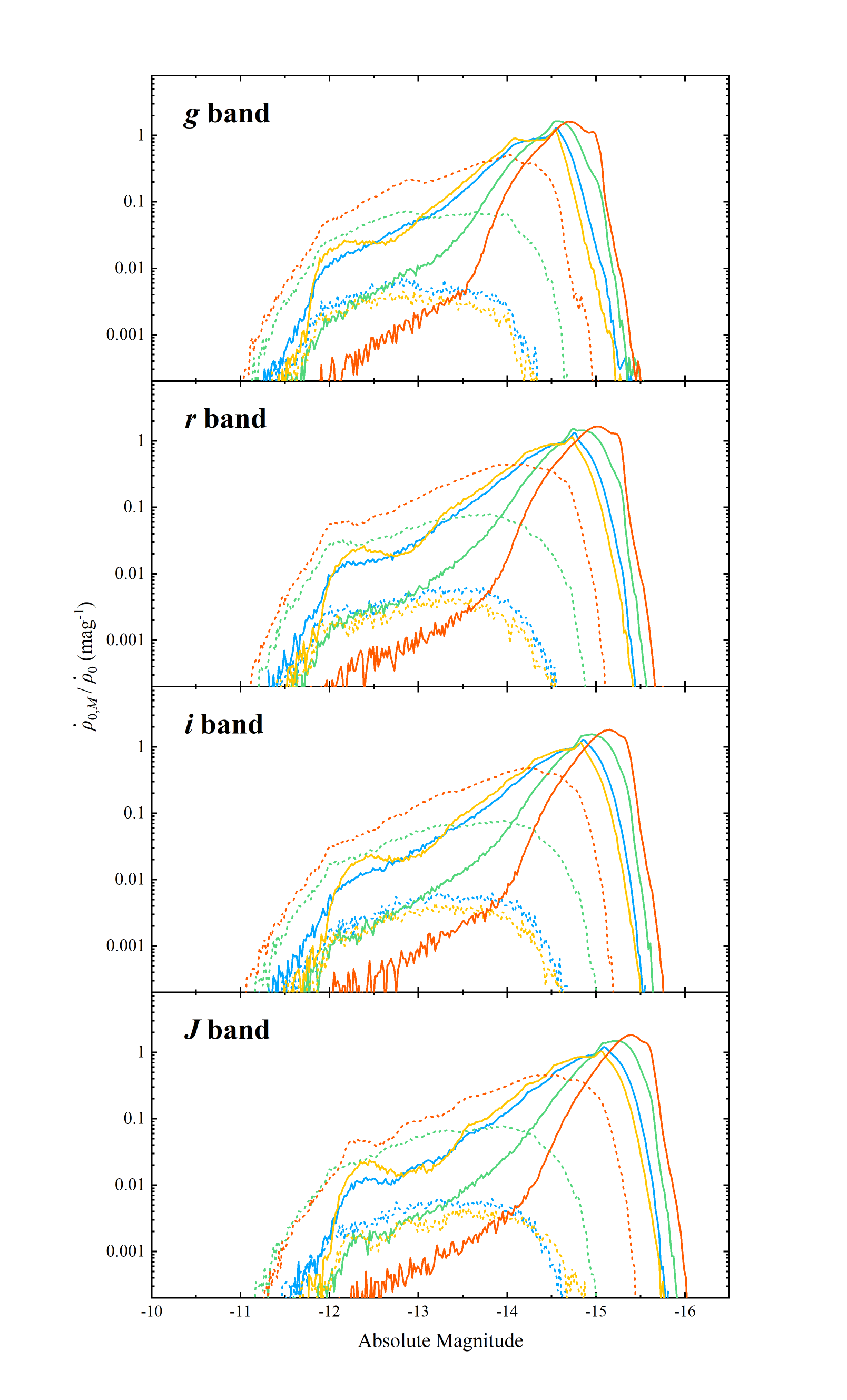}
    \caption{Kilonova luminosity functions in different bands. The label of colored lines are similar to those of lines in Figure \ref{fig:LF}.} 
    \label{fig:LFband}
\end{figure}

\subsection{Dependence on BH--NS System Parameters\label{Sec:LFSystemParameter}}

In Section \ref{Sec:BHNSdistribution}, we presented the variable parameter distributions for the NS EoS, NS mass, and BH spin in BH--NS systems. Hereafter, we consider 6 groups of parameter distributions containing a low-mass NS, 3 kinds of NS EoSs (AP4, DD2, and Ms1), and 2 kinds of BH spin distributions (Low $\chi_{\rm BH}$ and High $\chi_{\rm BH}$). In addition, for the DD2 EoS, we assume another two sets of massive NS distributions, so we make a total of 8 groups of BH--NS system parameter distributions to simulate the kilonova luminosity functions. Each group of simulation is named following the convention of `EoS$-$NS mass distribution$-$BH spin distribution' which is shown in the label of Figure \ref{fig:LF}. 

Following Equation (\ref{Eq: LFNumber}), we simulate $1\times10^6$ BH--NS mergers in the universe and map the relevant $\dot{\rho}_{0,M}/\dot{\rho}_0$ as a function of $M$ ($\dot{\rho}_{0,\log L}/\dot{\rho_0}$ as a function of $L$) shown in Figure \ref{fig:LF} for these eight different distributions of system parameters. Altogether, the results indicate that the kilonova absolute magnitude $M$ at $0.5\,{\rm days}$ postmerger for BH--NS mergers is mainly distributed in the range of $\sim-10$ to $\sim-15.5$ (or $\sim3\times10^{39}\,{\rm erg}\,{\rm s}^{-1}$ to $\sim5\times10^{41}\,{\rm erg}\,{\rm s}^{-1}$ for bolometric luminosity $L$). {The multi-band luminosity functions have similar results with bolometric luminosity functions as shown in Figure \ref{fig:LFband}. Since BH--NS kilonovae could be lanthanide-rich \citep[e.g.,][]{zhu2020,kawaguchi2020}, the infrared magnitudes would be brighter by $\sim0.5-1\,{\rm mag}$ than the optical magnitudes.}

The diversity of parameter distributions could be directly reflected in the distribution of the absolute magnitudes. Obviously, among these three variable parameters, the BH spin distribution has the largest impact on the kilonova luminosity distribution. As mentioned in Section \ref{Sec:EjectaMassAndBinarySystem}, whether the NS is directly plunged into the BH or tidally disrupted by the BH is determined by the comparison between two radii: one is the radius at which the tidal disruption occurs $R_{\rm tidal}$ (which is related to NS mass $M_{\rm NS}$ and NS compactness $C_{\rm NS}$), and the other is the ISCO radius $R_{\rm ISCO}$ (which is dependent on BH mass $M_{\rm BH}$ and dimensionless spin parameter $\chi_{\rm BH}$). If $R_{\rm tidal} \gtrsim R_{\rm ISCO}$, the NS would plunge entirely into the BH without kilonova emission. If $R_{\rm tidal} \lesssim R_{\rm ISCO}$, the NS would be partially disrupted to so that neutron-rich materials can be ejected to power a bright kilonova. For a fixed NS mass and EoS, the BH with a lower spin has a larger $R_{\rm ISCO}$ which cause the NS directly plunged for most cases. This is why the overall luminosity functions $\dot{\rho}_{0,M}/\dot{\rho}_0$ for low BH spin distribution shown in Figure \ref{fig:LF} and \ref{fig:LFband} have lower values compared to those for high BH spin distribution. Tidal disruption would be significant if the BH has a higher spin, so that more material could be ejected to power a brighter kilonova. For fixed BH parameters, a smaller NS mass and a stiffer EoS are required to have tidal disruption. Because a less massive NS is less compact and a stiffer EoS can support a larger NS radius, introducing a larger $R_{\rm tidal}$ makes it easier to tidally disrupt the NS and release more neutron-rich material. As shown in Figure \ref{fig:LF} and \ref{fig:LFband}, BH--NS kilonovae tend to be more luminous if the NS has a lower mass with a stiffer EoS.

One can conclude that the kilonova absolute magnitude at $0.5\,{\rm days}$ after a BH--NS merger is mainly distributed in the range of $\sim -10$ to $\sim -15.5$, corresponding to the bolometric luminosity $L$ from $\sim3\times10^{39}\,{\rm erg}\,{\rm s}^{-1}$ to $\sim5\times10^{41}\,{\rm erg}\,{\rm s}^{-1}$. The kilonova luminosity function is highly dependent on the BH--NS system parameter distributions. In the case of less massive NSs, stiffer EoSs, and higher BH spins in particular, NSs can be tidally disrupted significantly while BH--NS merger kilonovae have more luminous distributions.

\subsection{Viewing Angle Effect}

\begin{figure*}
    \centering
    \includegraphics[width = 0.49\linewidth , trim = 70 100 45 10, clip]{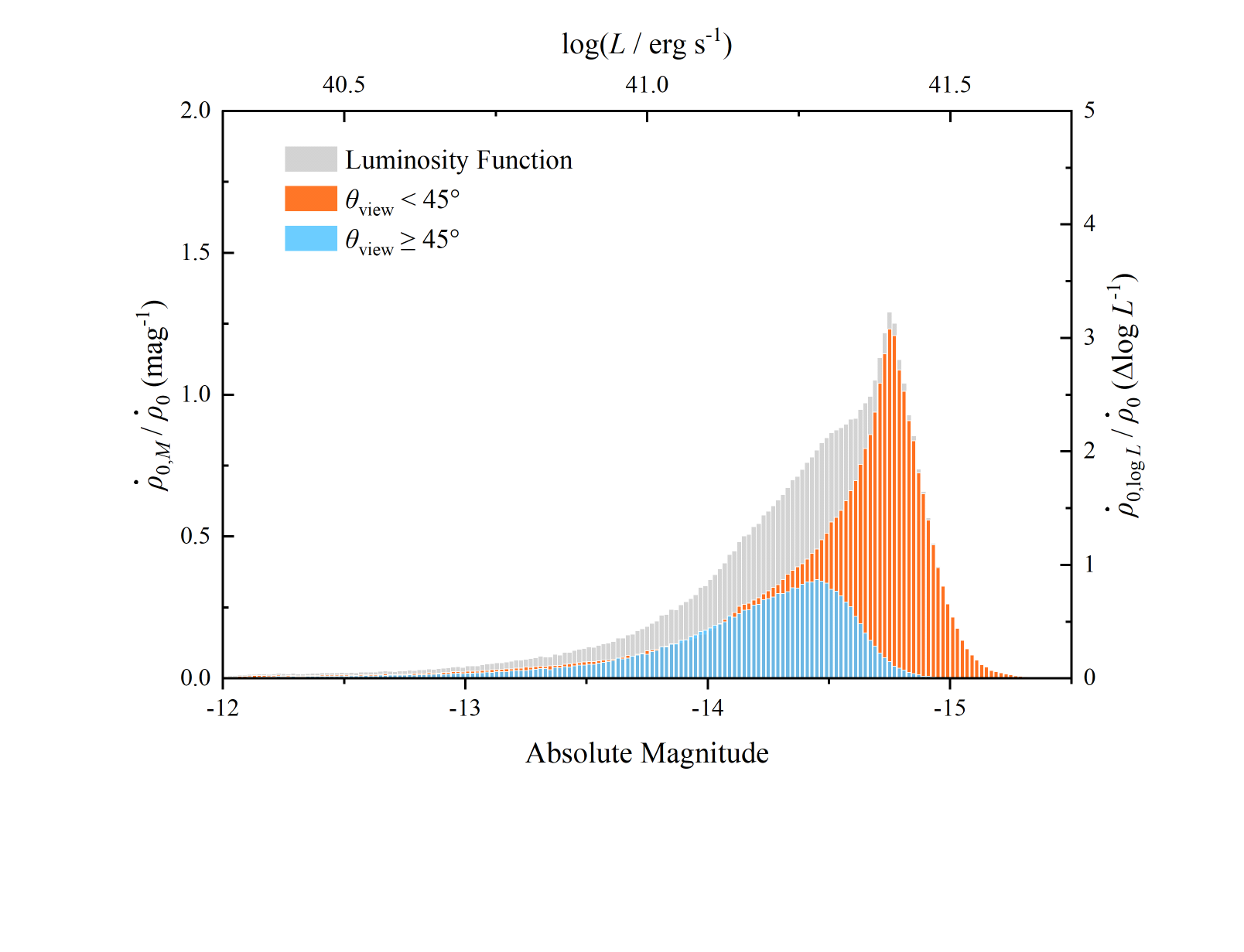}
    \includegraphics[width = 0.49\linewidth , trim = 70 100 45 10, clip]{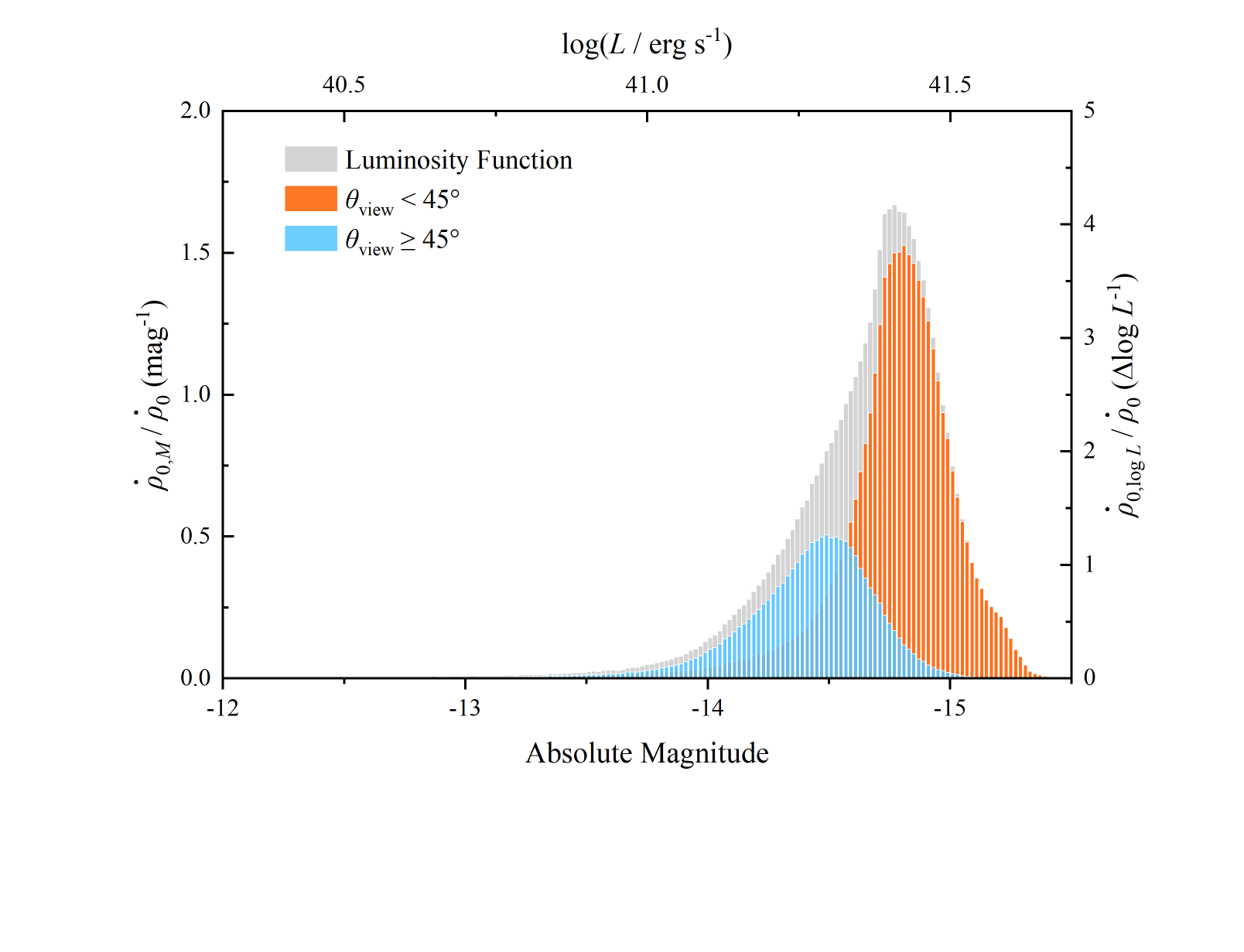}
    \caption{Examples of BH--NS kilonova luminosity function. The gray histograms in the left panel and right panel denote the luminosity functions for two different parameter distributions, i.e., AP4$-$Low $M_{\rm NS}-$High $\chi_{\rm BH}$ and DD2$-$Low $M_{\rm NS}-$High $\chi_{\rm BH}$. The red and blue histograms in each panel represent the luminosity functions with latitudinal viewing angle $\theta_{\rm view} < 45 ^\circ$ and $\theta_{\rm view} \geq 45 ^ \circ$, respectively. } 
    \label{fig:LFViewingAngle}
\end{figure*}

Observers tend to model the luminosity function for a certain type of transient by its observed luminosity. In \citetalias{zhu2020}, we showed that the variation of the observed luminosity for radioactivity-powered kilonova emission from BH--NS mergers is mainly caused by the change of the projected photosphere area of the dynamical ejecta, which causes a difference of only a factor of $\sim(2-3)$ of the observed luminosity between the face-on view ($\theta_{\rm view}\sim 0^\circ$) with the largest projected photosphere area and the edge-on view ($\theta_{\rm view}\sim 90^\circ$) with the smallest projected photosphere area. Since  kilonova emission is highly viewing-angle-dependent, its luminosity function should contain information of the viewing angle distribution. 

In Figure \ref{fig:LFViewingAngle}, we select two two different BH--NS system parameter distributions from Figure \ref{fig:LF}. The example luminosity functions contain the potential viewing angle distribution information of anisotropic kilonova emission. To see how viewing angles contribute to the final luminosity function, we also separately mark the contributions from latitudinal viewing angles  $\theta_{\rm view}<45^\circ$ and $\theta_{\rm view}\geq45^\circ$, respectively, in Figure \ref{fig:LFViewingAngle}. One can see that the kilonova luminosity function is a superposition of a high-luminosity subset with smaller latitudinal viewing angles and low-luminosity subset with larger latitudinal viewing angles. The ratio between the peak luminosities of the two subsets is $\sim 1.4$, corresponding to the difference in absolute magnitude by $\sim 0.5\,{\rm mag}$.

\section{Detectability for GW-triggered Target-of-opportunity Observations \label{Sec:ToO}}

\begin{figure*}
    \centering
    \includegraphics[width = 0.99\linewidth , trim = 0 150 0 0, clip]{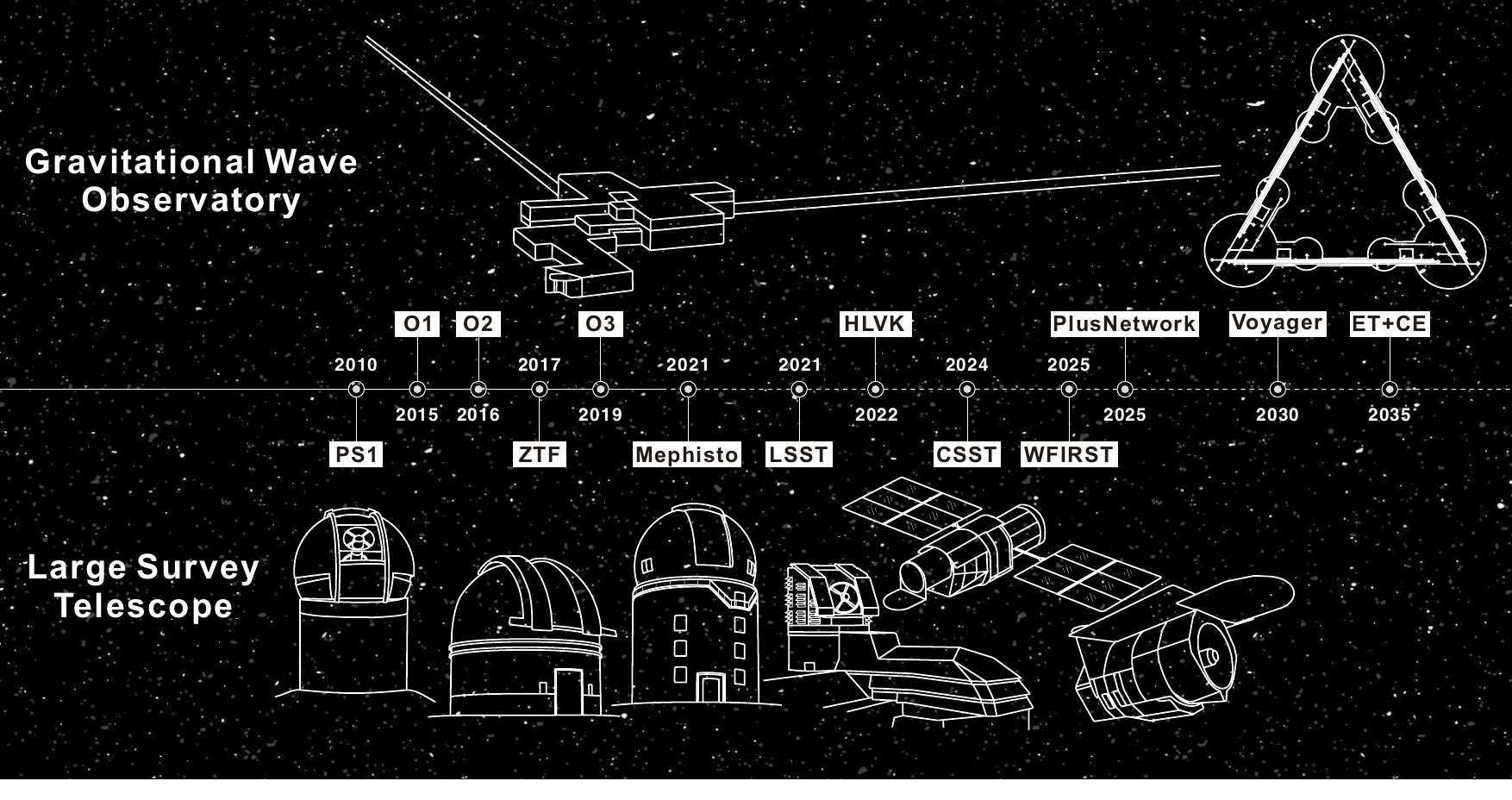}
    \caption{Timeline when GW observatories and large optical survey telescopes start operation.} 
    \label{fig:Timeline}
\end{figure*}

Figure \ref{fig:Timeline} shows the timeline of GW observations (above the line), i.e. the starting times of the past LIGO and LVC observing runs and future networks of 2nd, 2.5th, and 3rd generation detector (see Section \ref{sec:GWDetectionMethod} for details), and EM observations (below the line), i.e. the starting times of large survey telescopes including the Panoramic Survey Telescope and Rapid Response System \citep[PS1,][]{kaiser2010}, the Zwicky Transient Facility \citep[ZTF,][]{graham2019}, the Multi-channel Photometric Survey Telescope\footnote{See \url{http://www.swifar.ynu.edu.cn/kxyj_Science/wyjyz_Telescope_Development.htm}.} \citep[Mephisto,][]{er2020}, the Large Synoptic Survey Telescope \citep[LSST,][]{lsst2009}, the Chinese Space Station Telescope \citep[CSST,][]{zhan2011}, and the Wide Field Infrared Survey Telescope\footnote{WFIRST has been newly named as Nancy Grace Roman Space Telescope.}\citep[WFIRST,][]{spergel2015}. With the upgrade and iteration of GW observatories, numerous compact binary mergers from the distant universe will be discovered. Meanwhile, many planned large survey telescopes will soon come online, which can discover EM transients that are too faint for the currently operating survey projects. This section is to present a detailed calculation of the BH--NS GW detectability of the GW detectors in the next 15 years. We also simulate the follow-up EM detectability at different epochs of GW detector networks. In addition, some implications for the past and future GW-triggered target-of-opportunity observations are given in this section.

\subsection{GW Detectability}

\subsubsection{Method\label{sec:GWDetectionMethod}}

\begin{figure*}[thpb]
    \centering
	\includegraphics[width = 0.32\linewidth , trim = 60 30 80 60]{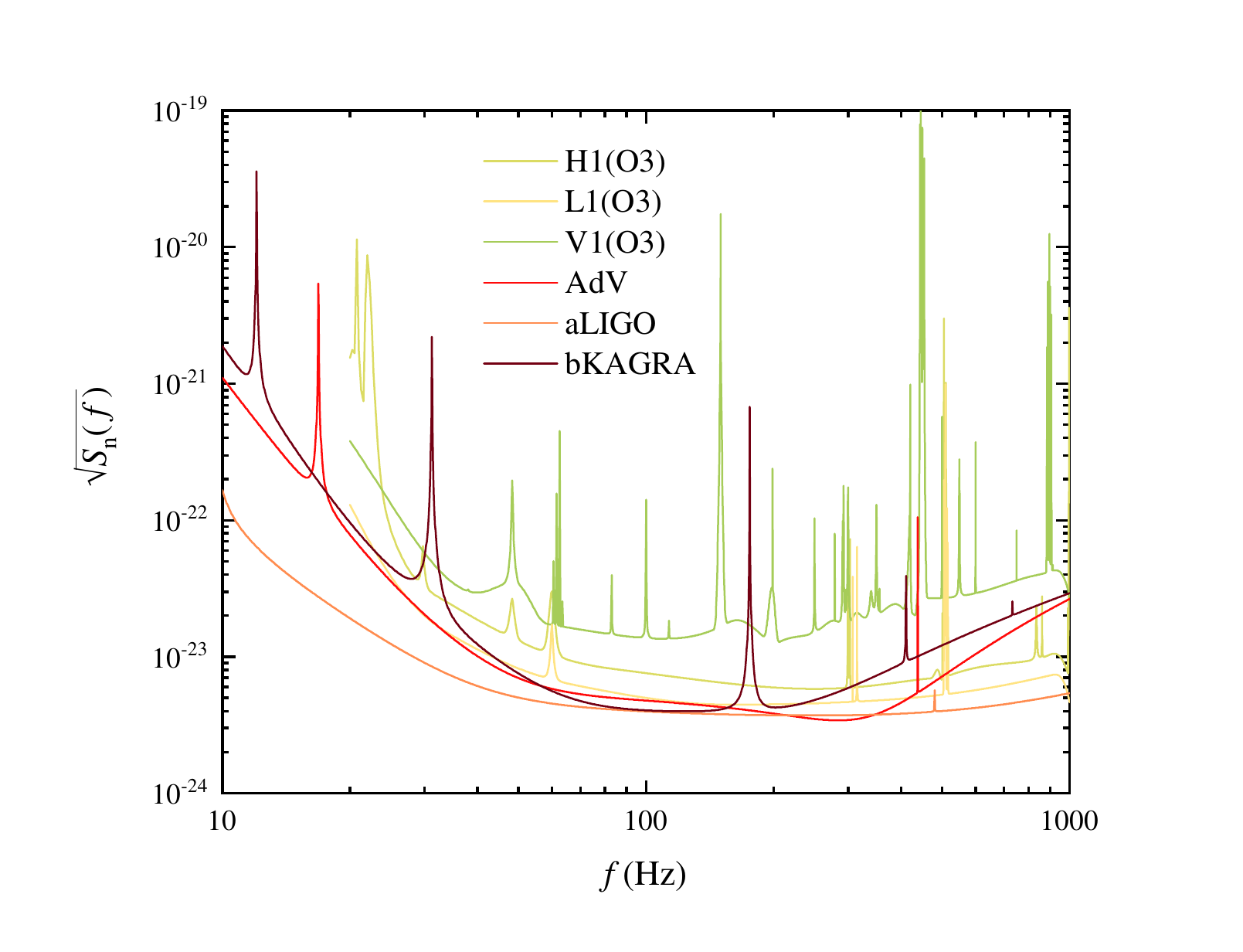}
	\includegraphics[width = 0.32\linewidth , trim = 60 30 80 60]{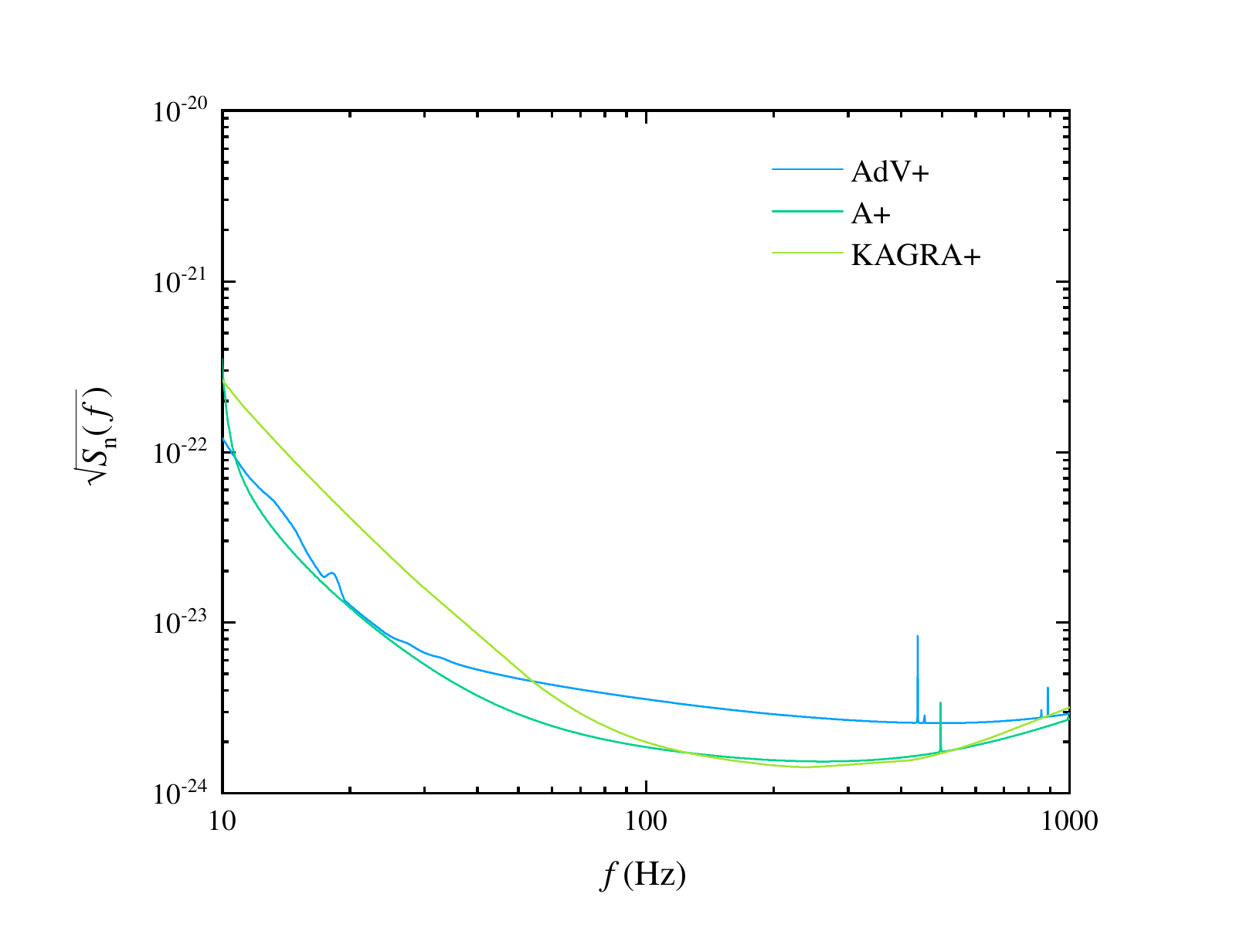}
	\includegraphics[width = 0.32\linewidth , trim = 60 30 80 60]{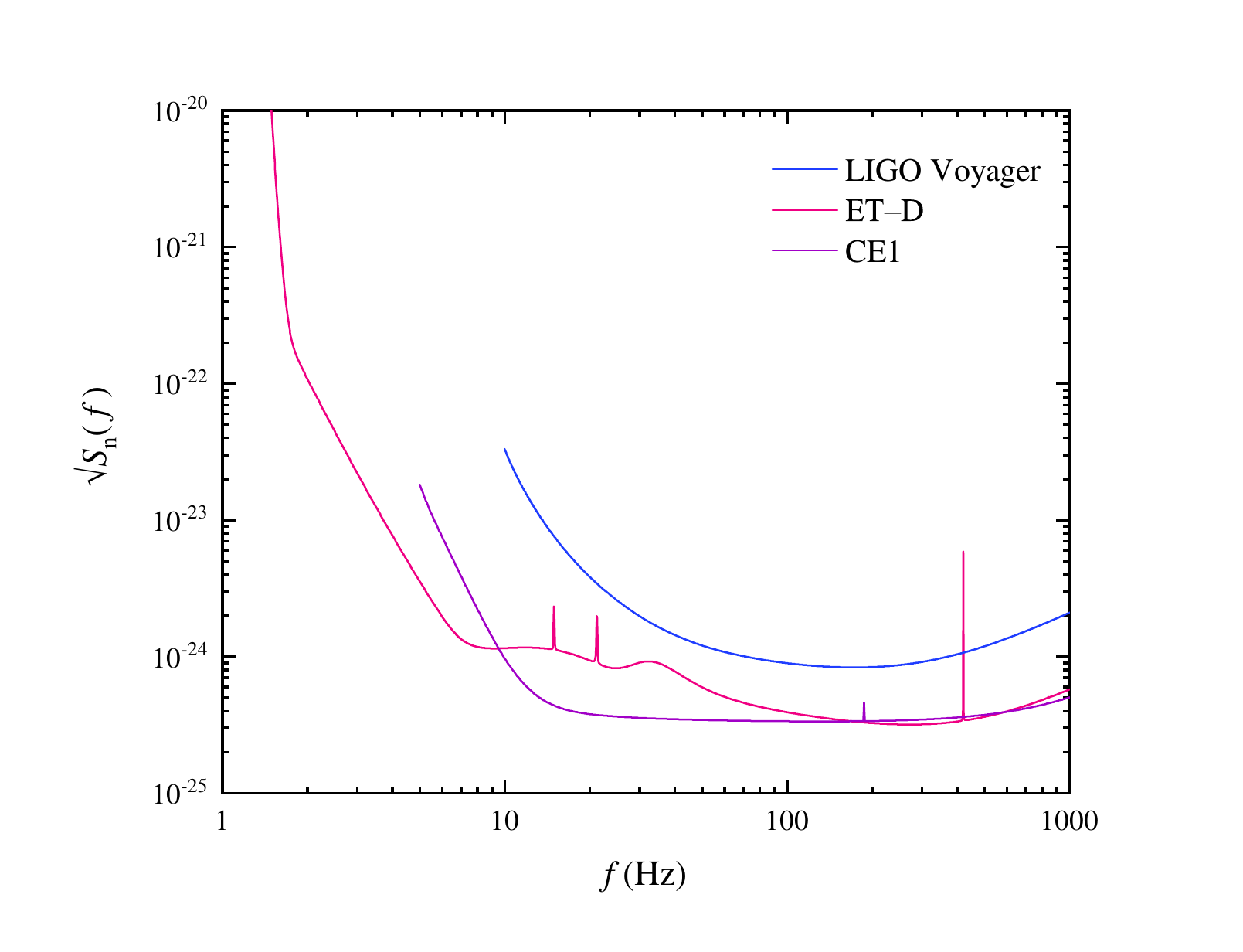}
    \caption{Left panel: the design sensitivity curves of 2nd generation GW detectors and O3 sensitivity curves. {The O3 sensitivity curves for H1, L1, and V1 are referred to GW190814 (see text).} Middle panel: the design sensitivity curves of 2.5th generation GW detectors. Right panel: the design sensitivity curves of 3rd generation GW detectors.}
    \label{fig:asd}
\end{figure*}

{In this work, we would like to predict the detection capabilities of different GW detector networks in the foreseeable future for BH--NS systems.} Accurate GW templates are needed to make our prediction results accurate. We use the phenomenological model \texttt{IMRPhenomPv2\_NRTidal}. Here we ignore the influence of higher-order modes \citep{ruiz2008multiple}, and only consider the GW signals during the BH--NS inspiral phase. We also ignore the precession effect.

At present, ground-based GW detectors include two Advanced LIGO detectors (H1 and L1) in the United States \citep{harry2010advanced,aasi2015advanced}, Advanced Virgo detector (V1) in Europe \citep{acernese2014advanced} and KAGRA detector (K1) in Japan \citep{aso2013interferometer,akutsu2018kagra}. It is expected that in the fourth observation run (O4), LIGO, Virgo and KAGRA will reach their respective design sensitivity \citep{abbott2020prospects}. In the following, we refer to the era during which these four detectors reach the design sensitivity as the ``HLVK era''\footnote{As the location of the third LIGO detector LIGO-India \citep{unnikrishnan2013indigo} has not been determined yet, this work does not consider LIGO-India.}, which is expected to begin in 2022. The detector sensitivities during this era, expressed as the amplitude spectral densities (ASD), are presented in the left panel of Figure \ref{fig:asd}, which are adopted from the public data\footnote{\url{https://dcc.ligo.org/LIGO-P1200087-v42/public}}. We also plot GW190814's sensitivity curves\footnote{\url{https://dcc.ligo.org/P2000183/public}} as representatives for O3 observations. As the 2nd generation detectors are about to reach the design sensitivity, people have begun to study how to further improve the sensitivity on the basis of the existing 2nd generation detectors. These upgrades are called 2.5th generation detectors. For example, the subsequent upgrade of Advanced LIGO is called Advanced LIGO Plus (A+) \citep{miller2015prospects}, and Advanced Virgo also has Advanced Virgo Plus (AdV+) upgrade plan \citep{abbott2018prospects}. KAGRA will also be upgraded to KAGRA+ \citep{michimura2020prospects} \footnote{KAGRA+ has multiple upgrade plans. For convenience, we select the sensitivity of the long-term upgrade plan ``KAGRAplusCombined'' \citep{michimura2014example} as the sensitivity of KAGRA+.}. The detector network composed of these 2.5th generation detectors is referred to ``PlusNetwork'' in the following. The sensitivity curves are shown in the middle panel of Figure \ref{fig:asd}, which are from public data\footnote{\url{https://gwdoc.icrr.u-tokyo.ac.jp/cgi-bin/DocDB/ShowDocument?docid=9537}}. After 2030, the 3rd generation GW detectors are expected to replace the 2nd generation and the 2.5th generation detectors. The 3rd generation detector plans currently proposed include the Einstein Telescope (ET) in Europe \citep{punturo2010einstein,punturo2010third,maggiore2020science}, Cosmic Explorer (CE) in the United States \citep{reitze2019cosmic}, and LIGO Voyager \citep{adhikari2020cryogenic} improved upon LIGO A+. Their design sensitivity curves are shown in the right panel of Figure \ref{fig:asd}. Since the location of ET and CE has not yet been determined, according to the convention \citep{vitale2017parameter,vitale2018characterization}, we place ET at the current Virgo detector position and CE at the current H1 position. Since LIGO Voyager is directly using the existing LIGO site, we place two LIGO Voyagers at the sites of H1 and L1 respectively. The sensitivity curves of ET and CE used in this paper come from the official websites\footnote{\url{http://www.et-gw.eu/index.php/etsensitivities}}\footnote{\url{https://dcc.cosmicexplorer.org/cgi-bin/DocDB/ShowDocument?docid=T2000017}}\footnote{Noting that we have selected the sensitivity of ``ET-D'' and ``CE1''}. LIGO Voyager's sensitivity curve comes from the public data\footnote{\url{https://dcc.ligo.org/LIGO-T1500293/public}}. We set the low cut-off frequency ($f_{\mathrm{min}}$) of all the 2nd and 2.5th generation detectors to $10\,{\rm Hz}$ \citep{miller2015prospects}, $20\,{\rm Hz}$ for O3. For the 3rd generation gravitational wave detectors, we set $f_{\mathrm{min}}$ of LIGO Voyager to $10\,{\rm Hz}$ \citep{adhikari2020cryogenic},  $f_{\mathrm{min}}$ of CE to $5\,{\rm Hz}$ \citep{reitze2019cosmic}, and  $f_{\mathrm{min}}$ of ET to $1\,{\rm Hz}$ \citep{punturo2010einstein}.

We simulate multiple sets of GW signal parameters according to different degrees of NS tidal deformation $\Lambda$, BH spin $\chi_{\rm BH}$ and NS mass $M_{\rm NS}$ according to the method presented in Section \ref{Sec:EjectaMassAndBinarySystem}. According to these parameters, we can obtain the redshifted masses in the geocentric coordinate system. By using the \texttt{IMRPhenomPv2\_NRTidal} template, we can simulate the GW waveform in the corresponding geocentric coordinate system, and then project it to different detectors to obtain the strain signal detected by each detector. The optimal SNR is calculated as
\begin{align}
\rho_{\text{opt}}^{2}=4 \int_{f_{\text{min}}}^{f_{\text{max}}} \frac{|\tilde{h}(f)|^{2}}{S_{n}(f)} d f=\int_{f_{\text{min}}}^{f_{\text{max}}} \frac{(2|\tilde{h}(f)| \sqrt{f})^{2}}{S_{n}(f)} d \ln (f), \label{eq:optimal_snr}
\end{align}
where $S_{n}(f)$ is the one-sided power spectral density (PSD) of the GW detector, which is the square of ASD. Here $f_{\text {max}}$ is set to $2048\,{\rm Hz}$, and $\tilde{h}(f)$ is the strain signal in the frequency domain. We can use the optimal SNR to approximate the matched filtering SNR of the GW signal detected by each detector, and then calculate the network SNR of the entire detector network, i.e., {the root sum squared of the SNR of all detectors.} When the network SNR is greater than the threshold 8, we expect that the corresponding GW signal is detected. In order to consider the best and worst situations in different eras, we consider ``all detectors in the corresponding era have reached the designed sensitivity and work normally'' as the optimal situation. In the mean time, we define ``only the detector with the worst sensitivity in the corresponding era works'' as the worst situation. For the 2nd generation detector network, the best case is that H1, L1, V1 and K1 all work normally, we abbreviate it as ``HLVK'', while the worst case is that only K1 works normally. In the same way, the best case for the 2.5th generation detector network is that A+, AdV+ and KAGRA+ all work normally, which we abbreviate as ``PlusNetwork'', and the worst case is that only AdV+ works. We discuss LIGO Voyager separately. ``ET+CE'' represents the best case of the 3rd generation era, and a single ET is considered as the worst case.

Since the GW from inspiral of compact binaries will reach the Earth earlier, early warning through GW to the EM community could be a feasible way of detecting more EM counterparts. There have been preliminary researches of early warning \citep{chan2018binary,sachdev2020early,nitz2020,2020ApJ...898L..39K,2020arXiv201012407S}. We will calculate the duration of BH--NS GW signals in different detector networks. This duration can be regarded as the upper limit of the early warning time\footnote{The matched filtering SNR will increase with the square root of signal duration, and the actual detection requires the matched filtering SNR to be higher than a certain threshold, so it must be shorter than the total duration of the GW signal.}. Here we use the 3.5 post-Newtonian order expression for the chirp time $T_{\text {chirp }}$ \citep[see Equation (E1) in][]{allen2012findchirp} to calculate the signal duration. Note that we ignore the spin effect and EoS of NS in the calculation of the chirp time $T_{\text {chirp}}$.

\subsubsection{GW Detection Rates, Detectable Distance, and Signal Duration \label{Sec:GWResults}}

\begin{figure*}
    \centering
    \includegraphics[width = 0.99\linewidth , trim = 100 270 140 20, clip]{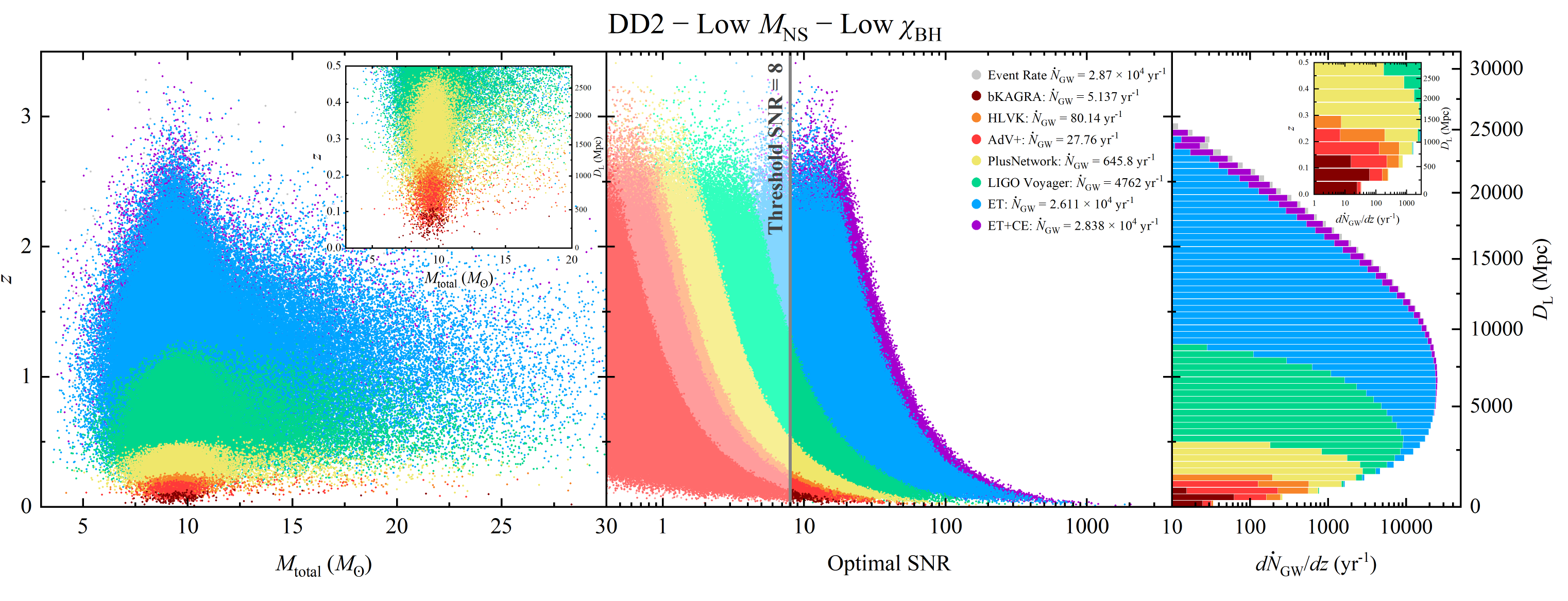}
    \caption{Detectability of BH--NS mergers by various detector networks for the example DD2$-$Low $M_{\rm NS}-$Low $\chi_{\rm BH}$ BH--NS signals in different GW detection eras. The left panels show the signals that can be detected by different detectors and detector networks (optimal SNR greater than 8). The crimson and red dots represent the detectable signals for K1 and AdV+, while the orange, yellow, green, blue, and violet dots are the GW signals  detectable at HLVK, PlusNetwork, LIGO Voyager, and ET+CE networks eras, respectively. The gray dots in the background represent undetectable signals. The small panels in the upper right corner are enlarged images of the low redshift area. The middle panels show the distributions of all simulated signals on the ``optimal SNR-redshift'' plane, with a detection threshold. The right side of the threshold are the GW signals that can be detected. The upper right corner is the converted annual detection rates. The right panels show the distributions of BH--NS detection rates with redshift, and the insets are zoom-in pictures in the low redshift region.}
    \label{fig:DD2GWresults}
\end{figure*}

\begin{deluxetable*}{cccccccc}[htpb]
\tablecaption{GW Detection Rate\label{table:gwDetectionRate}}
\tablecolumns{8}
\tablewidth{0pt}
\tablehead{
\colhead{Case} &
\colhead{bKAGRA} &
\colhead{HLVK} &
\colhead{AdV+} & 
\colhead{PlusNetwork} &
\colhead{LIGO Voyager} &
\colhead{ET} &
\colhead{ET+CE}
}
\startdata
\multirow{2}{*}{AP4$-$Low $M_{\rm NS}-$Low $\chi_{\rm BH}$} & $4.79$ & $77.5$ & $26.4$ & $640$ & $4.78\times10^3$ & $2.61\times10^4$ & $2.84\times10^4$ \\
{} & $(0.0167\%)$ & $(0.270\%)$ & $(0.0919\%)$ & $(2.23\%)$ & $(16.6\%)$ & $(91.0\%)$ & $(98.9\%)$ \\
\hline
\multirow{2}{*}{AP4$-$Low $M_{\rm NS}-$High $\chi_{\rm BH}$} & $6.43$ &	$93.4$	& $31.2$ &	$786$ & $5.41\times10^3$ & $2.64\times10^4$ & $2.84\times10^4$ \\
{} & $(0.0224\%)$ & $(0.325\%)$ & $(0.108\%)$ & $(2.74\%)$ & $(18.8\%)$ & $(91.9\%)$ & $(99.0\%)$ \\
\hline
\multirow{2}{*}{DD2$-$Low $M_{\rm NS}-$Low $\chi_{\rm BH}$} & $5.14$ & $80.1$	& $27.8$ & $646$	& $4.76\times10^3$ & $2.61\times10^4$ & $2.84\times10^4$ \\
{} & $(0.0179\%)$ & $(0.279\%)$	& $(0.0967\%)$ & $(2.25\%)$ &	$(16.6\%)$	& $(91.0\%)$ & $(98.9\%)$ \\
\hline
\multirow{2}{*}{DD2$-$Low $M_{\rm NS}-$High $\chi_{\rm BH}$} & $5.28$ & $88.2$ & $29.3$ & $782$	& $5.41\times10^3$ & $2.64\times10^4$ & 	$2.84\times10^4$ \\
{} & $(0.0184\%)$ &	$(0.307\%)$ & $(0.102\%)$ & $(2.72\%)$ &	$(18.8\%)$ & $(91.9\%)$	& $(99.0\%)$ \\
\hline
\multirow{2}{*}{DD2$-$High $M_{\rm NS}-$Low $\chi_{\rm BH}$} & $6.46$ & $108$ & $36.3$ & $876$	& $5.90\times10^3$ & $2.68\times10^4$ &	$2.85\times10^4$ \\
{} & $(0.0225\%)$ & $(0.375\%)$ & $(0.127\%)$ & $(3.05\%)$ &	$(20.6\%)$ & $(93.4\%)$	& $(99.3\%)$ \\
\hline
\multirow{2}{*}{DD2$-$High $M_{\rm NS}-$High $\chi_{\rm BH}$} & $7.89$ & $122$ & $40.1$ & $1.06\times10^3$ & $6.66\times10^3$ & $2.70\times10^4$ &	$2.85\times10^4$ \\
{} & $(0.0275\%)$ & $(0.426\%)$	& $(0.140\%)$ & $(3.69\%)$ &	$(23.2\%)$ & $(94.2\%)$ & $(99.4\%)$ \\
\hline
\multirow{2}{*}{Ms1$-$Low $M_{\rm NS}-$Low $\chi_{\rm BH}$} & $5.14$	& $80.6$ & $25.8$ & 	$639$ & 	$4.76\times10^3$ & $2.61\times10^4$ & $2.84\times10^4$ \\
{} & $(0.0179\%)$ & $(0.281\%)$	& $(0.0899\%)$ & 	$(2.23\%)$ & $(16.6\%)$ & $(91.0\%)$ & $(98.9\%)$ \\
\hline
\multirow{2}{*}{Ms1$-$Low $M_{\rm NS}-$High $\chi_{\rm BH}$} & $6.11$ & $91.3$ & $32.8$ & $787$	& $5.41\times10^3$ & $2.64\times10^4$ & $2.84\times10^4$ \\
{} & $(0.0213\%)$ & $(0.318\%)$ & $(0.114\%)$ & $(2.74\%)$ & $(18.9\%)$ & $(91.9\%)$ & $(99.0\%)$
\enddata
\tablecomments{The values represent the simulated BH--NS merger detection rates by GW detectors, in units of ${\rm yr}^{-1}$, while the numbers in brackets are corresponding detectable proportions in the amount of BH--NS merging per year in the Universe $(\sim2.87\times10^4\,{\rm yr}^{-1})$.}
\end{deluxetable*}

\begin{deluxetable*}{ccccccccc}[htpb]
\tablecaption{Detectable Distance\label{table:gwDetectionDistances}}
\tablecolumns{9}
\tablewidth{0pt}
\tabletypesize{\scriptsize}
\tablehead{
\colhead{Case} &
\colhead{Parameter} &
\colhead{bKAGRA} &
\colhead{HLVK} &
\colhead{AdV+} & 
\colhead{PlusNetwork} &
\colhead{LIGO Voyager} &
\colhead{ET} &
\colhead{ET+CE}
}
\startdata
\multirow{4}{*}{AP4$-$Low $M_{\rm NS}-$Low $\chi_{\rm BH}$} & $z$ & $0.069^{+0.030}_{-0.032}$ & $0.146^{+0.062}_{-0.062}$ & $0.118^{+0.054}_{-0.056}$ & $0.28^{+0.11}_{-0.12}$ & $0.54^{+0.31}_{-0.24}$ & $0.97^{+0.55}_{-0.48}$ & $1.00^{+0.57}_{-0.50}$\\
{} & $(D_L{\rm /100\,Mpc})$ & $(3.2^{+1.5}_{-1.5})$ & $(7.1^{+3.4}_{-3.2})$ & $(5.7^{+2.9}_{-2.8})$ & $(14.9^{+6.8}_{-6.9})$ & $(32^{+23}_{-16})$ & $(65^{+49}_{-37})$ & $(68^{+51}_{-39})$ \\
\cline{2-9}
{} & $z_{\rm max}$ & $0.121$ & $0.282$ & $0.219$ & 	$0.54$ &	$1.33$ &  $3.66$ & $3.66$ \\
{} & $(D_{L,{\rm max}}{\rm /100\,Mpc})$ & $(5.8)$ & $(14.9)$ & $(11.2)$ & $(31.8)$ & $(97)$ & $(330)$ & $(330)$ \\
\hline
\multirow{4}{*}{AP4$-$Low $M_{\rm NS}-$High $\chi_{\rm BH}$} & $z$ & $0.069^{+0.041}_{-0.029}$ & $0.155^{+0.061}_{-0.070}$ & $0.121^{+0.057}_{-0.061}$ & $0.30^{+0.12}_{-0.13}$ & $0.57^{+0.33}_{-0.26}$ & $0.97^{+0.56}_{-0.48}$ & $1.01^{+0.57}_{-0.50}$\\
{} & $(D_L{\rm /100\,Mpc})$ & $(3.2^{+2.0}_{-1.4})$  & $(7.6^{+3.4}_{-3.6})$ & $(5.8^{+3.1}_{-3.1})$ & $(15.9^{+7.5}_{-7.4})$ & $(33^{+26}_{-17})$ & $(66^{+49}_{-37})$ & $(69^{+51}_{-39})$\\
\cline{2-9}
{} & $z_{\rm max}$ & $0.130$ & $0.318$ &	$0.228$ &	$0.57$	& $1.45$ & $3.40$ & $3.53$ \\
{} & $(D_{L,{\rm max}}{\rm /100\,Mpc})$ & $(6.3)$ & $(17.1)$ & $(11.7)$ & $(33.9)$ & $(108)$ & $(302)$ & $(316)$\\
\hline
\multirow{4}{*}{DD2$-$Low $M_{\rm NS}-$Low $\chi_{\rm BH}$} & $z$ & $0.076^{+0.030}_{-0.038}$ & $0.147^{+0.059}_{-0.065}$ & $0.116^{+0.058}_{-0.055}$ & $0.28^{+0.11}_{-0.12}$ & $0.54^{+0.30}_{-0.24}$ & $0.97^{+0.55}_{-0.48}$ & $1.01^{+0.57}_{-0.50}$\\
{} & $(D_L{\rm /100\,Mpc})$ & $(3.5^{+1.5}_{-1.8})$ & $(7.2^{+3.2}_{-3.3})$ & $(5.6^{+3.1}_{-2.7})$ & $(14.8^{+6.9}_{-6.9})$ & $(32^{+23}_{-16})$ & $(65^{+49}_{-37})$ & $(68^{+51}_{-39})$\\
\cline{2-9}
{} & $z_{\rm max}$ & $0.130$	& $0.268$ & $0.222$ & $0.54$ &	$1.43$ & $3.23$	& $3.56$ \\
{} & $(D_{L,{\rm max}}{\rm /100\,Mpc})$ & $(6.3)$ & $(14.0)$	& $(11.4)$ & $(31.9)$ &	$(106)$	& $(284)$ & $(320)$ \\
\hline
\multirow{4}{*}{DD2$-$Low $M_{\rm NS}-$High $\chi_{\rm BH}$} & $z$ & $0.073^{+0.034}_{-0.034}$ & $0.155^{+0.061}_{-0.066}$ & $0.120^{+0.060}_{-0.054}$ & $0.30^{+0.12}_{-0.13}$ & $0.57^{+0.33}_{-0.25}$ & $0.97^{+0.56}_{-0.48}$ & $1.01^{+0.57}_{-0.50}$\\
{} & $(D_L{\rm /100\,Mpc})$ & $(3.4^{+1.7}_{-1.6})$ & $(7.6^{+3.4}_{-3.4})$ & $(5.8^{+3.2}_{-2.7})$ & $(16.0^{+7.3}_{-7.3})$ & $(34^{+26}_{-17})$ & $(66^{+49}_{-37})$ & $(69^{+51}_{-39})$\\
\cline{2-9}
{} & $z_{\rm max}$ & $0.133$	& $0.302$ & $0.230$ & 	$0.56$ &	$1.46$ &  $3.46$ & $3.50$ \\
{} & $(D_{L,{\rm max}}{\rm /100\,Mpc})$ & $(6.5)$ & $(16.1)$ & $(11.8)$ & $(33.3)$ & $(109)$ & $(309)$ & $(313)$ \\
\hline
\multirow{4}{*}{DD2$-$High $M_{\rm NS}-$Low $\chi_{\rm BH}$} & $z$ & $0.076^{+0.036}_{-0.035}$ & $0.161^{+0.064}_{-0.069}$ & $0.127^{+0.065}_{-0.058}$ & $0.31^{+0.12}_{-0.13}$ & $0.58^{+0.34}_{-0.26}$ & $0.98^{+0.56}_{-0.48}$ & $1.01^{+0.57}_{-0.50}$\\
{} & $(D_L{\rm /100\,Mpc})$ & $(3.6^{+1.8}_{-1.7})$ & $(8.0^{+3.6}_{-3.6})$ & $(6.1^{+3.5}_{-2.9})$ & $(16.5^{+7.6}_{-7.7})$ & $(35^{+26}_{-18})$ & $(66^{+49}_{-37})$ & $(69^{+51}_{-39})$\\
\cline{2-9}
{} & $z_{\rm max}$ & $0.142$ & $0.318$ &	$0.259$ & $0.58$	& $1.42$ & $3.34$ & $3.36$ \\
{} & $(D_{L,{\rm max}}{\rm /100\,Mpc})$& $(6.9)$ & $(17.2)$ & $(13.5)$ & $(34.7)$ &	$(105)$ & $(297)$	& $(309)$ \\
\hline
\multirow{4}{*}{DD2$-$High $M_{\rm NS}-$High $\chi_{\rm BH}$} & $z$ &
$0.079^{+0.037}_{-0.035}$ & $0.168^{+0.068}_{-0.071}$ & $0.133^{+0.064}_{-0.062}$ & $0.33^{+0.13}_{-0.14}$ & $0.61^{+0.37}_{-0.27}$ &
$0.98^{+0.56}_{-0.48}$ & 
$1.01^{+0.57}_{-0.50}$\\
{} & $(D_L{\rm /100\,Mpc})$ & $(3.7^{+1.9}_{-1.7})$ & $(8.3^{+3.9}_{-3.7})$ & $(6.4^{+3.5}_{-3.1})$ & $(17.6^{+8.3}_{-8.2})$ & $(37^{+29}_{-19})$ & $(67^{+50}_{-37})$ & $(69^{+51}_{-39})$\\
\cline{2-9}
{} & $z_{\rm max}$ & $0.159$ & $0.349$ &	$0.275$ & 	$0.60$	& $1.64$ & $3.35$ & $3.50$ \\
{} & $(D_{L,{\rm max}}{\rm /100\,Mpc})$ & $(7.8)$ & $(19.1)$	& $(14.5)$ & $(38.9)$ &	$(125)$ & $(298)$ & $(314)$ \\
\hline
\multirow{4}{*}{Ms1$-$Low $M_{\rm NS}-$Low $\chi_{\rm BH}$} & $z$ & $0.069^{+0.034}_{-0.035}$ & $0.149^{+0.058}_{-0.066}$ & $0.116^{+0.054}_{-0.056}$ & $0.28^{+0.11}_{-0.12}$ & $0.54^{+0.30}_{-0.24}$ & $0.97^{+0.55}_{-0.48}$ & $1.00^{+0.57}_{-0.50}$\\
{} & $(D_L{\rm /100\,Mpc})$ & $(3.2^{+1.7}_{-1.7})$ & $(7.3^{+3.2}_{-3.4})$ & $(5.6^{+2.9}_{-2.8})$ & $(14.8^{+6.9}_{-6.9})$ & $(32^{+23}_{-16})$ & $(65^{+49}_{-37})$ & $(68^{+51}_{-39})$\\
\cline{2-9}
{} & $z_{\rm max}$ & $0.134$	& $0.285$ & $0.232$ & $0.53$ &	$1.30$ & $3.51$ & $3.52$ \\
{} & $(D_{L,{\rm max}}{\rm /100\,Mpc})$ & $(6.5)$ & $(15.1)$ & $(12.0)$ & 	$(31.0)$ & $(94)$ & $(315)$ & $(315)$ \\
\hline
\multirow{4}{*}{Ms1$-$Low $M_{\rm NS}-$High $\chi_{\rm BH}$} & $z$ & $0.072^{+0.030}_{-0.036}$ & $0.152^{+0.064}_{-0.066}$ & $0.120^{+0.059}_{-0.056}$ & $0.30^{+0.20}_{-0.13}$ & $0.57^{+0.33}_{-0.25}$ & $0.97^{+0.55}_{-0.48}$ & $1.01^{+0.57}_{-0.50}$\\
{} & $(D_L{\rm /100\,Mpc})$ & $(3.4^{+1.5}_{-1.7})$ & $(7.5^{+3.5}_{-3.4})$ & $(5.8^{+3.2}_{-2.8})$ & $(16^{+7.3}_{-7.5})$ & $(34^{+25}_{-17})$ & $(66^{+49}_{-37})$ & $(69^{+51}_{-39})$\\
\cline{2-9}
{} & $z_{\rm max}$ & $0.127$ & $0.299$ & $0.244$ &	$0.57$ & $1.44$ & $3.42$ & $3.65$ \\
{} & $(D_{L,{\rm max}}{\rm /100\,Mpc})$ & $(6.1)$ & $(16.0)$ & $(12.6)$ & $(34)$ &	$(107)$ & $(305)$ &	$(330)$
\enddata
\tablecomments{The values are the BH--NS merger maximum detectable redshift and median detectable distances with consideration of 90\% detectable interval by GW detectors in our simulations, while the corresponding luminosity distances in units of ${\rm Mpc}$ are shown in brackets.}
\end{deluxetable*}

In this section, we show the simulated detection results of GW signals for 8 sets of parameter distributions of BH--NS system. For each set of parameter distribution, we simulate $1\times10^6$ BH--NS GW signals. Here, we take the simulation results of DD2$-$Low $M_{\rm NS}-$Low $\chi_{\rm BH}$ as an example to illustrate the detectability of BH--NS signals with GW detectors in different eras. One can see that for the 2nd generation GW detector network, the worst case is that only bKAGRA is running so that the observed BH--NS detection rate is $\sim5\,{\rm yr}^{-1}$. The best case is that the entire HLVK network is observing so that the detection rate is $\sim80\,{\rm yr}^{-1}$. For the 2.5th generation detector network, the worst case is that only AdV+ is in operation, with a detection rate of $\sim30\,{\rm yr}^{-1}$. This is better than the result of a single 2nd generation GW detector, but not as good as the HLVK detector network. The optimal situation is that PlusNetwork is fully operating, with a detection rate of $\sim650\,{\rm yr}^{-1}$. For the 3rd generation detector network, LIGO Voyager can increase the detection rate to $\sim4,800\,{\rm yr}^{-1}$. This result is better than the 2nd generation and 2.5th generation detector networks, but is far lower than the newly designed 3rd generation detectors. For a single ET, the detection rate is $\sim2.6 \times 10^{4}\,{\rm yr}^{-1}$, while for the ET+CE network, the detection rate is $\sim2.84 \times 10^{4}\,{\rm yr}^{-1}$. These results are very close to the total BH--NS event rate ($\sim2.87 \times 10^{4}$ per year for our models), indicating that ET and ET+CE can basically detect almost all BH--NS GW events in the entire Universe. As shown in right panels of Figure \ref{fig:DD2GWresults}, as the sensitivity of the GW detector increases, BH--NS events at higher redshifts gradually dominate the detected event rate. The event rate of ET and ET+CE is mainly dominated by the BH--NS mergers at $z \sim 1$, {which is near the most probable redshift where BH--NS mergers occurred in the universe (see the right panel of Figure \ref{fig:DD2GWresults}). }

All of our simulation results for GW detection rates are summarized in Table \ref{table:gwDetectionRate}. Through comparison, we can find that the simulation results of each group are the same in order of magnitude, but the effects of each variable on the detection rate of BH--NS mergers are different. High spin of the BHs can increase the detection rate of BH--NS mergers, but the increase is less than an order of magnitude. {The improvement brought by a heavier NS mass is larger than that by a higher black hole spin, because the amplitude of the GW signal is proportional to the chirp mass \citep{allen2012findchirp}}, and the NS EoS has almost no effect on the detection rate of BH--NS mergers. Besides, the median detectable distances (redshift) with a $90\%$ detectable interval, and the maximum detectable distances (redshift) for our simulated GW siginals are given in Table \ref{table:gwDetectionDistances}. Our results for ``HLVK era" are generally consist with LVC's prospect results in O4 \citep{abbott2020prospects}.

\begin{figure*}
    \centering
	\includegraphics[width = 0.49\linewidth , trim = 60 70 170 60, clip]{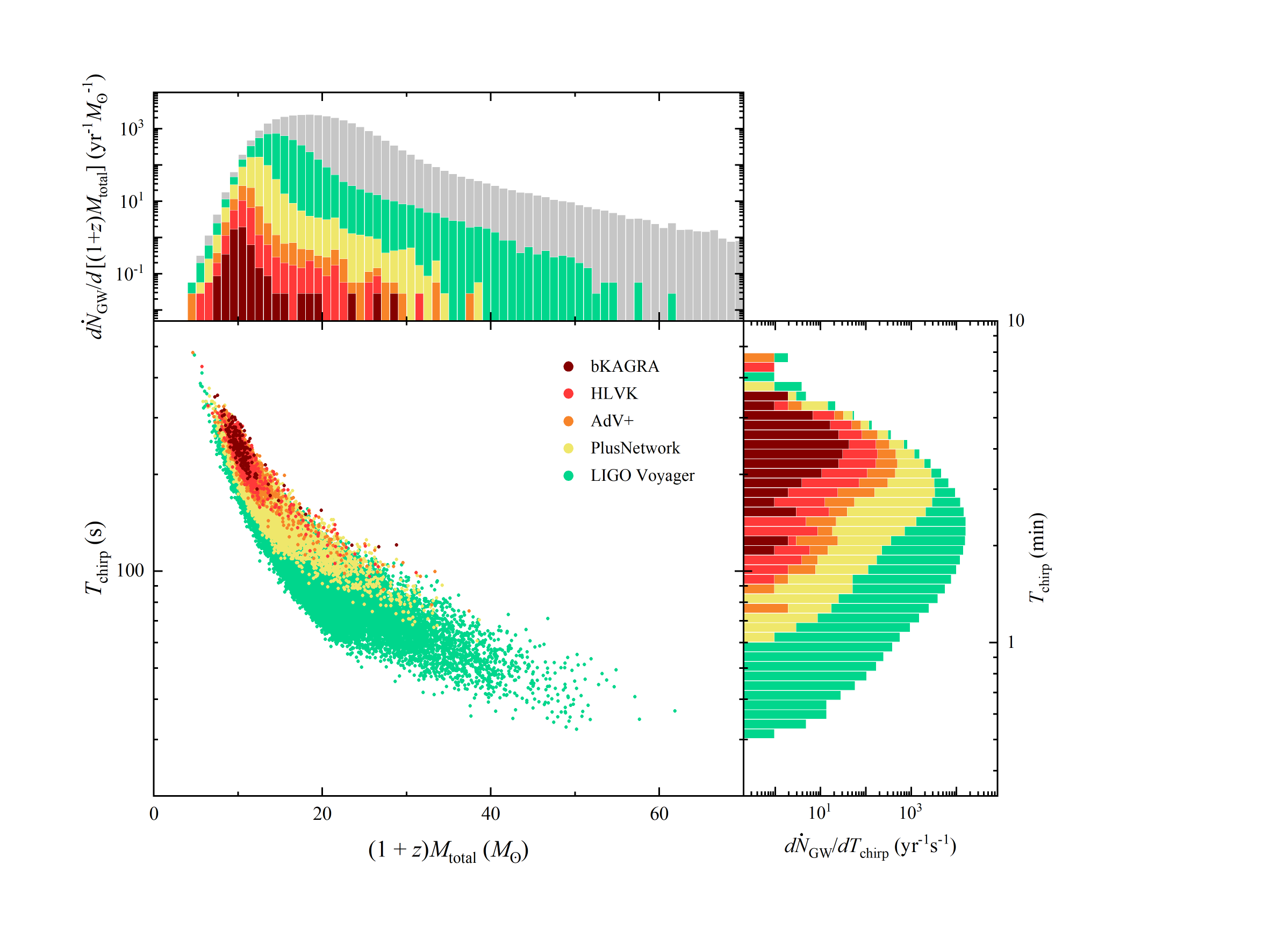}
	\includegraphics[width = 0.49\linewidth , trim = 60 70 170 60, clip]{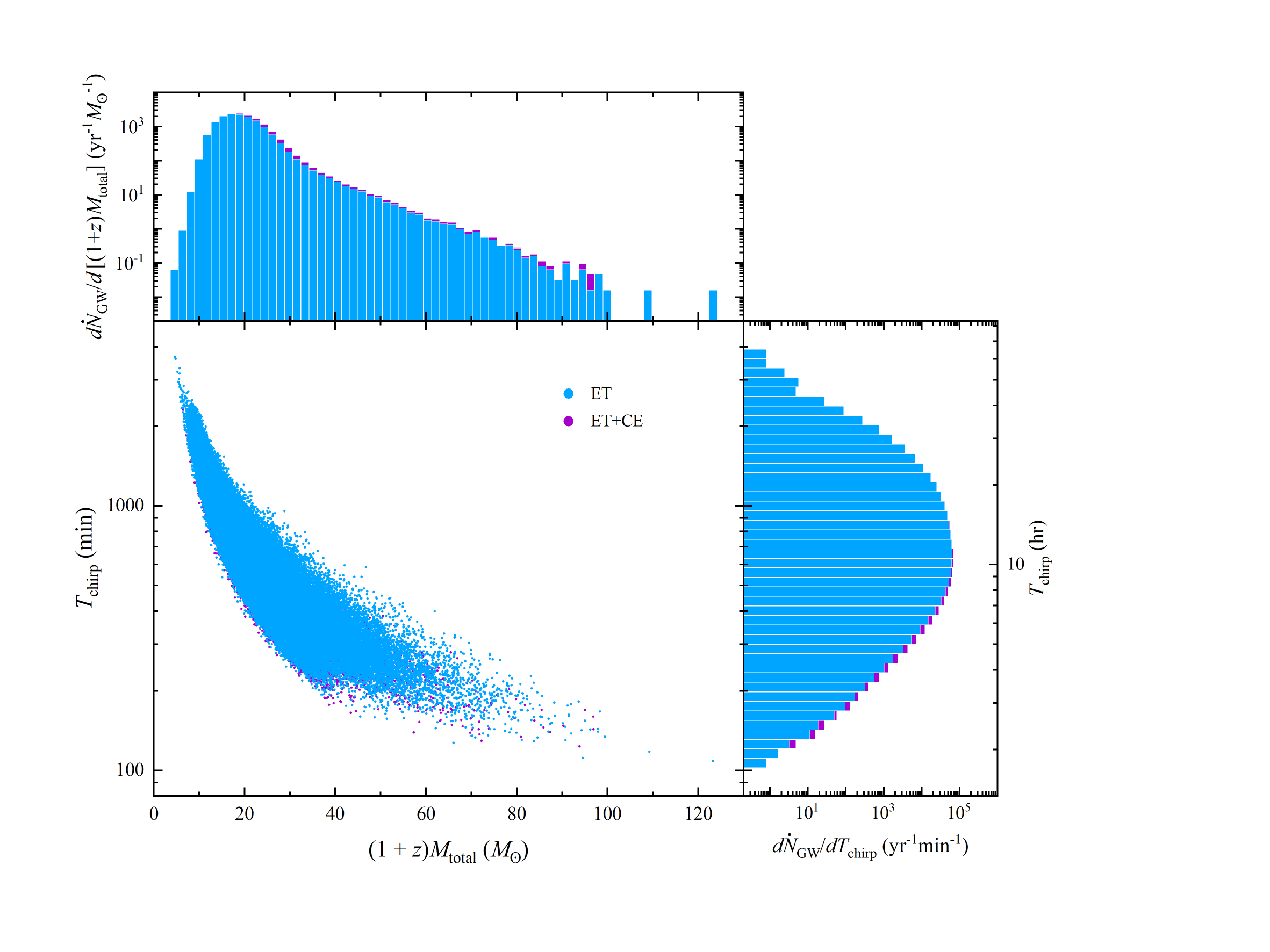}
    \caption{Left panels: the chirp time for the example DD2$-$Low $M_{\rm NS}$ BH--NS GW signals in 2nd/2.5th generation GW detectors and LIGO Voyager. Right panels: The chirp time for the DD2$-$Low $M_{\rm NS}$ BH--NS GW signals in ET and ET+CE. The gray histograms represent the distributions of the geocentric frame total mass for overall BH--NS events merged in the Universe every year.}
    \label{fig:DD2LowMassChirptime}
\end{figure*}

\begin{deluxetable*}{ccccccccc}[htpb]
\tablecaption{Chirp Time and Total Mass in Geocentric Frame of Detectable GW Signals (DD2)\label{table:gwDetectionChirptime}}
\tabletypesize{\footnotesize}
\tablecolumns{9}
\tablewidth{0pt}
\tablehead{
\colhead{Case} &
\colhead{Parameter} &
\colhead{bKAGRA} &
\colhead{HLVK} &
\colhead{AdV+} & 
\colhead{PlusNetwork} &
\colhead{LIGO Voyager} &
\colhead{ET} &
\colhead{ET+CE}
}
\startdata
\multirow{6}{*}{Low $M_{\rm NS}$}  & $T_{\rm chirp}/{\rm min}$ & $4.04^{+0.64}_{-0.70}$ & $3.60^{+0.73}_{-0.65}$ & $3.78^{+0.72}_{-0.69}$ & $3.00^{+0.76}_{-0.64}$ & $2.19^{+0.87}_{-0.67}$ & $11.4^{+7.1}_{-4.2}\times60$ & $11.1^{+7.2}_{-4.1}\times60$\\
{} & $T_{\rm chirp,min}/{\rm min}$ & 1.99  & 1.25 & 1.61 & 1.01 & 0.54 & $1.8\times60$ & $1.8\times60$ \\
{} & $T_{\rm chirp,max}/{\rm min}$ & 5.88  & 7.97 & 7.22 & 7.97 & 7.97 & $61.0\times60$ & $61.0\times60$ \\
\cline{2-9}
{} & $(1 + z)M_{\rm total}/{M_\odot}$ & $10.2^{+1.4}_{-1.2}$ & $11.0^{+2.0}_{-1.4}$ & $10.6^{+1.8}_{-1.3}$ & $12.2^{+2.4}_{-1.8}$ & $14.7^{+4.2}_{-2.9}$ & $18.7^{+6.7}_{-5.0}$ & $19.0^{+6.8}_{-5.1}$\\
{} & $[(1 + z)M_{\rm total}]_{\rm min}/{M_\odot}$ & 7.2 & 4.7 & 5.7 & 4.7 & 4.7 & 4.7 & 4.7\\
{} & $[(1 + z)M_{\rm total}]_{\rm max}/{M_\odot}$ & 28.8 & 37.5 & 31.1 & 38.6 & 61.8 & 123 & 123\\
\hline
\multirow{6}{*}{High $M_{\rm NS}$}  & $T_{\rm chirp}/{\rm min}$ & $3.34^{+0.60}_{-0.48}$ & $2.99^{+0.57}_{-0.56}$ & $3.12^{+0.59}_{-0.57}$ & $2.45^{+0.64}_{-0.52}$ & $1.77^{+0.74}_{-0.56}$ & $9.5^{+6.0}_{-3.5}\times60$ & $9.3^{+6.0}_{-3.4}\times60$\\
{} & $T_{\rm chirp,min}/{\rm min}$ & 1.45 & 1.18 & 1.18 & 0.90 & 0.37 & $1.5\times60$ & $1.5\times60$\\
{} & $T_{\rm chirp,max}/{\rm min}$ & 5.16 & 6.56 & 5.17 & 6.56 & 6.56 & $50.2\times60$ & $50.2\times60$\\
\cline{2-9}
{} & $(1 + z)M_{\rm total}/{M_\odot}$ & $10.5^{+1.6}_{-1.1}$ & $11.4^{+2.2}_{-1.4}$ & $11.0^{+2.1}_{-1.4}$ & $12.8^{+2.6}_{-1.9}$ & $15.5^{+4.7}_{-3.2}$ & $19.3^{+6.8}_{-5.1}$ & $19.6^{+6.9}_{-5.3}$\\
{} & $[(1 + z)M_{\rm total}]_{\rm min}/{M_\odot}$ & 7.5 & 5.5 & 7.5 & 5.4 & 5.4 & 5.4 & 5.4 \\
{} & $[(1 + z)M_{\rm total}]_{\rm max}/{M_\odot}$ & 25.3 & 33.8 & 30.6 & 42.2 & 63.9 & 118 & 118\\
\enddata
\tablecomments{We show median chirp time $T_{\rm chirp}$ with $90\%$ chirp time interval, minimum chirp time, maximum chirp time, median geocentric frame total mass with $90\%$ interval, minimum geocentric frame total mass, and maximum geocentric frame total mass for different GW detection eras.}
\end{deluxetable*}

{We calculate  $T_{\text {chirp}}$ as shown in Figure \ref{fig:DD2LowMassChirptime}.} All the BH--NS signals that can be detected by each detector or detector network are drawn in the figure. In the left panel, we can see the general trend that the chirp time decreases as the geocentric frame total mass increases. This is because the GW frequency of the ISCO of the binary star system is inversely proportional to the total mass \citep{wu2020measuring}. The larger the total mass, the closer the ISCO frequency is to $f_{\mathrm{min}}$. For the 2nd and 2.5th generations of GW detectors, although $f_{\mathrm{min}}$ is $10\,{\rm Hz}$, due to the increase in detector's sensitivity, more short GW signals can have sufficiently large SNR, that is, large mass BH--NS is easier to detect, which is why the median value of $T_{\text {chirp}}$ on the right side of each panel in Figure \ref{fig:DD2LowMassChirptime} decreases as the sensitivity of the detector increases. In the right panel of Figure \ref{fig:DD2LowMassChirptime}, the $T_{\text {chirp}}$ of ET and ET+CE are shown. For ET and ET+CE, the $f_{\mathrm{min}}$ is reduced from $10\,{\rm Hz}$ to $1\,{\rm Hz}$, which greatly increases the signal duration. All $T_{\text {chirp}}$ values are shown in Table \ref{table:gwDetectionChirptime}. From the table, we can see that for the 2nd and 2.5th generation detectors and LIGO Voyager, $T_{\text {chirp}}$ does not exceed 10 minutes, which means that the time for early warning is very limited. For ET and ET+CE, $T_{\text {chirp}}$ has reached the order of hours or even a few days. There will be enough preparation time for EM observations in these cases.

\subsection{EM Detectability}

\subsubsection{Method}

Target-of-opportunity observations after GW triggers would be an optimal strategy to search for associated EM signals, especially for kilonova, in the future. The BH--NS GW detectabilities for the networks of the 2nd, 2.5th, and 3rd generation GW detectors have been studied in detail in Section \ref{Sec:GWResults}. Based on these results, we now discuss the EM detection probabilities for GW-triggered target-of-opportunity observations under ideal follow-up conditions in this section.  {We consider three situations for the detection of EM signals:}

{1. Only kilonova emission is detected from BH-NS mergers.}

{2. Orphan afterglow emission detection --- Gravitational Wave High-energy Electromagnetic Counterpart All-sky Monitor (GECAM) launched in late 2020 can cover all-sky to detect GRBs \citep[e.g.,][]{song2019,yu2021}. GW-associated GRBs would essentially not be missed and on-axis jet afterglows and kilonovae of these events will likely be detected. {In the following, we do not focus on these joint GW/sGRB detections but focus on GW triggers without a sGRB association. In the absence of a sGRB (i.e., $\theta_{\rm view}\gtrsim\theta_{\rm c}$), as-yet undiscovered off-axis jet orphan afterglow \citep[e.g.,][]{vanerten2010} could establish an EM link with a GW trigger.} Therefore, the detectability of orphan afterglows from BH--NS mergers for target-of-opportunity observations is also predicted together with kilonovae.}

{3. Total emission --- In fact, the observed optical EM flux should be a superposition of the emissions from both the kilonova and the afterglow {(either on-axis or off-axis afterglow depending on the viewing direction)}, so we also give the detection probabilities of the total emission in this section.}

As Advanced LIGO and Virgo will upgrade while additional interferometers (KAGRA and LIGO-India) come online in the future, the detection horizon would be several hundred Mpc for BNS and BH--NS mergers with the localization accuracy reaching to $\sim{10}\,{\rm deg}^2$ by a network of these 2nd GW detectors \citep{abbott2018prospects}. In contrast, higher localization accuracy would be obtained by networking of 3rd generation GW detectors \citep{zhao2018,chan2018,hall2019,maggiore2020}. In these regimes, present and future wide-field-of-view survey projects {can be able to cover the sky localization} given by GW detections in few pointings and achieve deep detection depths with relatively short exposure integration times. The detection of GW170817/AT\,2017gfo was {made in a $i$-band filter} and subsequently confirmed by larger telescopes \citep{coulter2017}. Similar to this case, we assume that one can use only one filter to make follow-up searches after GW triggers. With quick response times of GW events follow-up observations, we also assume that the survey telescopes would not miss the peaks of kilonovae in each band. Therefore, by assuming that all GW detectors in each era can operate normally  {(i.e., the most optimal situation for different generation detector networks)}, for a given $X$-band limiting magnitude $m_{X,{\rm limit}}$, the detection rates for target-of-opportunity observations can be expressed as
\begin{equation}
\label{Eq: ToOObservation}
    \dot{N}_X = \int_0^{z_{\rm max}}\frac{\dot{\rho}_{<m_{X,{\rm limit}}}(z)}{1 + z}\frac{dV(z)}{dz}dz,
\end{equation}
where $z_{\rm max}$ is the maximum detectable distance and  $\dot{\rho}_{<m_{X,{\rm limit}}}(z)$ represents the cosmological event rate densities of the BH--NS mergers that can be triggered by GW detectors and be detected by telescopes in $X$ band. Here, the astronomical magnitude system we adopted is the AB magnitude system, i.e., $m_\nu = -2.5\log(F_\nu/3631\,{\rm Jy})$.

\subsubsection{Results for Target-of-opportunity Observations after GW Triggers\label{Sec:ToODetectionRatesResults}}

\begin{figure*}
    \centering
    \includegraphics[width = 0.49\linewidth , trim = 40 20 20 40, clip]{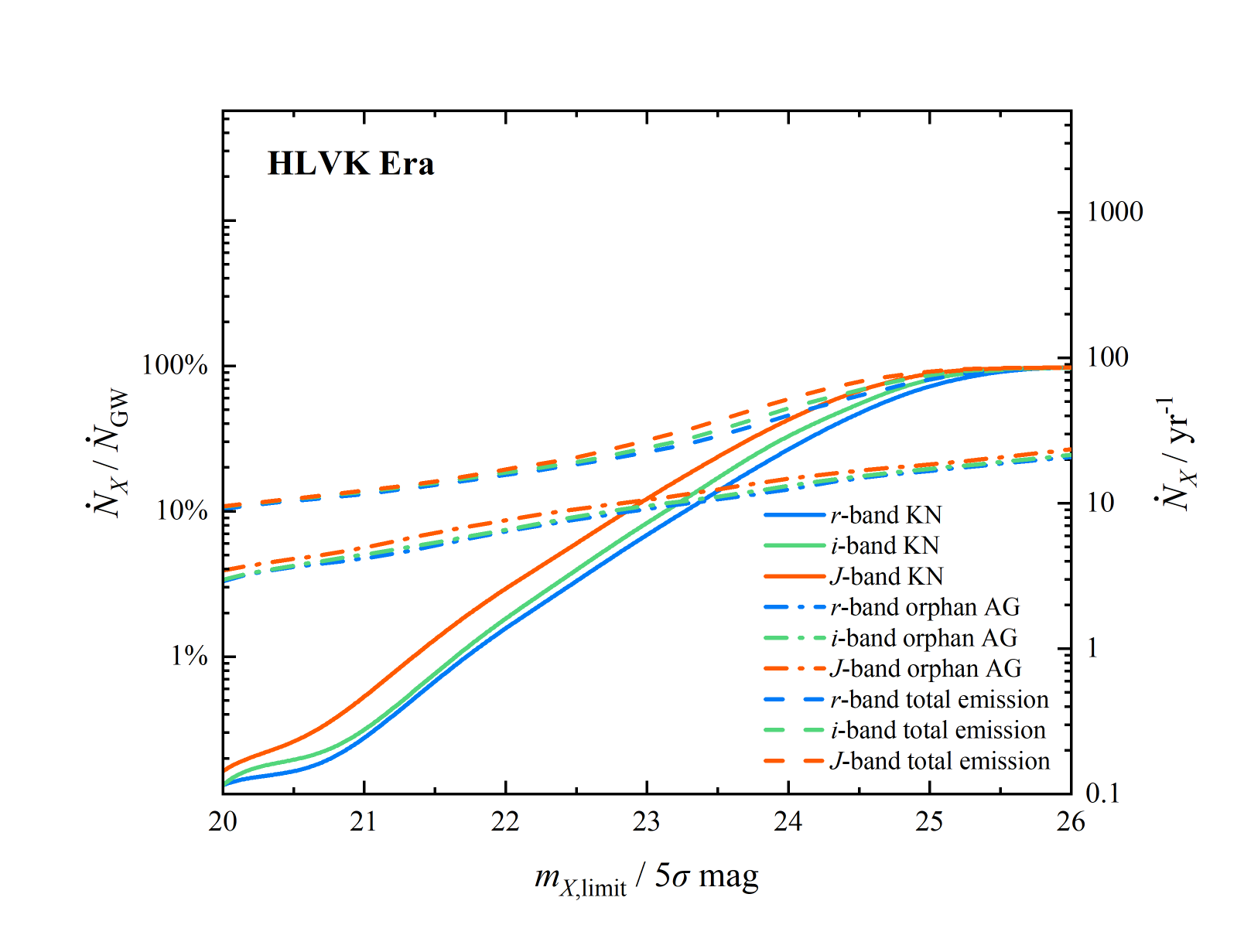}
    \includegraphics[width = 0.49\linewidth , trim = 40 20 20 40, clip]{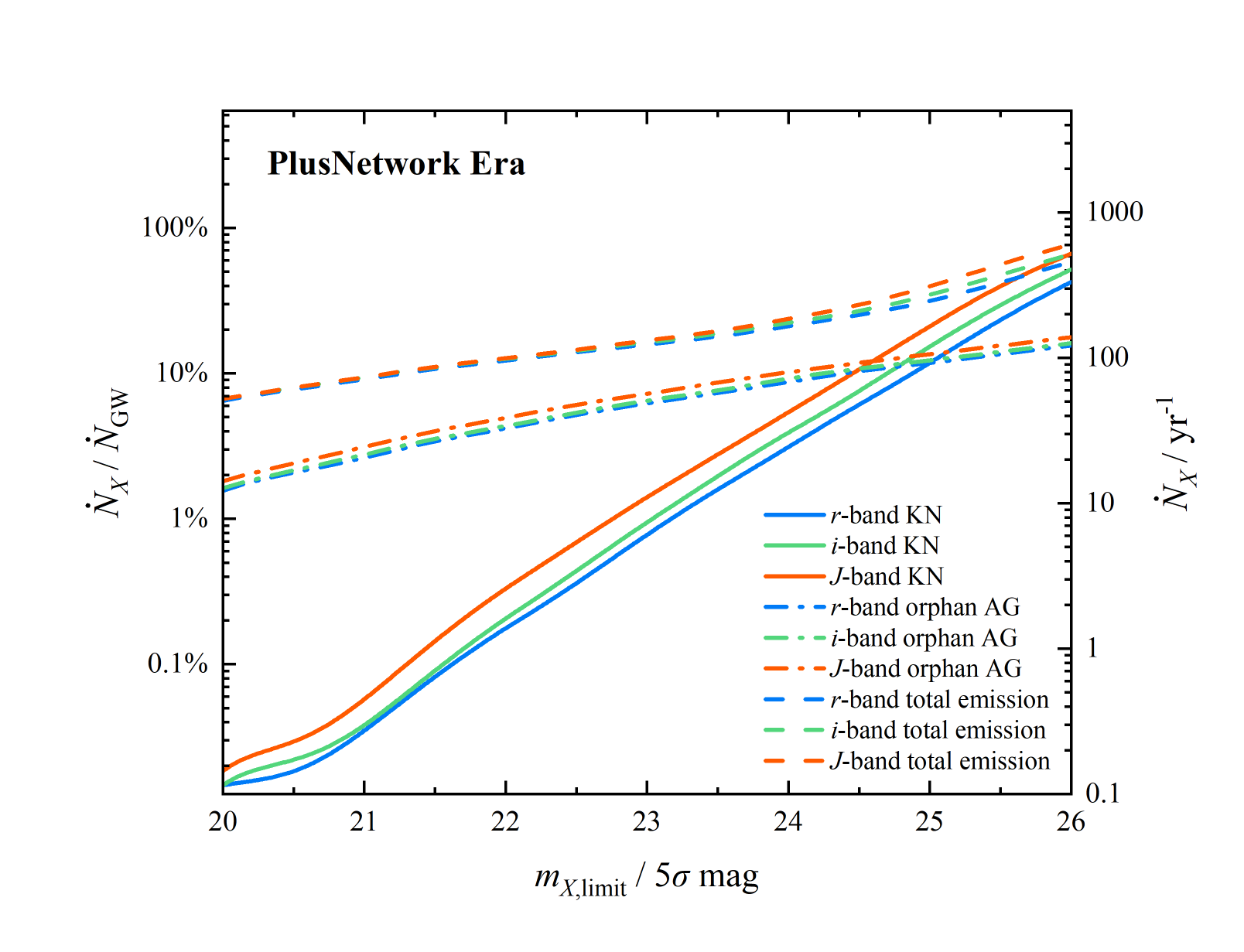}
    \includegraphics[width = 0.49\linewidth , trim = 40 20 20 40, clip]{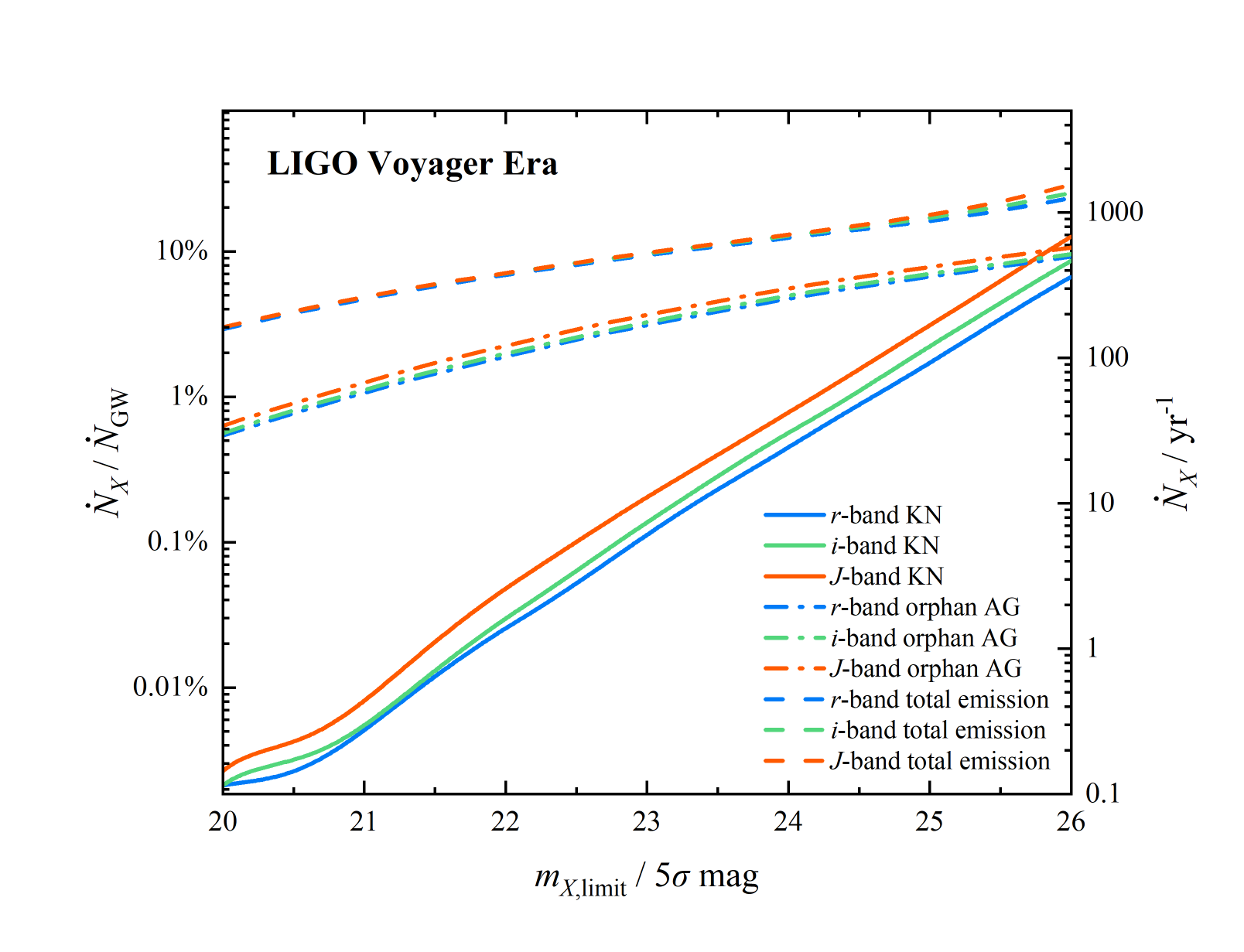}
    \includegraphics[width = 0.49\linewidth , trim = 40 20 20 40, clip]{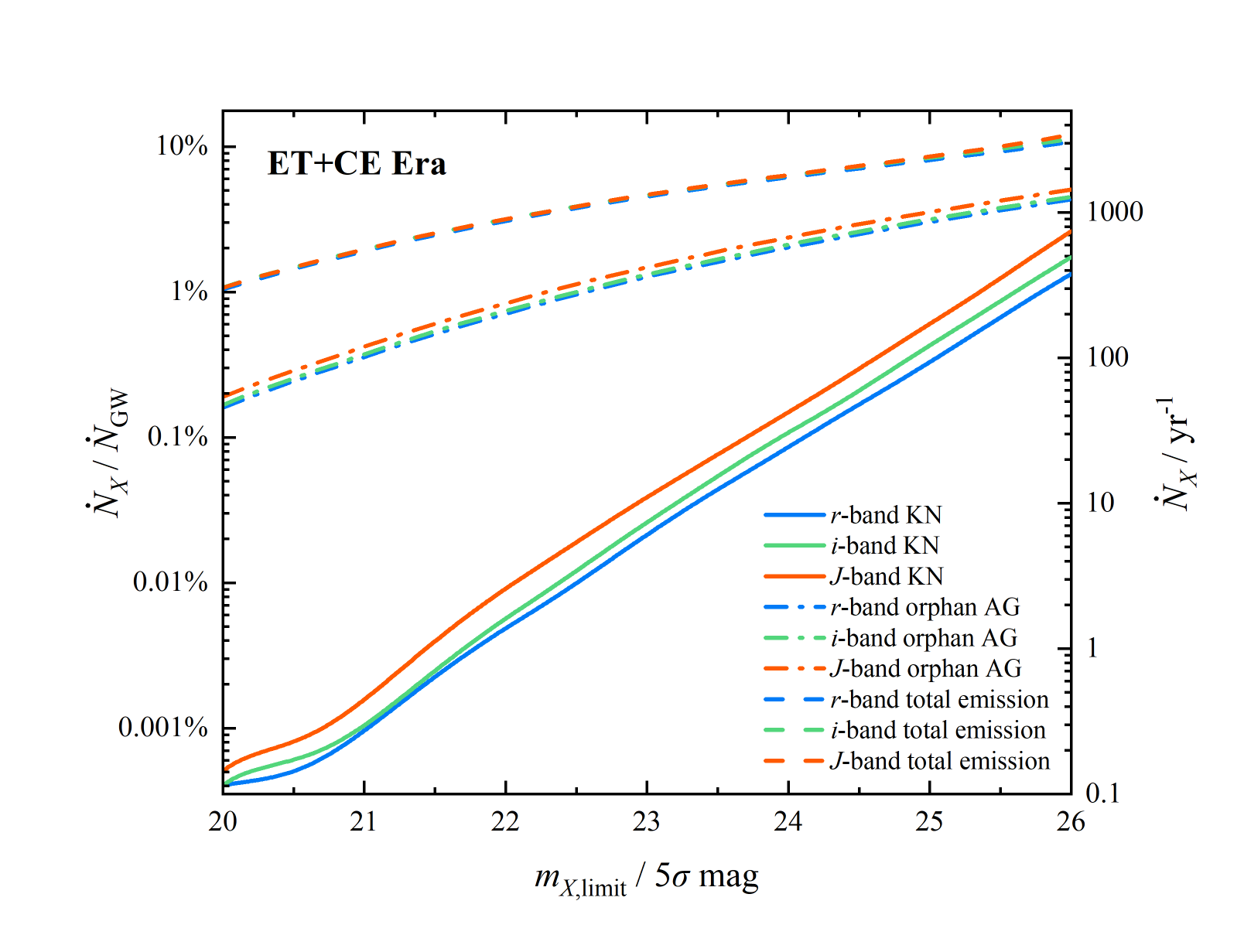}
    \caption{Detection fractions as functions of $5\sigma$ limiting magnitude and search filter for target-of-opportunity observations during four GW detection eras, i.e., HLVK, PlusNetwork, LIGO Voyager, and ET+CE eras. The solid, dashed-doted, and dashed lines denote the detection fractions of kilonova, orphan afterglow, and total emission, respectively. The blue, green, and red lines represent the detection fractions in $r$, $i$, and $J$ bands, respectively. The right axis of each panel represent the detection rates. The example parameter distributions of BH--NS system are adopted for the DD$2-$Low $M_{\rm NS}-$High $\chi_{\rm BH}$ model.} 
    \label{fig:ToO}
\end{figure*}

\begin{figure*}
    \centering
    \includegraphics[width = 0.99\linewidth , trim = 40 15 20 50, clip]{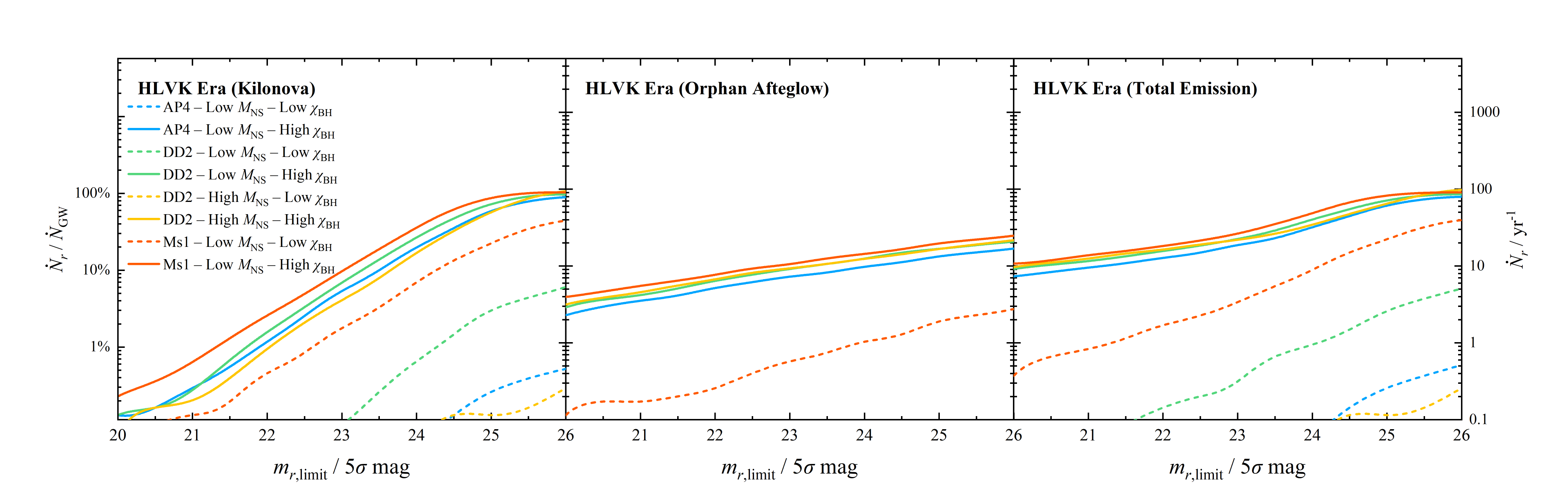}
    \includegraphics[width = 0.99\linewidth , trim = 40 15 20 50, clip]{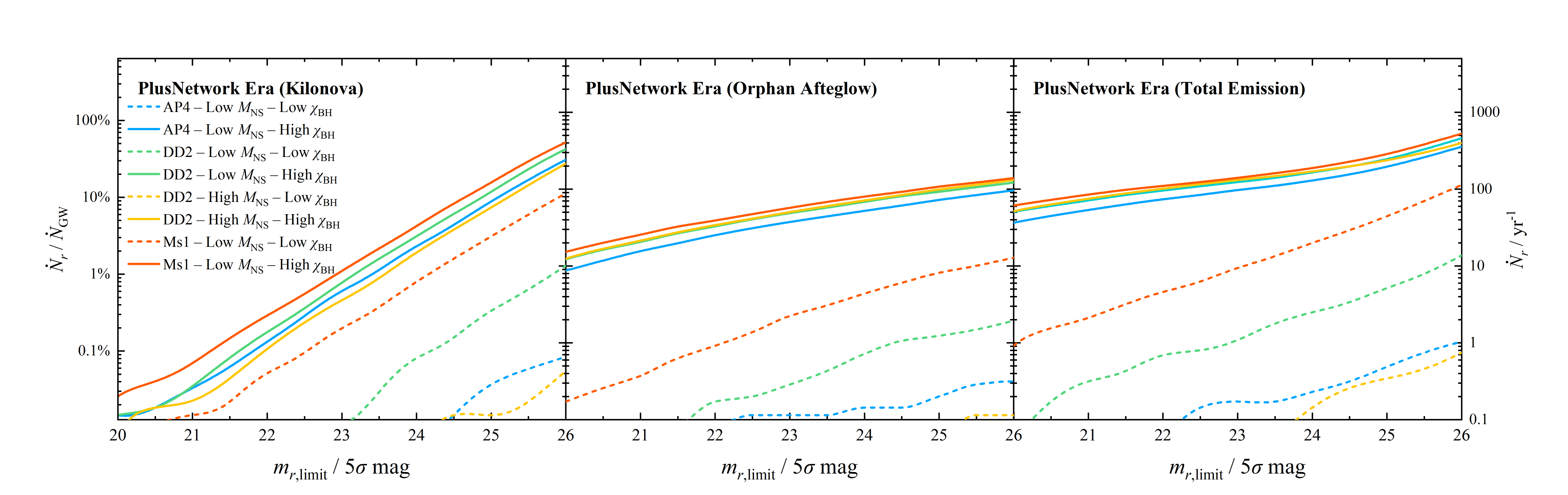}
    \includegraphics[width = 0.99\linewidth , trim = 40 15 20 50, clip]{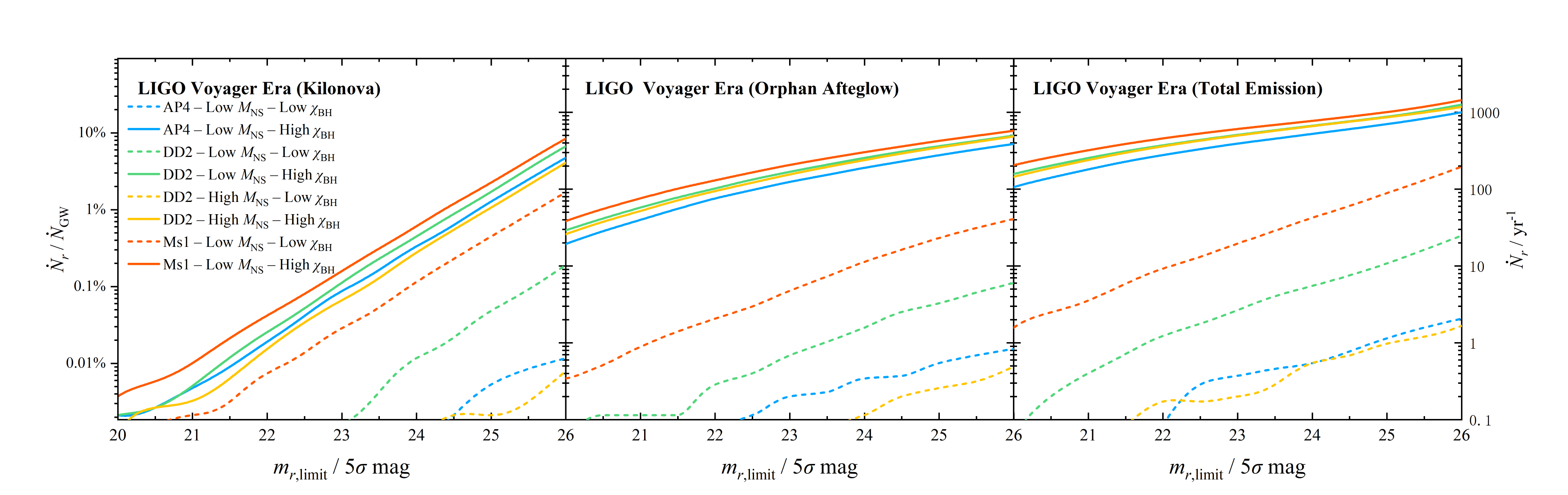}
    \includegraphics[width = 0.99\linewidth , trim = 40 15 20 50, clip]{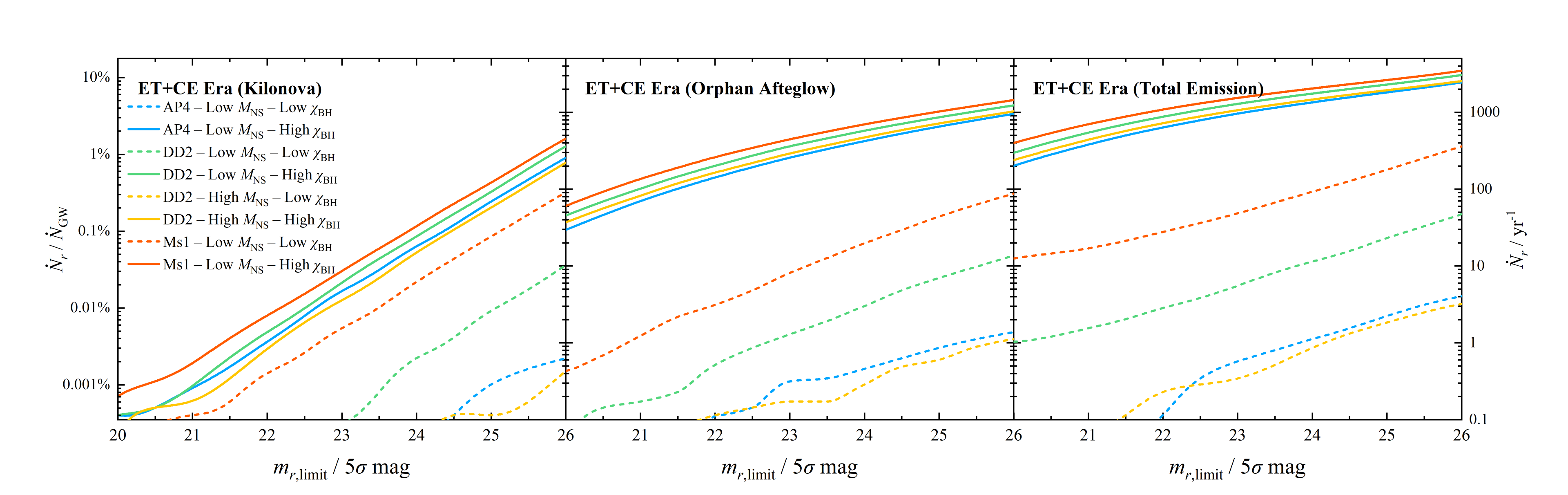}
    \caption{$r$-band EM detection fractions of GW-triggered events as a function of $r$-band $5\sigma$ limiting magnitude. The panels from left to right represent the detection fractions of kilonova, orphan afterglow, and total emission, respectively. The panels from top to bottom denote GW detection at four eras, i.e., HLVK, PlusNetwork, LIVO Voyager, and ET+CE eras. The right axis of each panel represent the detection rates. The line convention is the same as that in Figure \ref{fig:LF}.}
    \label{fig:ToO_r}
\end{figure*}

\begin{figure}
    \centering
    \includegraphics[width = 0.99\linewidth , trim = 40 280 720 60, clip]{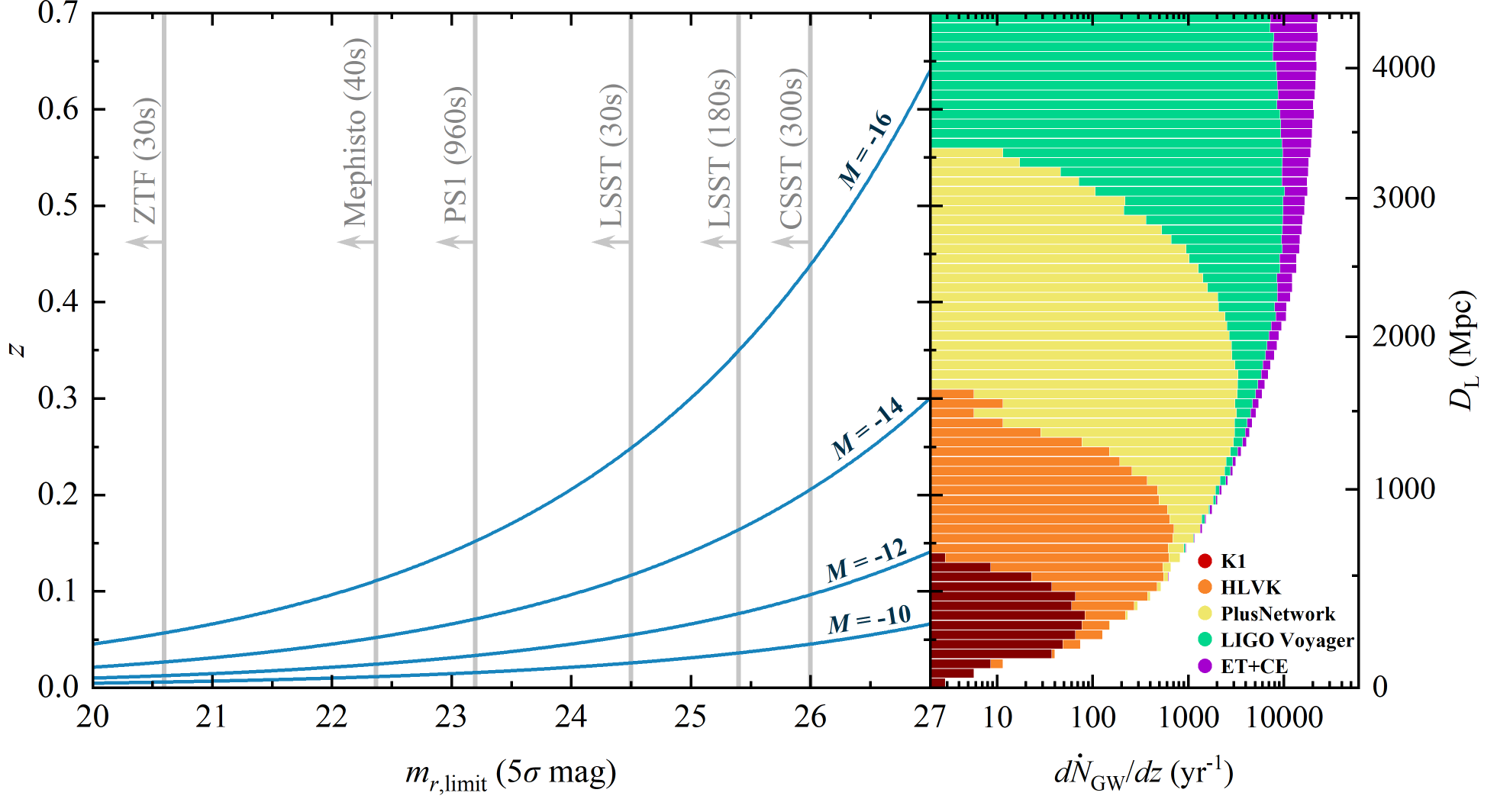}
    \caption{Left panel: the maximum search depth as a function of $5\sigma$ limiting magnitude for different peak absolute magnitudes of BH--NS merger kilonovae (blue lines). The gray lines represent the $r$-band $5\sigma$ limiting magnitude of present and planned survey projects, including ZTF, Mephisto, PS1, LSST, and CSST, with different exposure times. Right panel: GW detection rates per unit redshift  depending on distance for four GW detection eras. The example parameter distributions are adopted for the DD$2-$Low $M_{\rm NS}-$High $\chi_{\rm BH}$ model. }
    \label{fig:SearchDistance}
\end{figure}

We firstly consider specific parameter distribution for the DD2$-$Low $M_{\rm NS}-$High $\chi_{\rm BH}$ model as an example. Figure \ref{fig:ToO} shows the simulated detection fractions of kilonova, orphan afterglow, and total emission as a function of search limiting magnitude in three example observed filter bands, i.e., $riJ$ bands.  If the contribution from the afterglow is ignored, in the same search limiting magnitude for each filter, one can discover more kilonovae using redder bands since BH--NS merger kilonovae are optically dim but possibly infrared bright (\citetalias{zhu2020}). More specifically, kilonovae discovered in $J$ band are almost more than double of those discovered in $r$ band. However, in view that most survey projects always have a higher sensitivity in the optical band and that kilonovae lightcurves in the optical band evolves faster and reaches the peak with a shorter time, it seems that there is no significant difference between searching in optical and in infrared. Because the kilonova reaches the peak earlier in optical, a good strategy would be to follow up early in blue bands to learn the early-stage emission of the kilonova. 

In \citetalias{zhu2020}, we indicated that the AB absolute magnitudes of BH--NS merger kilonova are fainter than $\sim-15\,{\rm mag}$ and $\sim-16\,{\rm mag}$ in the optical and infrared bands, respectively. As shown in Figure \ref{fig:SearchDistance}, since the search limiting magnitudes of present and planned wide-field survey projects are $m_{\rm limit} \lesssim 26\,{\rm mag}$, almost all detectable BH--NS merger kilonovae would be discovered at a distance of $z\lesssim0.3$ which can cover the GW detection space in the HLVK era. For the GW events that are detected during the future eras except the HLVK era, one can hardly find more kilonovae by survey projects. As a result, there appears to be little difference in the amount of detectable kilonovae at different GW detection eras as shown in Figure \ref{fig:ToO}. Even so, for those BH--NS mergers whose jets beam toward us or slightly deviate from us, their sGRBs and afterglows could still be detected. For the wide-field survey projects whose search limiting magntidues are relatively shallow, Figure \ref{fig:ToO} reveals that it has a low probability of discovering BH--NS merger kilonovae after GW triggers. However, these survey projects can find bright on-axis jet afterglows by following up detections of sGRBs or orphan afterglows with a relatively higher probability than detecting kilonovae. Their relatively larger fields of view can make them more efficiently to localize the source, facilitating continued follow-up observations using large telescopes. In addition, the detection fractions of orphan afterglow and total emission would rise rapidly with the increasing search limiting magnitude and the continuous upgrading of the GW detectors. Although afterglow emission is beamed, it is the most probable detected EM signal in association of BH--NS GW signals in the future.

Since the detection fractions of kilonova, afterglow, and total emission are essentially not influenced by the search filter, we only show the $r$-band detection fractions for different parameter distributions of BH--NS system in Figure \ref{fig:ToO_r}. Similar to the discussion for the kilonova luminosity function in Section \ref{Sec:LFSystemParameter}, the detection fractions are also highly dependent on the parameter distributions, especially for the BH spin parameter. The BH--NS mergers can usually generate bright kilonovae and afterglows if the spins of the primary BHs are high. This would be beneficial for target-of-opportunity follow-up observations. However, for those GW-triggered BH--NS mergers that have a massive NS component with soft EoS and a low-spin primary BH, it is barely possible to detect their EM counterparts.

\subsubsection{Implications for Past and Future GW-triggered Target-of-opportunity observations}

Despite great efforts, follow-up observations to O3 LVC events revealed no creditable BH--NS merger kilonova. So, why did this happen? One possibility is that there was actually no EM counterpart produced from most BH--NS mergers. If the BH--NS system distributions tend to have low-spin primary BHs and massive NSs with a soft EoS, for a large portion of BH--NS mergers the NSs would directly plunge into the BHs without being tidally disrupted, so that no sGRB or kilonova is expected from these systems\footnote{Some weak EM signals could be produce if the NS or BH have a non-negligible charge \citep{zhang2019,dai2019}.}. Another possibility is that we could not detect the EM counterparts because the search limiting magnitude for follow-up survey observations is too shallow.

\begin{figure}
    \centering
    \includegraphics[width = 0.99\linewidth , trim = 50 30 95 60, clip]{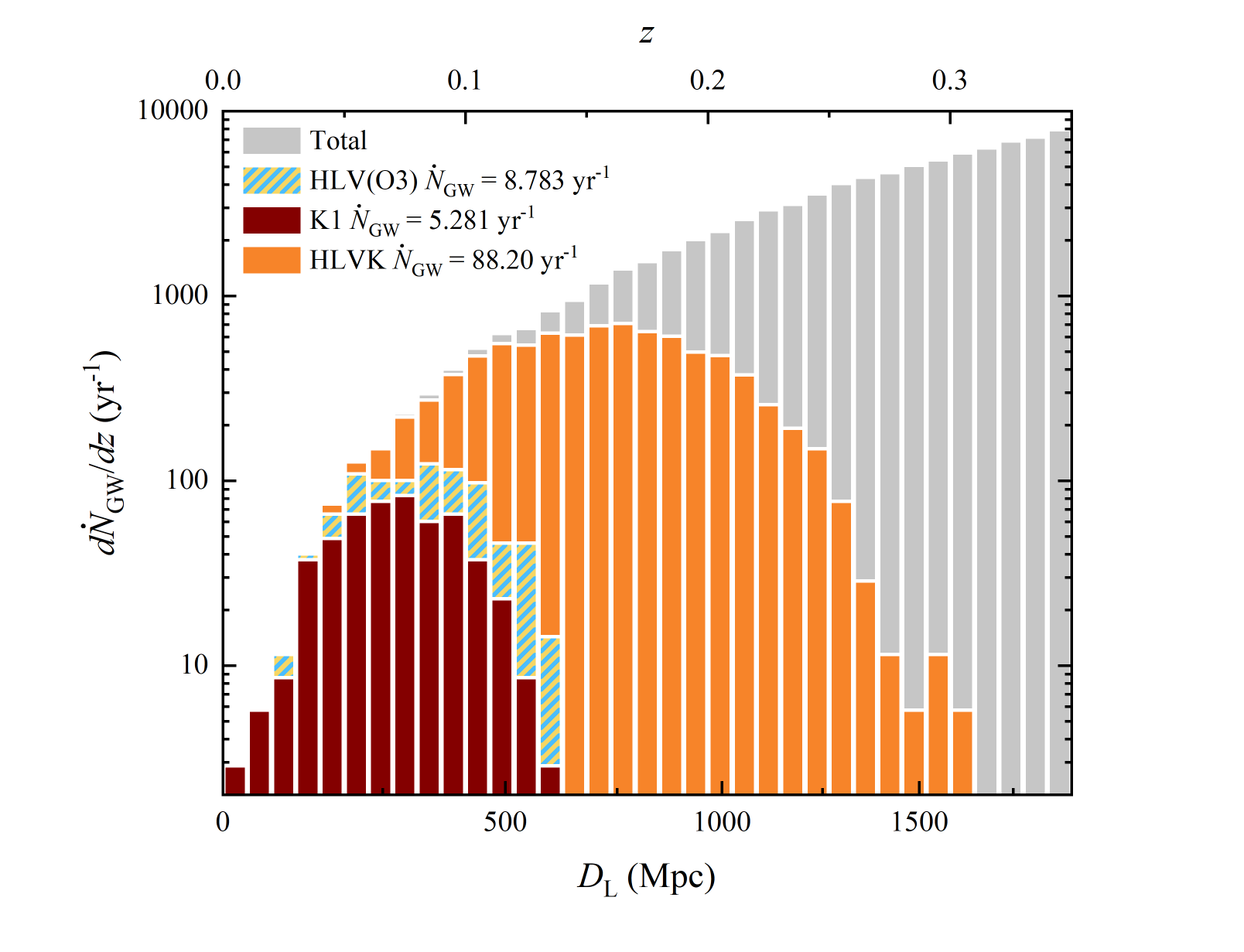}
    \caption{Comparison between the detectabilities for the network HLV (O3, striped histograms), for single K1 (carmine histograms), and for the network of 2nd generation GW detectors (orange histograms). Here we assume the example parameter distributions for the DD$2-$Low $M_{\rm NS}-$High $\chi_{\rm BH}$ model.}
    \label{fig:ComparsionO3andO4}
\end{figure}

Our simulation results show that the median detectable distance of targeted BH--NS mergers GW events for a single and a network of 2nd generation GW detectors are $\sim300\,{\rm Mpc}$ and $\sim700\,{\rm Mpc}$, respectively  (see Table \ref{table:gwDetectionDistances}). For  comparison, Figure \ref{fig:ComparsionO3andO4} shows that the detection rate and detectable distance for HLV (O3) are approximately the same as those for the case that only a bKAGRA is running. This is basically consistent with the detection rate and the distance distribution of BH--NS merger candidates detected during LVC O3 \citep[e.g.,][]{anand2020,antier2020b,coughlin2020b,compertz2020,kasliwal2020}. In Section \ref{Sec:LF}, {we have presented that the kilonova absolute magnitude at $0.5\,{\rm days}$ after a BH--NS merger} is mainly distributed in the range of $\sim-10$ to $\sim-15.5$. In view that the limiting magnitude of the follow-up wide-field survey projects is almost $\lesssim 21\,{\rm mag}$ \citep[e.g.][]{antier2020b,compertz2020,coughlin2020b,kasliwal2020,wyatt2020}, the maximum detectable distance for BH--NS kilonovae would be $\lesssim200\,{\rm Mpc}$ which can hardly cover the horizon of GW-triggered BH--NS merger events that O3 found (as shown in Figure \ref{fig:SearchDistance}). However, although BH--NS merger kilonovae can hardly be detected for the present search depths, Figure \ref{fig:ToO} and Figure \ref{fig:ToO_r} reveal that there are great opportunities to discover on-axis afterglows associated with sGRBs or orphan afterglows if the BH components have a high-spin distribution. In order to cover the distance range for searching BH--NS kilonovae for the network of 2nd generation GW detectors as complete as possible, a search limiting magnitude $m_{\rm limit}\gtrsim23 - 24$ is required as shown in Figure \ref{fig:SearchDistance}. Present survey projects could reach this search limiting magnitude by increasing exposure times and number of simultaneous exposures. However, the GW candidates during O3 had very large localization areas with an average of thousands of square degrees \citep{antier2020b}. Increasing exposure times makes it hard for the present survey projects to cover such large localization areas. Therefore, during the HLVK era, we recommend that survey projects may search for jet afterglows after GW triggers with a relatively shallow search limiting magnitude. If BH--NS mergers have a high location precision, reaching a limiting magnitude to $m_{\rm limit}\gtrsim23 - 24$ can be made, which gives a higher probability to discover associated kilonovae. 

At the late period of the HLVK era, LSST \citep{lsst2009} with its 8-meter-class telescope will come online by 2021 and begin full science operation by 2023. {It can detect kilonova emission} of almost all BH--NS merger GW signals to be triggered by the 2nd generation detector network (see Figure \ref{fig:SearchDistance}). Due to its large field of view ($9.6\,{\rm deg}^2$) and deep search limiting magnitude (e.g., $m_{r,{\rm limit}} = 24.7\,{\rm mag}$ with $30\,{\rm s}$ exposure time), LSST can be used to achieve follow-up detections and identifications of kilonovae with a high efficiency \citep{cowperthwaite2019,setzer2019}. If the localization of BH--NS merger GW signals is only a few hundred square degrees, Mephisto \citep{er2020}, which will see the first light by 2021 and is expected to  operate in 2022, can allow a deep enough volumetric coverage of BH--NS merger events within a short period of time. Its capability of simultaneous imaging in three bands allows to obtain color information that is conducive to rapidly identifying kilonovae and afterglows. In addition, the CSST \citep{zhan2011}, which also has a wide and deep survey capability, is planned to be launched in 2024. For those GW signals with sky positions to a few square degrees \citep{abbott2018prospects}, CSST with a $1.1\,{\rm deg}^2$ effective field-of-view can also efficiently detect the EM counterparts of the GW events. Its good performance in ultraviolet band can provide essential early observations that allow one to study the initial emission properties of kilonovae.

\section{Detectability for Time-domain Survey Observations\label{Sec:Time_domain}}

\begin{deluxetable*}{cccccc}[tb!]
\tablecaption{Summary Technical Information for Each Survey\label{table:Telescope}}
\tabletypesize{\small}
\tablecolumns{6}
\tablewidth{0pt}
\tablehead{
\colhead{Survey} &
\colhead{Filters} &
\colhead{Survey Depth/$5\sigma$ mag} &
\colhead{Field of View/$\rm deg^2$} &
\colhead{Areal Survey Rate\tablenotemark{a}/${\rm deg}^2\,{\rm day}^{-1}$} &
\colhead{Reference}
}
\startdata
PS1 & $grizy$ & $23.3,\,23.2,\,23.1,\,22.3,\,21.4$ & $7$ & $49.9$ & (1)\\
ZTF & $gri$ & $20.8,\,20.6,\,19.9$ & $47.7$ & $11280$ & (2) \\
Mephisto & $uvgriz$ & $21.6, 21.7, 22.6, 22.4, 22.0, 21.1$ & $3.14$ & $1356$ & (3)\\
LSST & $ugrizy$ & $23.9, 25.0, 24.7, 24.0, 23.3, 22.1$ & $9.6$ & $3240$ & (4) \\
WFIRST & $RZYJHF$ & $26.2, 25.7, 25.6, 25.5, 25.4, 24.9$ & $0.281$ & $8.29$ & (5) \\
CSST & $NUVugrizy$ & $25.4,\,25.4,\,26.3,\,26.0,\,25.9,\,25.2,\,24.4$ & $1.1$ & $32.7$ & (6)\\
\enddata
\tablenotetext{a}{Areal survey rate for each survey is calculated with the assumptions of an average $6\,{\rm hr}$ observing time per night and two exposure bands for each visit. Since Mephisto can achieve simultaneously imaging in three bands, we calculate its area survey rate by assuming single exposure for each visit. }
\tablecomments{Reference: (1) \cite{chambers2016}; (2) \cite{bellm2019,masci2019}; (3) \cite{er2020,lei2021}; (4) \cite{lsst2009}; (5) \cite{hounsell2018,scolnic2018}; (6) \cite{cao2018,gong2019}.}
\end{deluxetable*}

Even though GW-triggered follow-up observations is the best way to achieve an optimal searching for kilonovae, the time in survey projects that can be assigned to target-of-opportunity follow-up observations of GW triggers is limited. For example, only a time investment of $\sim1\%-2\%$ of the total survey time for LSST is designated to follow-up observations \citep{margutti2018ToO}. Therefore, it is also important to discuss how many EM counterparts of BH--NS mergers can be discovered from time-domain survey observations. In this section, we show the simulation results of the detection rates of kilonova, orphan afterglow, and total emission from BH--NS mergers for time survey observations as functions of search filter, limiting magnitude, and cadence time. In practice, if one detects a fast-evolving transient, in order to confirm whether it is a kilonova, one needs to use its recorded color evolution to perform analyses. Therefore, for specific survey projects we discuss, including PS1, ZTF, Mephisto, LSST, WFIRST, and CSST, we assume that these survey projects would search each point in two filters that can be beneficial for identifications. Values of some technical parameters for these survey telescopes are given in Table \ref{table:Telescope}.

\subsection{Method}

\begin{figure}
    \centering
    \includegraphics[width = 0.99\linewidth , trim = 60 30 95 60, clip]{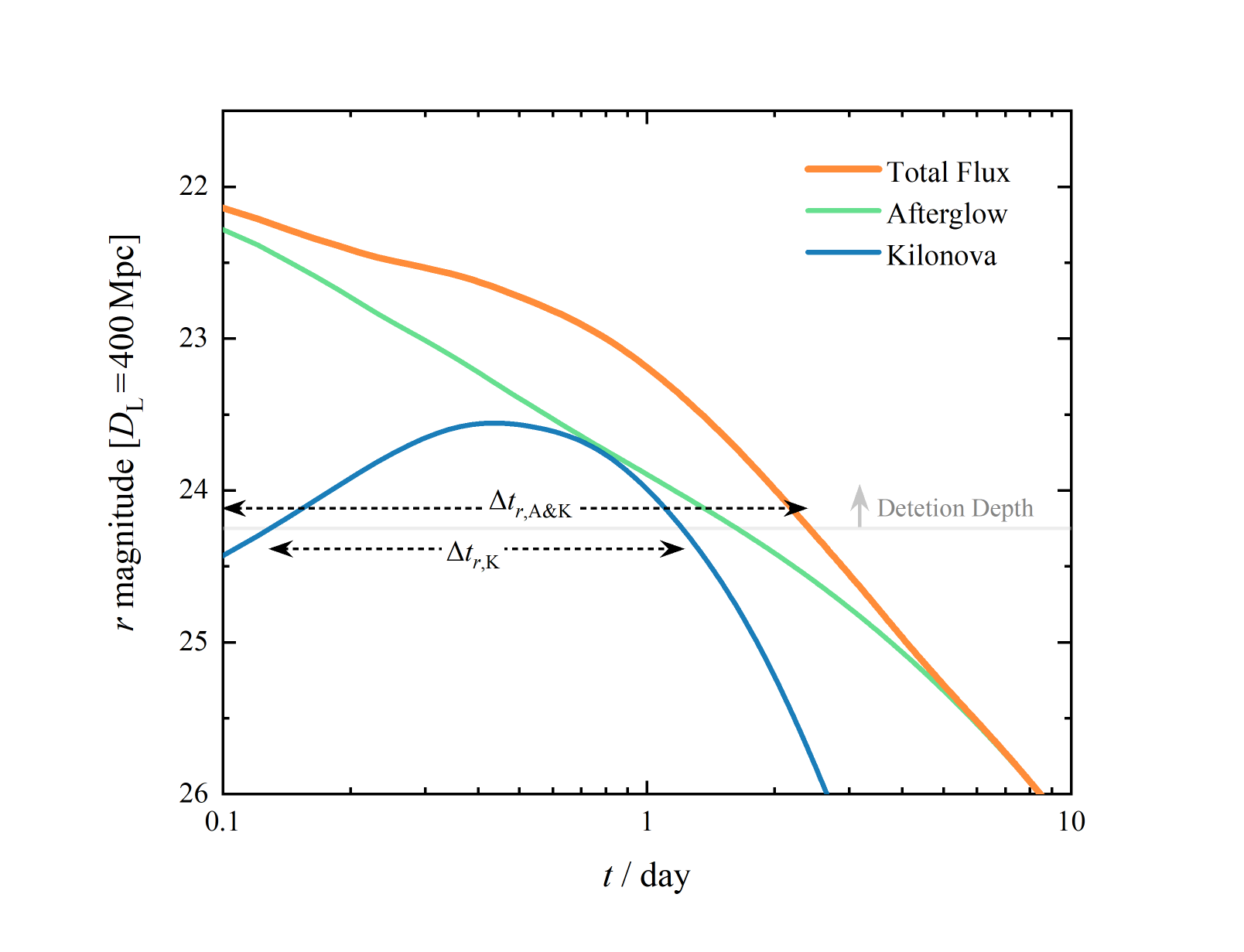}
    \caption{A schematic diagram of observed afterglow and kilonova lightcurves from the BH--NS merger exploded at low density environment, for observers with a distance of $D_{\rm L} = 400\,{\rm Mpc}$. The green, blue, and orange lines represent $r$-band afterglow emission, $r$-band kilonova emission, and total emission, respectively. For a given limiting magnitude of telescope, we define the timescales that {the kilonova emission and total emission are larger than the magnitude }are $\Delta t_{r,{\rm K}}$ and $\Delta t_{r,{\rm A}\&{\rm K}}$, respectively.}
    \label{fig:DeltaT}
\end{figure}

For the time-domain survey observations, the probability that a source is detected could be considered as the ratio of survey area within the time that the kilonova brightness is above limiting magnitude to the area of the celestial sphere. By defining $\omega$ as the areal survey rate (i.e., sky area covered per unit time) {which is dependent on specific survey telescope} and $\Delta t_{X,{\rm K}}$ is the time during which kilonova brightness is above limiting magnitude in $X$ band $m_{X,{\rm limit}}$ (see Figure \ref{fig:DeltaT} for a schematic diagram), one can obtain the survey area as $\omega\Delta t_{X,{\rm K}}$. {Therefore, the maximum probability for a source to be detected is $\omega\Delta t_{X,{\rm K}}/4\pi$.} However, high-cadence observations would restrict the survey area to $\omega t_{\rm cad}$, {where the cadence time $t_{\rm cad}$ is defined as the interval time between consecutive observations of the same sky area by a telescope.} {It means that the probability for a source being detected by the high-cadence search would be a constant, i.e., $\omega t_{\rm cad}/4\pi$}. We thus express the survey area as $\omega\Delta t$,  {where $\Delta t = \min{(t_{\rm cad}, \Delta t_{X,\rm  K})}$}. {By counting the detection probabilities of all cosmological sources}, one can write the detection rate for time-domain observations as
\begin{equation}
\label{Eq: BlindSurveyObservation}
    \dot{N}_X = \frac{\omega\Delta t}{4\pi}\int_0^{z_{\rm max}}\frac{\dot{\rho}_0f(z)}{1 + z}\frac{dV(z)}{dz}dz.
\end{equation}
If one considers both emission from kilonova and afterglow, $\Delta t_{X,{\rm K}}$ in Equation (\ref{Eq: BlindSurveyObservation}) can be substituted by $\Delta t_{X,{\rm A\&K}}$. We note that $\omega\Delta t$ is required to be smaller than $4\pi$. 

\subsection{EM Detection Rates for Time-domain Survey Observations}

\begin{figure*}
    \centering
    \includegraphics[width = 0.99\linewidth , trim = 40 10 70 20, clip]{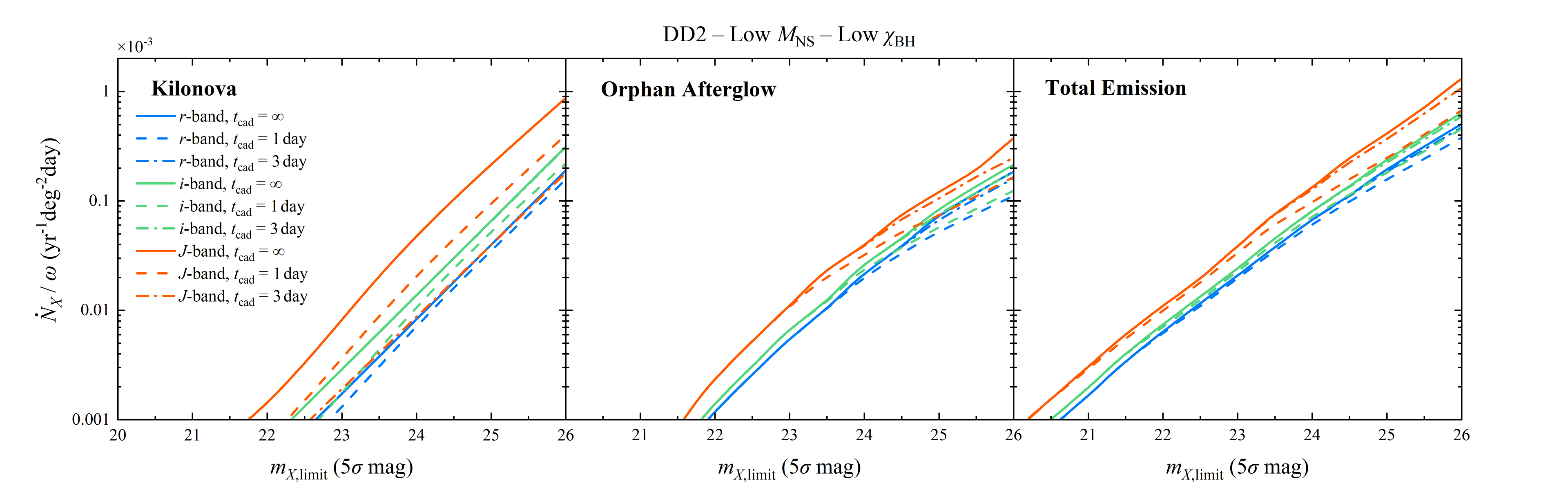}
    \includegraphics[width = 0.99\linewidth , trim = 40 10 70 10, clip]{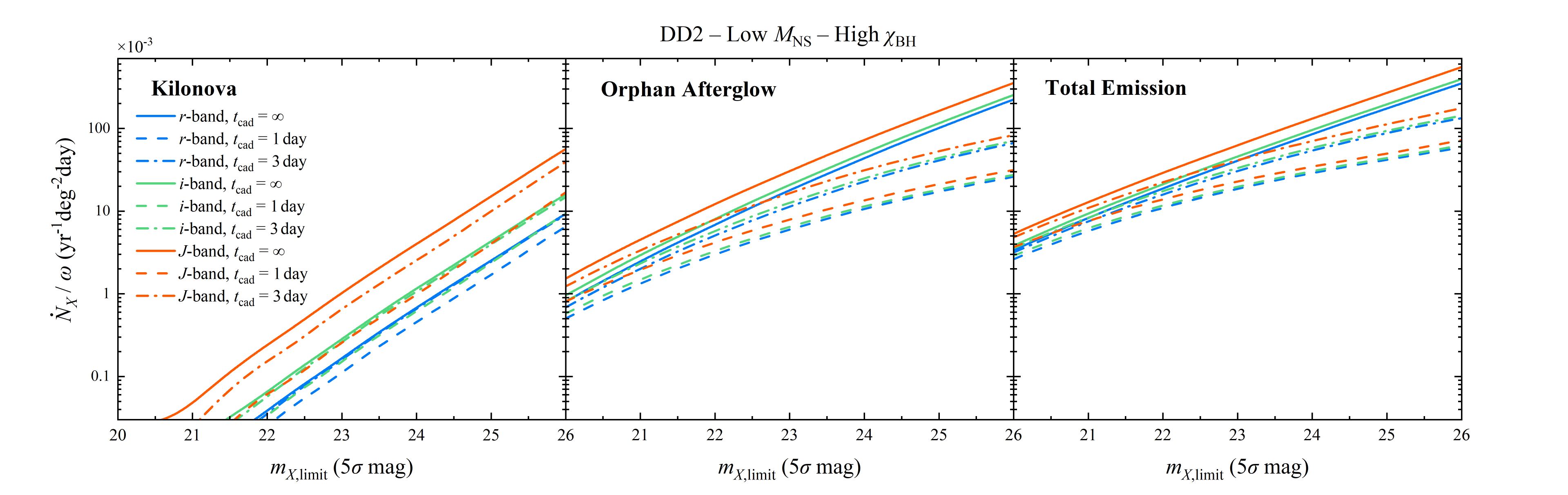}
    \caption{Detection rates per unit search rate of kilonovae (left panels), orphan afterglows (middle panels), and total emissions (right panels) as functions of $5\sigma$ limiting magnitude, cadence time, and search filter for time-domain survey observation. The solid, dashed, dashed-doted lines represent the detection rates with no cadence, $1\,{\rm day}$ cadence, and $3\,{\rm days} $ cadence, respectively. The panels from left to right represent the detection rates of kilonova, orphan afterglow, and total emission, respectively. We show two example parameters distributions of BH--NS system, i.e., DD2$-$Low $M_{\rm NS}-$Low $\chi_{\rm BH}$ for top panels, and DD2$-$Low $M_{\rm NS}-$High $\chi_{\rm BH}$ for bottom panels.} 
    \label{fig:Blind}
\end{figure*}

\begin{figure*}
    \centering
    \includegraphics[width = 0.99\linewidth , trim = 40 10 70 20, clip]{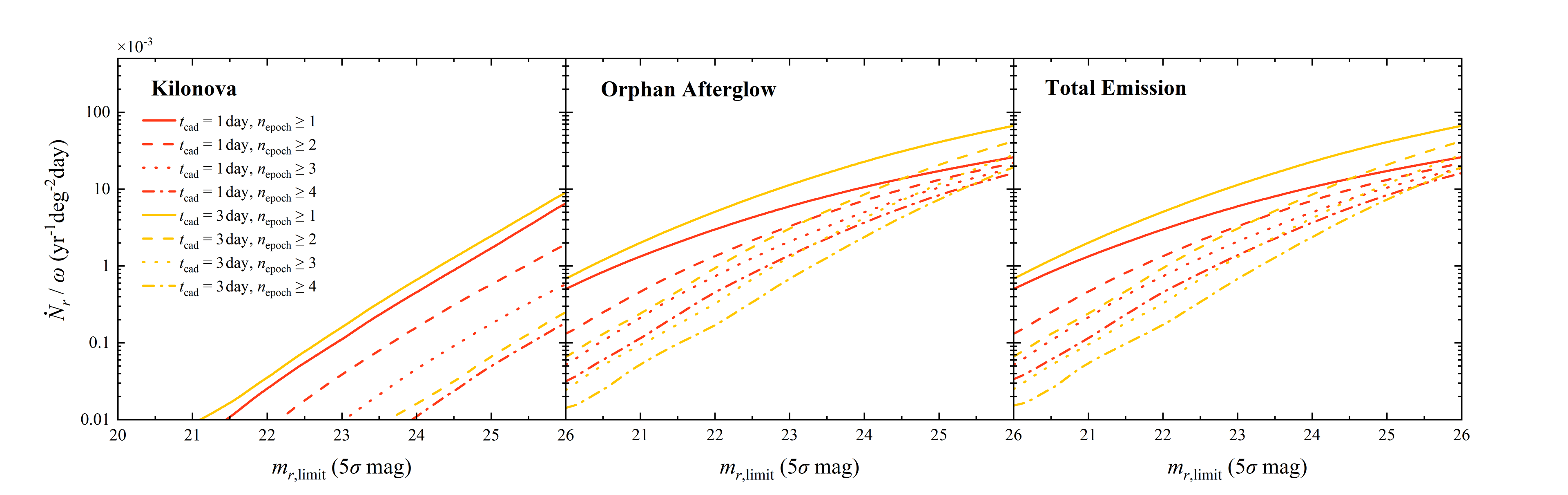}
    \includegraphics[width = 0.99\linewidth , trim = 40 10 70 20, clip]{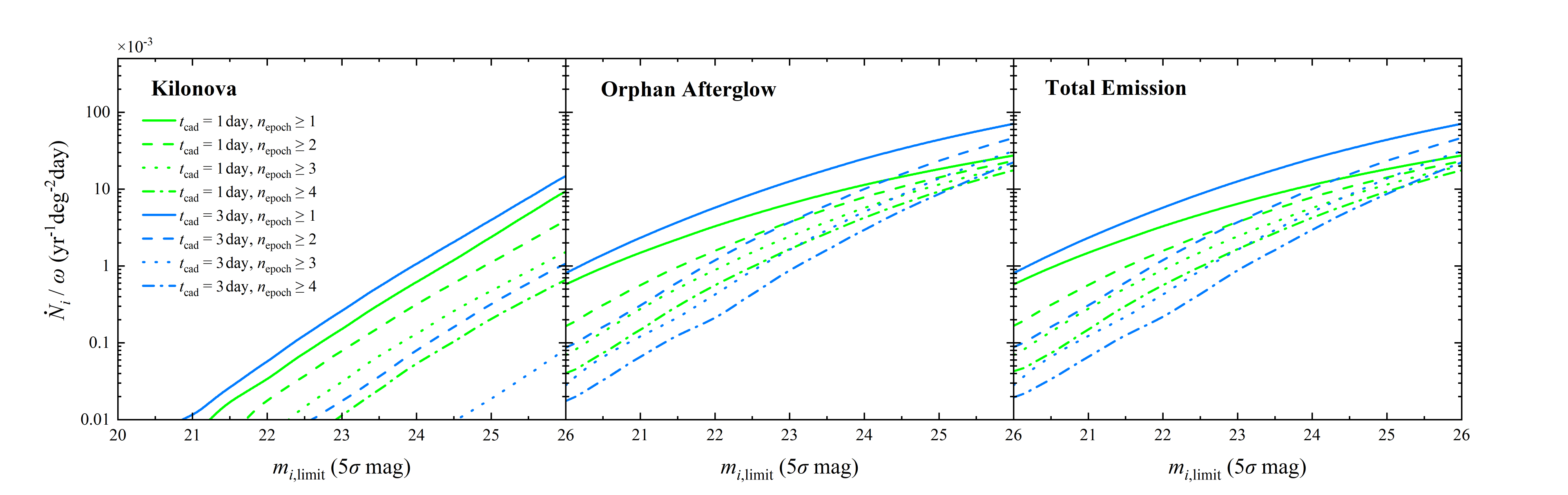}
    \includegraphics[width = 0.99\linewidth , trim = 40 10 70 20, clip]{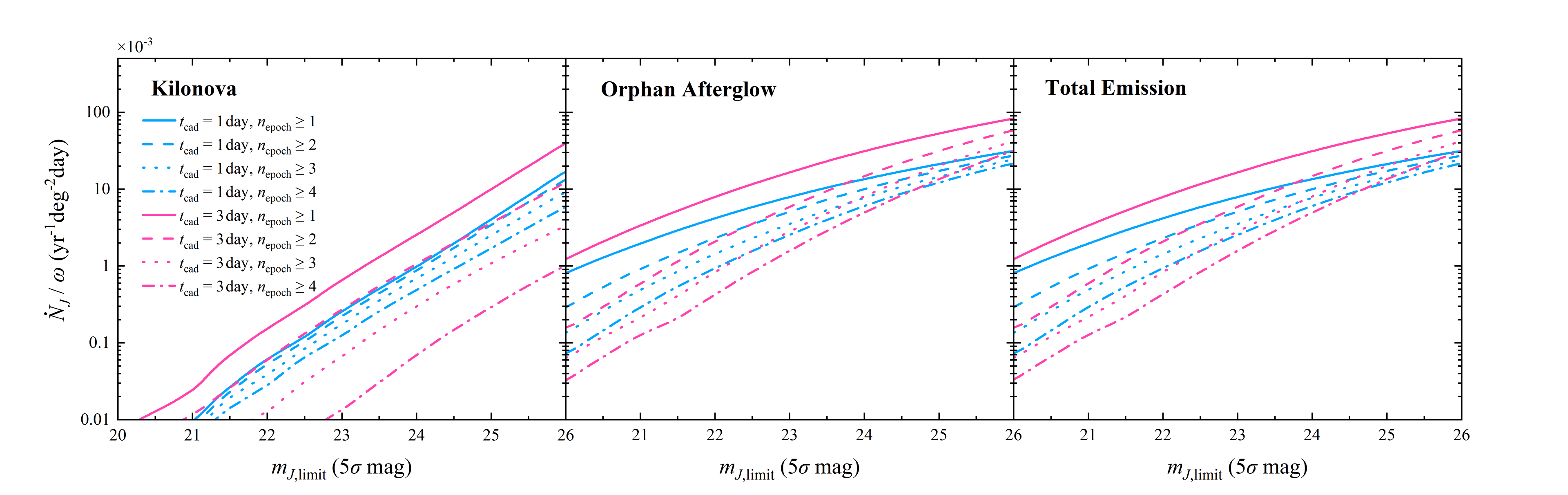}
    \caption{Detection rates per unit search rate with different detectable epoches for kilonovae (left panels), orphan afterglows (middle panels), and total emissions (right panels). The panels from top to bottom represent EM counterparts to be detected in $r$ band, $i$ band, and $J$ band, respectively. The observations can be detected in at least one, two, three, and four epoches are denoted as solid lines, dashed lines, dotted lines, and dashed-dotted lines. The red, green, and light blue lines represent the detection rates by adopting one-day cadence searching strategy, while yellow, mazarine, and purple lines represent the detection rates by adopting three-day cadence searching strategy. The example parameter distributions of BH--NS system are ${\rm DD}2-{\rm Low}\  M_{\rm NS}-{\rm High}\ \chi_{\rm BH}$.} 
    \label{fig:Blind_band}
\end{figure*}

By giving two example parameter distributions of BH--NS system, i.e., DD2$-$Low $M_{\rm NS}-$Low $\chi_{\rm BH}$ and DD2$-$Low $M_{\rm NS}-$High $\chi_{\rm BH}$, we show the detection rates per unit search rate (i.e., $\dot{N}_X / \omega$) of BH--NS merger kilonova, orphan afterglow, and total emission for time-domain observations with the consideration of different limiting magnitudes, search filters and cadence times in Figure \ref{fig:Blind}. In \citetalias{zhu2020}, we revealed that BH--NS merger kilonovae could be dim in optical but bright in infrared. Due to the large amount of lanthanide-rich ejecta formed after BH--NS mergers, its longer diffusion timescale makes the infrared kilonova emission evolve slower. Therefore, the time duration when the kilonova brightness is above the limiting magnitude would be longer in the infrared band. It is obvious that searching in redder filter bands can discover more sources for time-domain survey observations. Figure \ref{fig:Blind} shows that the detection rate of kilonova in $J$ band is $\sim8-10$ times more than in $r$ band. In contrast, the detection rate ratios between $J$-band and $r$-band for orphan afterglow and total emission are close to $1.5-2$.

Figure \ref{fig:Blind} also shows that high-cadence time-domain observations, e.g., $t_{\rm cad} = 1\,{\rm day}$, would lose detection rates significantly. This is because the shorter the search cadence, the smaller the sky area that telescopes can cover at the same time while the detection rates would be constrained. However, it makes kilonovae or afterglows detectable in a few epochs by survey projects, which is of great advantage for identification. Using a three-day cadence with limiting magnitude of $m_{X{\rm ,limit}} \gtrsim 23$ to search for orphan afterglows and total emissions would obtain the same conclusion, because afterglows, especially off-axis ones, are usually bright and last longer. If one searches kilonovae or afterglows using small telescopes with a relatively longer cadence time, e.g., $t_{\rm cad} = 3\,{\rm day}$, the detection rates would not decrease significantly since the cadence time is almost larger than the time when the brightness is above limiting magnitude. However, this would cause that most of sources can be only detected in one or two epochs, which makes it hard to identify them.

In view that the kilonova and afterglow emission from BH--NS mergers whose BH component has low spin is difficult to detect for time-domain searches, we only show the detection rates of kilonova, orphan afterglow, and total emission by setting the parameter distributions as those of the DD2$-$Low $M_{\rm NS}-$High $\chi_{\rm BH}$ model in Figure \ref{fig:Blind_band}. A three-day cadence, especially for searches in $r$ band and $i$ band, can barely detect kilonovae in only one epoch. However, if one adopts high-cadence searches, kilonovae are expected to be detected many times. Figure \ref{fig:Blind_band} shows that there are $\sim25\%$ detectable kilonovae that can be probed in at least 4 epochs by using a one-day cadence probed in $J$ band. Therefore, we suggest that the time-domain survey observations for kilonovae should adopt a one-day cadence search strategy in redder bands. A similar strategy is recommended for orphan afterglow and total emission searches with a relatively shallow limiting magnitude. As also shown in Figure \ref{fig:Blind_band}, the difference of orphan afterglow and total emission detection rates between high-cadence and low-cadence searches is small for deep survey projects.

\subsection{Detectability for Specific Survey Project}

\begin{figure*}
    \centering
    \includegraphics[width = 0.90\linewidth , trim = 40 120 20 40, clip]{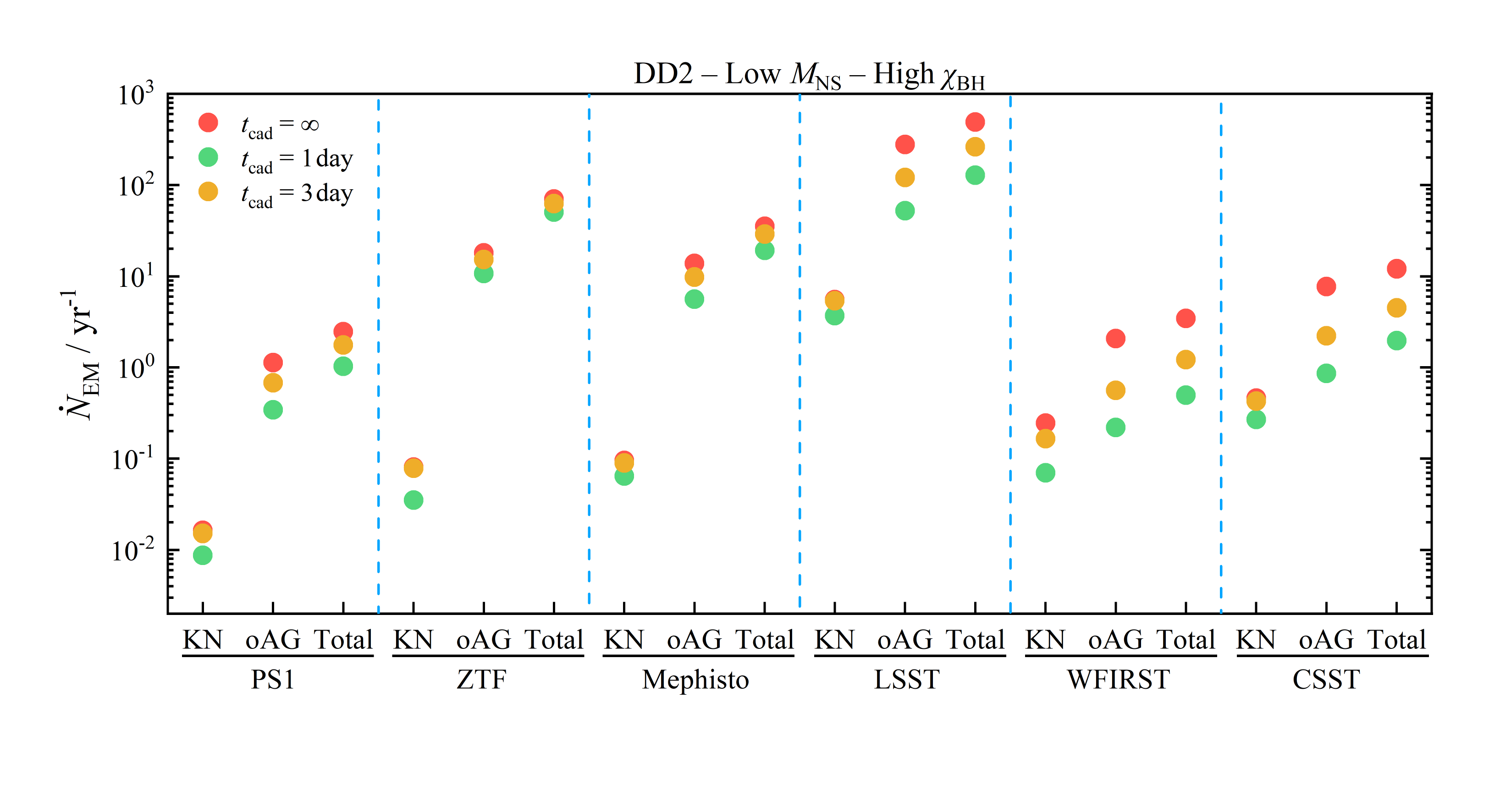}
    \caption{Detection rates of kilonvoa, orphan afterglow, and total emission for specific survey projects, including PS1, ZTF, Mephisto, LSST, WFIRST, and CSST. The green, yellow, and red circles represent the detection rates by adopting a one-day cadence strategy, three-day cadence strategy, and no cadence searching, respectively. The example parameter distributions of BH--NS system are those of the DD2$-$Low $M_{\rm NS}-$High $\chi_{\rm BH}$ model.} 
    \label{fig:Detectivity}
\end{figure*}

For time-domain survey observations, we simulate the detection rates of kilonova, orphan afterglow, and total emission based on the survey information for PS1, ZTF, Mephisto, LSST, WFIRST, and CSST, collected in Table \ref{table:Telescope}. The predicted values of detections for each survey by considering an example parameter distributions, i.e., DD2$-$Low $M_{\rm NS}-$High $\chi_{\rm BH}$, are exhibited in Figure \ref{fig:Detectivity}. The possibility for detecting EM counterparts from BH--NS mergers is extremely low if the BH component has a low spin distribution. 

Overall, for time-domain survey observations, the primary EM counterpart targets for BH--NS mergers would be on-axis jet afterglows and orphan afterglows. Compared with other survey projects, the relatively deep depth and wide field of view of LSST with a 8-meter-class telescope would detect most EM counterparts from BH-NS mergers. ZTF has the shallowest search depth, but could still discover many BH--NS merger afterglows due to its ultra wide field of view. Mephisto should have similar results to ZTF. For other survey projects, despite their deep search limiting magnitudes, the possibilities for searching EM counterparts from BH--NS mergers would be limited due to their relatively small field of view.

\section{Conclusions and Discussion \label{Sec:Conclusions}}

In this paper, based on our model proposed in the companion paper (\citetalias{zhu2020}), we have presented  simulated BH--NS merger kilonova luminosity function, detectability of GWs for different generations of GW detectors, joint-search  GW signals and optical EM counterparts, as well as time-domain survey blind-search detectability for BH--NS EM signals. Our simulated results provide important implications for targeted-of-opportunity observations of GW triggers and future time-domain survey searches. 

{\em Effect of parameter distributions} --- {We have discussed} the parameter distributions of BH--NS system in detail and discuss their effect on the kilonova luminosity function, GW signal detections, GW-triggered target-of-opportunity observations, and time-domain survey blind searches for BH--NS merger EM signals. Different parameter distributions show only small effects on the GW detections, detectable distances, and chirp times of BH--NS merger events. On the contrary, different distributions can significantly influence kilonova luminosity function and EM detections. In the case of primary BH having a high-spin distribution and NS component being less massive with a stiff EoS, the NS can be disrupted by the BH for almost all cases to power a bright kilonova and afterglow \citep[e.g.,][]{barbieri2019,barbieri2020,bhattacharya2019}. {Our simulated results have shown} that the kilonova luminosity function and EM detectability might rely more on the BH spin distribution. For those BH--NS mergers whose primary BH has low-spin, the NS would generally plunge into the BHs, or the NS would undergo tidal disruption but only a small amount of material can remain outside the post-merger BH remnant. If the BH spin in BH--NS systems is universally low, the brightness distribution of BH--NS merger kilonovae could be dim, and the detectable event rate would be low. Therefore, in such a case, the EM counterparts from BH--NS mergers could be difficult to discover. At present, the BHs predicted by several studies \citep[e.g.,][]{qin2018,fuller2019} and detected from GW observations \citep[e.g.,][]{abbott2019gwtc} usually rotate very slowly, while the BHs in some observed high-mass X-ray binaries (which could be progenitors of BH--NS systems) are born with a high spin \citep[e.g.,][]{mcclintock2011,miller2011}. In view of the little present knowledge about BH spin in BH--NS systems, the constraints by future GW detections could improve our understanding of BH--NS mergers and their EM signals. 

{\em Luminosity Function} --- By adopting the viewing-angle-dependent model of radioactivity-powered BH--NS merger kilonova model presented in \citetalias{zhu2020}, we for the first time present a detailed calculation of BH--NS merger kilonova luminosity function with the consideration of different parameter distributions. The predicted kilonova absolute magnitude at $0.5\,{\rm days}$ after a BH--NS merger is mainly distributed in the range of $\sim-10$ to $\sim -15.5$, corresponding to the bolometric luminosity from $\sim3\times10^{39}\, {\rm erg}\,{\rm s}^{-1}$ to $\sim5\times10^{41}\,{\rm erg}\,{\rm s}^{-1}$.  {The differences in luminosity of BH--NS kilonova between our model and other models \citep[e.g.,][]{kawaguchi2016,kawaguchi2020,barbieri2019,darbha2021} are within a factor of two, corresponding to uncertainties of  $\sim1\,{\rm mag}$ for the luminosity function.} The simulated luminosity functions potentially contain the viewing angle distribution information of the anisotropic kilonova emission. Besides, the possible additional energy injection from the remnant BH, e.g., due to fall-back accretion \citep{rosswog2007,kyutoku2015} or the Blandford-Payne mechanism \citep{blandford1982,ma2018}, would make the kilonova luminosity function more complex. Compared with BH--NS mergers, BNS mergers could form more polar-dominated lanthanide-free neutrino-driven wind ejecta \citep[e.g.,][]{grossman2014,perego2014,kasen2015,martin2015} whose mass could be dependent on the survival times of remnant NSs \citep[e.g.,][]{kasen2015,fujibayashi2017,gill2019}, so that lightcurves of BNS merger kilonovae would be diversified and highly viewing-angle-dependent. {Furthermore, the long lasting X-ray afterglows in sGRB events \citep{compertz2013,compertz2014,rowlinson2014,lv2015} possibly indicated continuous energy ejection from a long-lived remnant magnetar after some BNS mergers.} The spin-down of magnetars could also power X-ray tranisents \citep{zhang2013,sun2017,sun2019,xue2019,xiao2019}, give rise to a strong broadband afterglow-like emission upon interaction between the merger ejecta and the ambient medium \citep{gao2013b,liu2020} and heat the ejecta to enhance the brightness of the kilonovae \citep{yu2013,metzger2014a}. Such magnetar-powered kilonovae could be a few tens or hundreds times brighter than the radioactive-powered kilonovae. Therefore, it can be predicted that the kilonova luminosity function would be more complex for BNS mergers. Future extensive researches of diverse kilonovae from different GW events, constraints on the event rates of compact binary mergers, NS EoS, and NS mass distribution can lead to better constraints on the kilonova luminosity function. {Moreover,  it could help us to better understand the NS EoS and the spin distribution of BH component by building the luminosity function and jointly analysing with GW data in the future.} For future detected fast-evolving transients, one can make use of of their peak luminosity and inferred event rate to judge whether these events could be kilonovae.

{\em GW Detections} --- {We have made detailed calculations} on the detection capabilities of the 2nd, 2.5th, and 3rd generation detector networks in the near future for BH--NS signals. By setting the local event rate of BH--NS mergers as $\dot{\rho_0} = 35\,{\rm Gpc}^{-3}\,{\rm yr}^{-1}$, the GW detection rates of the worst and best cases for the 2nd generation GW detector network are $\sim5\,{\rm yr}^{-1}$ and $\sim 80\,{\rm yr}^{-1}$, respectively. The values of the worst and best detection rates for 2.5th generation GW detector network would increase to $\sim 30\,{\rm yr}^{-1}$ and $\sim650\,{\rm yr}^{-1}$, respectively. For the 3rd generation detector network, LIGO Voyager can detect $\sim4,800$ BH--NS merger events per year. A single ET and the ET+CE network are expected to detect all BH--NS merger events in the entire Universe, with the detection rates $\sim 2.6\times10^{4}\,{\rm yr}^{-1}$ and $2.84\times10^{4}\,{\rm yr}^{-1}$, respectively. As the sensitivity of the GW detector increases, BH--NS events at high redshifts gradually dominate of the detected events. For ET and ET+CE, the detection rate is mainly dominated by the BH--NS mergers at $z\sim1$. We also show the chirp time of different generation detector networks. For the 2nd generation, 2.5th generation detectors, and LIGO voyager, the median chirp time is distributed in the range of $\sim{2 - 4}\,{\rm min}$, and does not exceed $10\,{\rm min}$. Since the minimum frequency reduces from $10\,{\rm Hz}$ to $1\,{\rm Hz}$ for ET and ET+CE, the chirp time would reach the order of hours or even days with a median value close to $\sim10\,{\rm hr}$, which can provide an enough preparation time for EM observations. The detailed GW detection rates, median and maximum detectable distances, chirp time and total mass in geocentric frame of detectable GW signals can be consulted from Table \ref{table:gwDetectionRate}, Table \ref{table:gwDetectionDistances}, and Table \ref{table:gwDetectionChirptime}.  We also verify that the detection rate and detectable distance for HLV (O3) are approximately similar to those assuming only bKAGRA running, which are basically consistent with the detection rate and distance distribution of BH--NS merger candidates observed during LVC O3. We predict the detection rates of BH--NS mergers for different generation detector networks. The predicted median detectable distances and chirp times can provide important implications for future EM follow-up observations of GW triggers.

{\em GW-triggered target-of-opportunity observations} --- In view that the EM detectability is highly dependent on the BH--NS system parameter distributions, one can hardly give explicit predictions of the detection rates for GW-triggered target-of-opportunity observations. Here, we simulate the detectability of kilonovae, orphan afterglows, and total emissions with the consideration of different parameter distributions and provide some implications for the past and future EM follow-up observations. {We have discussed why no kilonova  from BH--NS merger candidates has been detected during LVC O3.} One possibility is that there is actually no EM signal produced from BH--NS mergers or the EM signals are too dim for us to detect, because the bulk of BH--NS systems possibly have low-spin primary BHs and massive NSs with a not very stiff EoS.  {In this case, most BH--NS events could be plunging events and no bright EM counterparts are expected in these mergers. This has also been predicted by other recent works  \citep[e.g.][]{zappa2019,Drozda2020}.} Alternatively, it may be that the previous follow-up observation searches have been too shallow to detect BH--NS merger kilonovae. In order to cover the median BH--NS merger detectable distance for the 2nd generation GW detector networks, {our simulated results have shown} that a limiting magnitude $m_{\rm limit}\sim23-24\,{\rm mag}$ is required. At the late period of the HLVK era, LSST \citep{lsst2009} with its 8-meter-class telescope will join the operation and would detect almost all BH--NS merger kilonovae with GW triggers. For those survey projects whose limiting magnitudes are $m_{\rm limit} \lesssim 23- 24\,{\rm mag}$, we suggest that they could search for afterglow signals after GW triggers. For the 2.5th and 3rd generation detector networks, the kilonovae are even fainter for GW-triggered BH--NS mergers. Nonetheless, follow-up observations may detect the associated sGRBs or afterglows and look for possible kilonova signatures in the lightcurves.

{\em Time-domain blind survey observations} --- Since survey telescopes only invest a relatively small percentage of time in GW trigger follow-up observations, {we have also discussed the detectability of BH--NS merger EM counterparts for time-domain blind survey observations.} In order to find more BH--NS merger kilonovae with multiple observations, a high-cadence search in redder filters is recommended. Similar conclusions has been obtained by \cite{almualla2020} in recent. Compared with kilonova emission, on-axis jet afterglows and orphan afterglows are more likely to be discovered by time-domain blind survey observations. Very recently, some on-axis jet afterglows and orphan afterglows has been detected through the blind survey searchings by ZTF \citep{andreoni2021}. The detectability for partial survey projects, including ZTF, PS1, WFIRST, Mephisto, LSST, and CSST, are discussed in detail. We indicate that LSST would provide a unique combination of deep searches and wide field-of-view survey observations, which will be well suited to detect more EM counterparts from BH--NS mergers.

\software{Matlab, \url{https://www.mathworks.com}; Python, \url{https://www.python.org}; LALSuite, \citep{lalsuite}}

\acknowledgments

{We thank Nicola Giacobbo, Michela Mapelli for providing population synthesis data, and Ping Chen, Xiao-Wei Duan, Micheal Coughlin, Mansi M. Kasliwal, Mattia Bulla, Francois Foucart for valuable comments, and an anonymous referee for constructive suggestions}. The work of J.P.Z is partially supported by the National Science Foundation of China under Grant No. 11721303 and the National Basic Research Program of China under grant No. 2014CB845800. Y.W.Y is supported by the National Natural Science Foundation of China under Grant No. 11822302, 11833003. Y.P.Y is supported by National Natural Science Foundation of China grant No. 12003028 and Yunnan University grant No.C176220100087. H.G. is supported by the National Natural Science Foundation of China under Grant No. 11722324, 11690024, 11633001, the Strategic Priority Research Program of the Chinese Academy of Sciences, Grant No. XDB23040100 and the Fundamental Research Funds for the Central Universities. L.D.L. is supported by the National Postdoctoral Program for Innovative Talents (grant No. BX20190044), China Postdoctoral Science Foundation (grant No. 2019M660515), and “LiYun” postdoctoral fellow of Beijing Normal University. Z.L is supported by the National Natural Science Foundation of China under Grant No. 11773003, U1931201. Z.J.C is supported by the National Natural Science Foundation of China (No. 11690023).

\bibliography{BHNS}

\begin{thebibliography}{}
\expandafter\ifx\csname natexlab\endcsname\relax\def\natexlab#1{#1}\fi
\providecommand{\url}[1]{\href{#1}{#1}}
\providecommand{\dodoi}[1]{doi:~\href{http://doi.org/#1}{\nolinkurl{#1}}}
\providecommand{\doeprint}[1]{\href{http://ascl.net/#1}{\nolinkurl{http://ascl.net/#1}}}
\providecommand{\doarXiv}[1]{\href{https://arxiv.org/abs/#1}{\nolinkurl{https://arxiv.org/abs/#1}}}

\bibitem[{Aasi {et~al.}(2015)Aasi, Abbott, Abbott, Abbott, Abernathy, Ackley,
  Adams, Adams, Addesso, Adhikari, {et~al.}}]{aasi2015advanced}
Aasi, J., Abbott, B., Abbott, R., {et~al.} 2015, Classical and quantum gravity,
  32, 074001

\bibitem[{{Abadie} {et~al.}(2010){Abadie}, {Abbott}, {Abbott}, {Abernathy},
  {Accadia}, {Acernese}, {Adams}, {Adhikari}, {Ajith}, {Allen}, {Allen},
  {Amador Ceron}, {Amin}, {Anderson}, {Anderson}, {Antonucci}, {Aoudia},
  {Arain}, {Araya}, {Aronsson}, {Arun}, {Aso}, {Aston}, {Astone}, {Atkinson},
  {Aufmuth}, {Aulbert}, {Babak}, {Baker}, {Ballardin}, {Ballmer}, {Barker},
  {Barnum}, {Barone}, {Barr}, {Barriga}, {Barsotti}, {Barsuglia}, {Barton},
  {Bartos}, {Bassiri}, {Bastarrika}, {Bauchrowitz}, {Bauer}, {Behnke}, {Beker},
  {Belczynski}, {Benacquista}, {Bertolini}, {Betzwieser}, {Beveridge},
  {Beyersdorf}, {Bigotta}, {Bilenko}, {Billingsley}, {Birch}, {Birindelli},
  {Biswas}, {Bitossi}, {Bizouard}, {Black}, {Blackburn}, {Blackburn}, {Blair},
  {Bland}, {Blom}, {Blomberg}, {Boccara}, {Bock}, {Bodiya}, {Bondarescu},
  {Bondu}, {Bonelli}, {Bork}, {Born}, {Bose}, {Bosi}, {Boyle}, {Braccini},
  {Bradaschia}, {Brady}, {Braginsky}, {Brau}, {Breyer}, {Bridges}, {Brillet},
  {Brinkmann}, {Brisson}, {Britzger}, {Brooks}, {Brown}, {Budzy{\'n}ski},
  {Bulik}, {Bulten}, {Buonanno}, {Burguet-Castell}, {Burmeister}, {Buskulic},
  {Byer}, {Cadonati}, {Cagnoli}, {Calloni}, {Camp}, {Campagna}, {Campsie},
  {Cannizzo}, {Cannon}, {Canuel}, {Cao}, {Capano}, {Carbognani}, {Caride},
  {Caudill}, {Cavagli{\`a}}, {Cavalier}, {Cavalieri}, {Cella}, {Cepeda},
  {Cesarini}, {Chalermsongsak}, {Chalkley}, {Charlton}, {Chassand e Mottin},
  {Chelkowski}, {Chen}, {Chincarini}, {Christensen}, {Chua}, {Chung}, {Clark},
  {Clark}, {Clayton}, {Cleva}, {Coccia}, {Colacino}, {Colas}, {Colla},
  {Colombini}, {Conte}, {Cook}, {Corbitt}, {Corda}, {Cornish}, {Corsi},
  {Costa}, {Coulon}, {Coward}, {Coyne}, {Creighton}, {Creighton}, {Cruise},
  {Culter}, {Cumming}, {Cunningham}, {Cuoco}, {Dahl}, {Danilishin},
  {Dannenberg}, {D'Antonio}, {Danzmann}, {Dari}, {Das}, {Dattilo}, {Daudert},
  {Davier}, {Davies}, {Davis}, {Daw}, {Day}, {Dayanga}, {De Rosa}, {DeBra},
  {Degallaix}, {del Prete}, {Dergachev}, {DeRosa}, {DeSalvo}, {Devanka},
  {Dhurandhar}, {Di Fiore}, {Di Lieto}, {Di Palma}, {Emilio}, {Di Virgilio},
  {D{\'\i}az}, {Dietz}, {Donovan}, {Dooley}, {Doomes}, {Dorsher}, {Douglas},
  {Drago}, {Drever}, {Driggers}, {Dueck}, {Dumas}, {Eberle}, {Edgar},
  {Edwards}, {Effler}, {Ehrens}, {Engel}, {Etzel}, {Evans}, {Evans}, {Fafone},
  {Fairhurst}, {Fan}, {Farr}, {Fazi}, {Fehrmann}, {Feldbaum}, {Ferrante},
  {Fidecaro}, {Finn}, {Fiori}, {Flaminio}, {Flanigan}, {Flasch}, {Foley},
  {Forrest}, {Forsi}, {Fotopoulos}, {Fournier}, {Franc}, {Frasca}, {Frasconi},
  {Frede}, {Frei}, {Frei}, {Freise}, {Frey}, {Fricke}, {Friedrich},
  {Fritschel}, {Frolov}, {Fulda}, {Fyffe}, {Gammaitoni}, {Garofoli}, {Garufi},
  {Gemme}, {Genin}, {Gennai}, {Gholami}, {Ghosh}, {Giaime}, {Giampanis},
  {Giardina}, {Giazotto}, {Gill}, {Goetz}, {Goggin}, {Gonz{\'a}lez},
  {Gorodetsky}, {Go{\ss}ler}, {Gouaty}, {Graef}, {Granata}, {Grant}, {Gras},
  {Gray}, {Greenhalgh}, {Gretarsson}, {Greverie}, {Grosso}, {Grote},
  {Grunewald}, {Guidi}, {Gustafson}, {Gustafson}, {Hage}, {Hall}, {Hallam},
  {Hammer}, {Hammond}, {Hanks}, {Hanna}, {Hanson}, {Harms}, {Harry}, {Harry},
  {Harstad}, {Haughian}, {Hayama}, {Heefner}, {Heitmann}, {Hello}, {Heng},
  {Heptonstall}, {Hewitson}, {Hild}, {Hirose}, {Hoak}, {Hodge}, {Holt},
  {Hosken}, {Hough}, {Howell}, {Hoyland}, {Huet}, {Hughey}, {Husa}, {Huttner},
  {Huynh-Dinh}, {Ingram}, {Inta}, {Isogai}, {Ivanov}, {Jaranowski}, {Johnson},
  {Jones}, {Jones}, {Jones}, {Ju}, {Kalmus}, {Kalogera}, {Kandhasamy},
  {Kanner}, {Katsavounidis}, {Kawabe}, {Kawamura}, {Kawazoe}, {Kells},
  {Keppel}, {Khalaidovski}, {Khalili}, {Khazanov}, {Kim}, {Kim}, {King},
  {Kinzel}, {Kissel}, {Klimenko}, {Kondrashov}, {Kopparapu}, {Koranda},
  {Kowalska}, {Kozak}, {Krause}, {Kringel}, {Krishnamurthy}, {Krishnan},
  {Kr{\'o}lak}, {Kuehn}, {Kullman}, {Kumar}, {Kwee}, {Landry}, {Lang}, {Lantz},
  {Lastzka}, {Lazzarini}, {Leaci}, {Leong}, {Leonor}, {Leroy}, {Letendre},
  {Li}, {Li}, {Lin}, {Lindquist}, {Lockerbie}, {Lodhia}, {Lorenzini},
  {Loriette}, {Lormand}, {Losurdo}, {Lu}, {Luan}, {Lubi{\'n}ski}, {Lucianetti},
  {L{\"u}ck}, {Lundgren}, {Machenschalk}, {MacInnis}, {Mackowski},
  {Mageswaran}, {Mailand }, {Majorana}, {Mak}, {Man}, {Mandel}, {Mandic},
  {Mantovani}, {Marchesoni}, {Marion}, {M{\'a}rka}, {M{\'a}rka}, {Maros},
  {Marque}, {Martelli}, {Martin}, {Martin}, {Marx}, {Mason}, {Masserot},
  {Matichard}, {Matone}, {Matzner}, {Mavalvala}, {McCarthy}, {McClelland},
  {McGuire}, {McIntyre}, {McIvor}, {McKechan}, {Meadors}, {Mehmet}, {Meier},
  {Melatos}, {Melissinos}, {Mendell}, {Men{\'e}ndez}, {Mercer}, {Merill},
  {Meshkov}, {Messenger}, {Meyer}, {Miao}, {Michel}, {Milano}, {Miller},
  {Minenkov}, {Mino}, {Mitra}, {Mitrofanov}, {Mitselmakher}, {Mittleman},
  {Moe}, {Mohan}, {Mohanty}, {Mohapatra}, {Moraru}, {Moreau}, {Moreno},
  {Morgado}, {Morgia}, {Morioka}, {Mors}, {Mosca}, {Moscatelli}, {Mossavi},
  {Mours}, {MowLowry}, {Mueller}, {Mukherjee}, {Mullavey},
  {M{\"u}ller-Ebhardt}, {Munch}, {Murray}, {Nash}, {Nawrodt}, {Nelson}, {Neri},
  {Newton}, {Nishizawa}, {Nocera}, {Nolting}, {Ochsner}, {O'Dell}, {Ogin},
  {Oldenburg}, {O'Reilly}, {O'Shaughnessy}, {Osthelder}, {Ottaway}, {Ottens},
  {Overmier}, {Owen}, {Page}, {Pagliaroli}, {Palladino}, {Palomba}, {Pan},
  {Pankow}, {Paoletti}, {Papa}, {Pardi}, {Pareja}, {Parisi}, {Pasqualetti},
  {Passaquieti}, {Passuello}, {Patel}, {Pedraza}, {Pekowsky}, {Penn},
  {Peralta}, {Perreca}, {Persichetti}, {Pichot}, {Pickenpack}, {Piergiovanni},
  {Pietka}, {Pinard}, {Pinto}, {Pitkin}, {Pletsch}, {Plissi}, {Poggiani},
  {Postiglione}, {Prato}, {Predoi}, {Price}, {Prijatelj}, {Principe},
  {Privitera}, {Prix}, {Prodi}, {Prokhorov}, {Puncken}, {Punturo}, {Puppo},
  {Quetschke}, {Raab}, {Rabaste}, {Rabeling}, {Radke}, {Radkins}, {Raffai},
  {Rakhmanov}, {Rankins}, {Rapagnani}, {Raymond}, {Re}, {Reed}, {Reed},
  {Regimbau}, {Reid}, {Reitze}, {Ricci}, {Riesen}, {Riles}, {Roberts},
  {Robertson}, {Robinet}, {Robinson}, {Robinson}, {Rocchi}, {Roddy},
  {R{\"o}ver}, {Rogstad}, {Rolland}, {Rollins}, {Romano}, {Romano}, {Romie},
  {Rosi{\'n}ska}, {Rowan}, {R{\"u}diger}, {Ruggi}, {Ryan}, {Sakata}, {Sakosky},
  {Salemi}, {Sammut}, {Sancho de la Jordana}, {Sandberg}, {Sannibale},
  {Santamar{\'\i}a}, {Santostasi}, {Saraf}, {Sassolas}, {Sathyaprakash},
  {Sato}, {Satterthwaite}, {Saulson}, {Savage}, {Schilling}, {Schnabel},
  {Schofield}, {Schulz}, {Schutz}, {Schwinberg}, {Scott}, {Scott}, {Searle},
  {Seifert}, {Sellers}, {Sengupta}, {Sentenac}, {Sergeev}, {Shaddock},
  {Shapiro}, {Shawhan}, {Shoemaker}, {Sibley}, {Siemens}, {Sigg}, {Singer},
  {Sintes}, {Skelton}, {Slagmolen}, {Slutsky}, {Smith}, {Smith}, {Smith},
  {Somiya}, {Sorazu}, {Speirits}, {Stein}, {Stein}, {Steinlechner},
  {Steplewski}, {Stochino}, {Stone}, {Strain}, {Strigin}, {Stroeer}, {Sturani},
  {Stuver}, {Summerscales}, {Sung}, {Susmithan}, {Sutton}, {Swinkels},
  {Talukder}, {Tanner}, {Tarabrin}, {Taylor}, {Taylor}, {Thomas}, {Thorne},
  {Thorne}, {Thrane}, {Th{\"u}ring}, {Titsler}, {Tokmakov}, {Toncelli},
  {Tonelli}, {Torres}, {Torrie}, {Tournefier}, {Travasso}, {Traylor}, {Trias},
  {Trummer}, {Tseng}, {Ugolini}, {Urbanek}, {Vahlbruch}, {Vaishnav}, {Vajente},
  {Vallisneri}, {van den Brand}, {Van Den Broeck}, {van der Putten}, {van der
  Sluys}, {van Veggel}, {Vass}, {Vaulin}, {Vavoulidis}, {Vecchio}, {Vedovato},
  {Veitch}, {Veitch}, {Veltkamp}, {Verkindt}, {Vetrano}, {Vicer{\'e}},
  {Villar}, {Vinet}, {Vocca}, {Vorvick}, {Vyachanin}, {Waldman}, {Wallace},
  {Wanner}, {Ward}, {Was}, {Wei}, {Weinert}, {Weinstein}, {Weiss}, {Wen},
  {Wen}, {Wessels}, {West}, {Westphal}, {Wette}, {Whelan}, {Whitcomb}, {White},
  {Whiting}, {Wilkinson}, {Willems}, {Williams}, {Willke}, {Winkelmann},
  {Winkler}, {Wipf}, {Wiseman}, {Woan}, {Wooley}, {Worden}, {Yakushin},
  {Yamamoto}, {Yamamoto}, {Yeaton-Massey}, {Yoshida}, {Yu}, {Yvert}, {Zanolin},
  {Zhang}, {Zhang}, {Zhao}, {Zotov}, {Zucker}, {Zweizig}, {LIGO Scientific
  Collaboration}, \& {Virgo Collaboration}}]{abadie2010}
{Abadie}, J., {Abbott}, B.~P., {Abbott}, R., {et~al.} 2010, Classical and
  Quantum Gravity, 27, 173001, \dodoi{10.1088/0264-9381/27/17/173001}

\bibitem[{Abbott {et~al.}(2019{\natexlab{a}})Abbott, Abbott, Abbott, Abraham,
  Acernese, Ackley, Adams, Adhikari, Adya, Affeldt, {et~al.}}]{abbott2019gwtc}
Abbott, B., Abbott, R., Abbott, T., {et~al.} 2019{\natexlab{a}}, Physical
  Review X, 9, 031040

\bibitem[{Abbott {et~al.}(2019{\natexlab{b}})Abbott, Abbott, Abbott, Acernese,
  Ackley, Adams, Adams, Addesso, Adhikari, Adya,
  {et~al.}}]{abbott2019properties}
---. 2019{\natexlab{b}}, Physical Review X, 9, 011001

\bibitem[{Abbott {et~al.}(2017{\natexlab{a}})Abbott, Abbott, Abbott, Acernese,
  Ackley, Adams, Adams, Addesso, Adhikari, Adya, {et~al.}}]{abbott2017gw170817}
Abbott, B.~P., Abbott, R., Abbott, T., {et~al.} 2017{\natexlab{a}}, \prl, 119,
  161101

\bibitem[{Abbott {et~al.}(2017{\natexlab{b}})Abbott, Abbott, Abbott, Acernese,
  Ackley, Adams, Adams, Addesso, Adhikari, Adya,
  {et~al.}}]{abbott2017gravitational}
---. 2017{\natexlab{b}}, \apjl, 848, L13

\bibitem[{Abbott {et~al.}(2017{\natexlab{c}})Abbott, Abbott, Abbott, Acernese,
  Ackley, Adams, Adams, Addesso, Adhikari, Adya,
  {et~al.}}]{abbott2017multimessenger}
---. 2017{\natexlab{c}}, \apjl, 848, L12, \dodoi{10.3847/2041-8213/aa91c9}

\bibitem[{Abbott {et~al.}(2018)Abbott, Abbott, Abbott, Abernathy, Acernese,
  Ackley, Adams, Adams, Addesso, Adhikari, {et~al.}}]{abbott2018prospects}
---. 2018, Living Reviews in Relativity, 21, 3

\bibitem[{Abbott {et~al.}(2020{\natexlab{a}})Abbott, Abbott, Abbott, Abernathy,
  Acernese, Ackley, Adams, Adams, Addesso, Adhikari,
  {et~al.}}]{abbott2020prospects}
---. 2020{\natexlab{a}}, Living Reviews in Relativity, 23, 3

\bibitem[{Abbott {et~al.}(2020{\natexlab{b}})Abbott, Abbott, Abraham, Acernese,
  Ackley, Adams, Adhikari, Adya, Affeldt, Agathos, {et~al.}}]{Abbott2020gwtc2}
Abbott, R., Abbott, T., Abraham, S., {et~al.} 2020{\natexlab{b}}, arXiv
  e-prints, arXiv:2010.14527.
\newblock \doarXiv{2010.14527}

\bibitem[{Acernese {et~al.}(2014)Acernese, Agathos, Agatsuma, Aisa, Allemandou,
  Allocca, Amarni, Astone, Balestri, Ballardin,
  {et~al.}}]{acernese2014advanced}
Acernese, F., Agathos, M., Agatsuma, K., {et~al.} 2014, Classical and Quantum
  Gravity, 32, 024001

\bibitem[{Adhikari {et~al.}(2020)Adhikari, Aguiar, Arai, Barr, Bassiri,
  Billingsley, Birney, Blair, Briggs, Brooks, {et~al.}}]{adhikari2020cryogenic}
Adhikari, R.~X., Aguiar, O., Arai, K., {et~al.} 2020, arXiv preprint
  arXiv:2001.11173

\bibitem[{{Ai} {et~al.}(2018){Ai}, {Gao}, {Dai}, {Wu}, {Li}, {Zhang}, \&
  {Li}}]{ai2018}
{Ai}, S., {Gao}, H., {Dai}, Z.-G., {et~al.} 2018, \apj, 860, 57,
  \dodoi{10.3847/1538-4357/aac2b7}

\bibitem[{{Akmal} \& {Pandharipande}(1997)}]{akmal1997}
{Akmal}, A., \& {Pandharipande}, V.~R. 1997, \prc, 56, 2261,
  \dodoi{10.1103/PhysRevC.56.2261}

\bibitem[{Akutsu {et~al.}(2018)Akutsu, Ando, Arai, Arai, Araki, Araya, Aritomi,
  Asada, Aso, Atsuta, {et~al.}}]{akutsu2018kagra}
Akutsu, T., Ando, M., Arai, K., {et~al.} 2018, arXiv preprint arXiv:1811.08079

\bibitem[{{Alexander} {et~al.}(2017){Alexander}, {Berger}, {Fong}, {Williams},
  {Guidorzi}, {Margutti}, {Metzger}, {Annis}, {Blanchard}, {Brout}, {Brown},
  {Chen}, {Chornock}, {Cowperthwaite}, {Drout}, {Eftekhari}, {Frieman}, {Holz},
  {Nicholl}, {Rest}, {Sako}, {Soares-Santos}, \& {Villar}}]{alexander2017}
{Alexander}, K.~D., {Berger}, E., {Fong}, W., {et~al.} 2017, \apjl, 848, L21,
  \dodoi{10.3847/2041-8213/aa905d}

\bibitem[{Allen {et~al.}(2012)Allen, Anderson, Brady, Brown, \&
  Creighton}]{allen2012findchirp}
Allen, B., Anderson, W.~G., Brady, P.~R., Brown, D.~A., \& Creighton, J.~D.
  2012, \prd, 85, 122006

\bibitem[{{Almualla} {et~al.}(2020){Almualla}, {Anand}, {Coughlin}, {Dietrich},
  {Guessoum}, {Sagu{\'e}s Carracedo}, {Ahumada}, {Andreoni}, {Antier}, {Bellm},
  {Bulla}, \& {Singer}}]{almualla2020}
{Almualla}, M., {Anand}, S., {Coughlin}, M.~W., {et~al.} 2020, arXiv e-prints,
  arXiv:2011.10421.
\newblock \doarXiv{2011.10421}

\bibitem[{{Anand} {et~al.}(2020){Anand}, {Coughlin}, {Kasliwal}, {Bulla},
  {Ahumada}, {Sagu{\'e}s Carracedo}, {Almualla}, {Andreoni}, {Stein},
  {Foucart}, {Singer}, {Sollerman}, {Bellm}, {Bolin}, {Caballero-Garc{\'\i}a},
  {Castro-Tirado}, {Cenko}, {De}, {Dekany}, {Duev}, {Feeney}, {Fremling},
  {Goldstein}, {Golkhou}, {Graham}, {Guessoum}, {Hankins}, {Hu}, {Kong},
  {Kool}, {Kulkarni}, {Kumar}, {Laher}, {Masci}, {Mr{\'o}z}, {Nissanke},
  {Porter}, {Reusch}, {Riddle}, {Rosnet}, {Rusholme}, {Serabyn},
  {S{\'a}nchez-Ram{\'\i}rez}, {Rigault}, {Shupe}, {Smith}, {Soumagnac},
  {Walters}, \& {Valeev}}]{anand2020}
{Anand}, S., {Coughlin}, M.~W., {Kasliwal}, M.~M., {et~al.} 2020, Nature
  Astronomy, \dodoi{10.1038/s41550-020-1183-3}

\bibitem[{{Andreoni} {et~al.}(2017){Andreoni}, {Ackley}, {Cooke}, {Acharyya},
  {Allison}, {Anderson}, {Ashley}, {Baade}, {Bailes}, {Bannister}, {Beardsley},
  {Bessell}, {Bian}, {Bland}, {Boer}, {Booler}, {Brandeker}, {Brown},
  {Buckley}, {Chang}, {Coward}, {Crawford}, {Crisp}, {Crosse}, {Cucchiara},
  {Cup{\'a}k}, {de Gois}, {Deller}, {Devillepoix}, {Dobie}, {Elmer}, {Emrich},
  {Farah}, {Farrell}, {Franzen}, {Gaensler}, {Galloway}, {Gendre}, {Giblin},
  {Goobar}, {Green}, {Hancock}, {Hartig}, {Howell}, {Horsley}, {Hotan},
  {Howie}, {Hu}, {Hu}, {James}, {Johnston}, {Johnston-Hollitt}, {Kaplan},
  {Kasliwal}, {Keane}, {Kenney}, {Klotz}, {Lau}, {Laugier}, {Lenc}, {Li},
  {Liang}, {Lidman}, {Luvaul}, {Lynch}, {Ma}, {Macpherson}, {Mao},
  {McClelland}, {McCully}, {M{\"o}ller}, {Morales}, {Morris}, {Murphy},
  {Noysena}, {Onken}, {Orange}, {Os{\l}owski}, {Pallot}, {Paxman}, {Potter},
  {Pritchard}, {Raja}, {Ridden-Harper}, {Romero-Colmenero}, {Sadler}, {Sansom},
  {Scalzo}, {Schmidt}, {Scott}, {Seghouani}, {Shang}, {Shannon}, {Shao},
  {Shara}, {Sharp}, {Sokolowski}, {Sollerman}, {Staff}, {Steele}, {Sun},
  {Suntzeff}, {Tao}, {Tingay}, {Towner}, {Thierry}, {Trott}, {Tucker},
  {V{\"a}is{\"a}nen}, {Krishnan}, {Walker}, {Wang}, {Wang}, {Wayth}, {Whiting},
  {Williams}, {Williams}, {Wolf}, {Wu}, {Wu}, {Yang}, {Yuan}, {Zhang}, {Zhou},
  \& {Zovaro}}]{andreoni2017}
{Andreoni}, I., {Ackley}, K., {Cooke}, J., {et~al.} 2017, \pasa, 34, e069,
  \dodoi{10.1017/pasa.2017.65}

\bibitem[{{Andreoni} {et~al.}(2021){Andreoni}, {Coughlin}, {Kool}, {Kasliwal},
  {Kumar}, {Bhalerao}, {Sagu{\'e}s Carracedo}, {Ho}, {Pang}, {Saraogi},
  {Sharma}, {Shenoy}, {Burns}, {Ahumada}, {Anand}, {Singer}, {Perley}, {De},
  {Fremling}, {Bellm}, {Bulla}, {Crellin-Quick}, {Dietrich}, {Drake}, {Duev},
  {Goobar}, {Graham}, {Kaplan}, {Kulkarni}, {Laher}, {Mahabal}, {Shupe},
  {Sollerman}, {Walters}, \& {Yao}}]{andreoni2021}
{Andreoni}, I., {Coughlin}, M.~W., {Kool}, E.~C., {et~al.} 2021, arXiv
  e-prints, arXiv:2104.06352.
\newblock \doarXiv{2104.06352}

\bibitem[{{Antier} {et~al.}(2020{\natexlab{a}}){Antier}, {Agayeva}, {Aivazyan},
  {Alishov}, {Arbouch}, {Baransky}, {Barynova}, {Bai}, {Basa}, {Beradze},
  {Bertin}, {Berthier}, {Bla{\v{z}}ek}, {Bo{\"e}r}, {Burkhonov}, {Burrell},
  {Cailleau}, {Chabert}, {Chen}, {Christensen}, {Coleiro}, {Cordier}, {Corre},
  {Coughlin}, {Coward}, {Crisp}, {Delattre}, {Dietrich}, {Ducoin}, {Duverne},
  {Marchal-Duval}, {Gendre}, {Eymar}, {Fock-Hang}, {Han}, {Hello}, {Howell},
  {Inasaridze}, {Ismailov}, {Kann}, {Kapanadze}, {Klotz}, {Kochiashvili},
  {Lachaud}, {Leroy}, {Le Van Su}, {Lin}, {Li}, {Lognone}, {Marron}, {Mo},
  {Moore}, {Natsvlishvili}, {Noysena}, {Perrigault}, {Peyrot}, {Samadov},
  {Sadibekova}, {Simon}, {Stachie}, {Teng}, {Thierry}, {Th{\"o}ne}, {Tillayev},
  {Turpin}, {de Ugarte Postigo}, {Vachier}, {Vardosanidze}, {Vasylenko},
  {Vidadi}, {Wang}, {Wang}, {Wei}, {Yan}, {Zhang}, {Zhang}, \&
  {Zhang}}]{antier2020a}
{Antier}, S., {Agayeva}, S., {Aivazyan}, V., {et~al.} 2020{\natexlab{a}},
  \mnras, 492, 3904, \dodoi{10.1093/mnras/stz3142}

\bibitem[{{Antier} {et~al.}(2020{\natexlab{b}}){Antier}, {Agayeva}, {Almualla},
  {Awiphan}, {Baransky}, {Barynova}, {Beradze}, {Bla{\v{z}}ek}, {Bo{\"e}r},
  {Burkhonov}, {Christensen}, {Coleiro}, {Corre}, {Coughlin}, {Crisp},
  {Dietrich}, {Ducoin}, {Duverne}, {Marchal-Duval}, {Gendre}, {Gokuldass},
  {Eggenstein}, {Eymar}, {Hello}, {Howell}, {Ismailov}, {Kann}, {Karpov},
  {Klotz}, {Kochiashvili}, {Lachaud}, {Leroy}, {Lin}, {Li}, {Ma{\v{s}}ek},
  {Mo}, {Menard}, {Morris}, {Noysena}, {Orange}, {Prouza}, {Rattanamala},
  {Sadibekova}, {Saint-Gelais}, {Serrau}, {Simon}, {Stachie}, {Th{\"o}ne},
  {Tillayev}, {Turpin}, {Postigo}, {Vasylenko}, {Vidadi}, {Was}, {Wang},
  {Zhang}, {Zhang}, \& {Zhang}}]{antier2020b}
{Antier}, S., {Agayeva}, S., {Almualla}, M., {et~al.} 2020{\natexlab{b}},
  \mnras, 497, 5518, \dodoi{10.1093/mnras/staa1846}

\bibitem[{{Arcavi} {et~al.}(2017){Arcavi}, {Hosseinzadeh}, {Howell}, {McCully},
  {Poznanski}, {Kasen}, {Barnes}, {Zaltzman}, {Vasylyev}, {Maoz}, \&
  {Valenti}}]{arcavi2017}
{Arcavi}, I., {Hosseinzadeh}, G., {Howell}, D.~A., {et~al.} 2017, \nat, 551,
  64, \dodoi{10.1038/nature24291}

\bibitem[{{Ascenzi} {et~al.}(2019){Ascenzi}, {Coughlin}, {Dietrich}, {Foley},
  {Ramirez-Ruiz}, {Piranomonte}, {Mockler}, {Murguia-Berthier}, {Fryer},
  {Lloyd-Ronning}, \& {Rosswog}}]{ascenzi2019}
{Ascenzi}, S., {Coughlin}, M.~W., {Dietrich}, T., {et~al.} 2019, \mnras, 486,
  672, \dodoi{10.1093/mnras/stz891}

\bibitem[{Aso {et~al.}(2013)Aso, Michimura, Somiya, Ando, Miyakawa, Sekiguchi,
  Tatsumi, Yamamoto, Collaboration, {et~al.}}]{aso2013interferometer}
Aso, Y., Michimura, Y., Somiya, K., {et~al.} 2013, \prd, 88, 043007

\bibitem[{{Barbieri} {et~al.}(2019){Barbieri}, {Salafia}, {Perego}, {Colpi}, \&
  {Ghirlanda}}]{barbieri2019}
{Barbieri}, C., {Salafia}, O.~S., {Perego}, A., {Colpi}, M., \& {Ghirlanda}, G.
  2019, \aap, 625, A152, \dodoi{10.1051/0004-6361/201935443}

\bibitem[{{Barbieri} {et~al.}(2020){Barbieri}, {Salafia}, {Perego}, {Colpi}, \&
  {Ghirlanda}}]{barbieri2020}
---. 2020, European Physical Journal A, 56, 8,
  \dodoi{10.1140/epja/s10050-019-00013-x}

\bibitem[{{Barnes} \& {Kasen}(2013)}]{barnes2013}
{Barnes}, J., \& {Kasen}, D. 2013, \apj, 775, 18,
  \dodoi{10.1088/0004-637X/775/1/18}

\bibitem[{{Bellm} {et~al.}(2019){Bellm}, {Kulkarni}, {Graham}, {Dekany},
  {Smith}, {Riddle}, {Masci}, {Helou}, {Prince}, {Adams}, {Barbarino},
  {Barlow}, {Bauer}, {Beck}, {Belicki}, {Biswas}, {Blagorodnova}, {Bodewits},
  {Bolin}, {Brinnel}, {Brooke}, {Bue}, {Bulla}, {Burruss}, {Cenko}, {Chang},
  {Connolly}, {Coughlin}, {Cromer}, {Cunningham}, {De}, {Delacroix}, {Desai},
  {Duev}, {Eadie}, {Farnham}, {Feeney}, {Feindt}, {Flynn}, {Franckowiak},
  {Frederick}, {Fremling}, {Gal-Yam}, {Gezari}, {Giomi}, {Goldstein},
  {Golkhou}, {Goobar}, {Groom}, {Hacopians}, {Hale}, {Henning}, {Ho}, {Hover},
  {Howell}, {Hung}, {Huppenkothen}, {Imel}, {Ip}, {Ivezi{\'c}}, {Jackson},
  {Jones}, {Juric}, {Kasliwal}, {Kaspi}, {Kaye}, {Kelley}, {Kowalski},
  {Kramer}, {Kupfer}, {Landry}, {Laher}, {Lee}, {Lin}, {Lin}, {Lunnan},
  {Giomi}, {Mahabal}, {Mao}, {Miller}, {Monkewitz}, {Murphy}, {Ngeow},
  {Nordin}, {Nugent}, {Ofek}, {Patterson}, {Penprase}, {Porter}, {Rauch},
  {Rebbapragada}, {Reiley}, {Rigault}, {Rodriguez}, {van Roestel}, {Rusholme},
  {van Santen}, {Schulze}, {Shupe}, {Singer}, {Soumagnac}, {Stein}, {Surace},
  {Sollerman}, {Szkody}, {Taddia}, {Terek}, {Van Sistine}, {van Velzen},
  {Vestrand}, {Walters}, {Ward}, {Ye}, {Yu}, {Yan}, \& {Zolkower}}]{bellm2019}
{Bellm}, E.~C., {Kulkarni}, S.~R., {Graham}, M.~J., {et~al.} 2019, \pasp, 131,
  018002, \dodoi{10.1088/1538-3873/aaecbe}

\bibitem[{{Berger}(2009)}]{berger2009}
{Berger}, E. 2009, \apj, 690, 231, \dodoi{10.1088/0004-637X/690/1/231}

\bibitem[{{Berger}(2014)}]{berger2014}
---. 2014, \araa, 52, 43, \dodoi{10.1146/annurev-astro-081913-035926}

\bibitem[{{Berger} {et~al.}(2013){Berger}, {Fong}, \& {Chornock}}]{berger2013}
{Berger}, E., {Fong}, W., \& {Chornock}, R. 2013, \apjl, 774, L23,
  \dodoi{10.1088/2041-8205/774/2/L23}

\bibitem[{{Bhattacharya} {et~al.}(2019){Bhattacharya}, {Kumar}, \&
  {Smoot}}]{bhattacharya2019}
{Bhattacharya}, M., {Kumar}, P., \& {Smoot}, G. 2019, \mnras, 486, 5289,
  \dodoi{10.1093/mnras/stz1147}

\bibitem[{{Blandford} \& {Payne}(1982)}]{blandford1982}
{Blandford}, R.~D., \& {Payne}, D.~G. 1982, \mnras, 199, 883,
  \dodoi{10.1093/mnras/199.4.883}

\bibitem[{{Blandford} \& {Znajek}(1977)}]{blandford1977}
{Blandford}, R.~D., \& {Znajek}, R.~L. 1977, \mnras, 179, 433,
  \dodoi{10.1093/mnras/179.3.433}

\bibitem[{{Brege} {et~al.}(2018){Brege}, {Duez}, {Foucart}, {Deaton}, {Caro},
  {Hemberger}, {Kidder}, {O'Connor}, {Pfeiffer}, \& {Scheel}}]{brege2018}
{Brege}, W., {Duez}, M.~D., {Foucart}, F., {et~al.} 2018, \prd, 98, 063009,
  \dodoi{10.1103/PhysRevD.98.063009}

\bibitem[{{Bulla} {et~al.}(2019){Bulla}, {Covino}, {Kyutoku}, {Tanaka},
  {Maund}, {Patat}, {Toma}, {Wiersema}, {Bruten}, {Jin}, \&
  {Testa}}]{bulla2019}
{Bulla}, M., {Covino}, S., {Kyutoku}, K., {et~al.} 2019, Nature Astronomy, 3,
  99, \dodoi{10.1038/s41550-018-0593-y}

\bibitem[{{Bulla} {et~al.}(2020){Bulla}, {Kyutoku}, {Tanaka}, {Covino},
  {Bruten}, {Matsumoto}, {Maund}, {Testa}, \& {Wiersema}}]{bulla2020}
{Bulla}, M., {Kyutoku}, K., {Tanaka}, M., {et~al.} 2020, arXiv e-prints,
  arXiv:2009.07279.
\newblock \doarXiv{2009.07279}

\bibitem[{{Cantiello} {et~al.}(2020){Cantiello}, {Jermyn}, \&
  {Lin}}]{cantiello2020}
{Cantiello}, M., {Jermyn}, A.~S., \& {Lin}, D. N.~C. 2020, arXiv e-prints,
  arXiv:2009.03936.
\newblock \doarXiv{2009.03936}

\bibitem[{{Cao} {et~al.}(2018){Cao}, {Gong}, {Meng}, {Xu}, {Chen}, {Guo}, {Li},
  {Liu}, {Xue}, {Cao}, {Fu}, {Zhang}, {Wang}, \& {Zhan}}]{cao2018}
{Cao}, Y., {Gong}, Y., {Meng}, X.-M., {et~al.} 2018, \mnras, 480, 2178,
  \dodoi{10.1093/mnras/sty1980}

\bibitem[{{Chambers} {et~al.}(2016){Chambers}, {Magnier}, {Metcalfe},
  {Flewelling}, {Huber}, {Waters}, {Denneau}, {Draper}, {Farrow}, {Finkbeiner},
  {Holmberg}, {Koppenhoefer}, {Price}, {Rest}, {Saglia}, {Schlafly}, {Smartt},
  {Sweeney}, {Wainscoat}, {Burgett}, {Chastel}, {Grav}, {Heasley}, {Hodapp},
  {Jedicke}, {Kaiser}, {Kudritzki}, {Luppino}, {Lupton}, {Monet}, {Morgan},
  {Onaka}, {Shiao}, {Stubbs}, {Tonry}, {White}, {Ba{\~n}ados}, {Bell},
  {Bender}, {Bernard}, {Boegner}, {Boffi}, {Botticella}, {Calamida},
  {Casertano}, {Chen}, {Chen}, {Cole}, {Deacon}, {Frenk}, {Fitzsimmons},
  {Gezari}, {Gibbs}, {Goessl}, {Goggia}, {Gourgue}, {Goldman}, {Grant},
  {Grebel}, {Hambly}, {Hasinger}, {Heavens}, {Heckman}, {Henderson}, {Henning},
  {Holman}, {Hopp}, {Ip}, {Isani}, {Jackson}, {Keyes}, {Koekemoer}, {Kotak},
  {Le}, {Liska}, {Long}, {Lucey}, {Liu}, {Martin}, {Masci}, {McLean}, {Mindel},
  {Misra}, {Morganson}, {Murphy}, {Obaika}, {Narayan}, {Nieto-Santisteban},
  {Norberg}, {Peacock}, {Pier}, {Postman}, {Primak}, {Rae}, {Rai}, {Riess},
  {Riffeser}, {Rix}, {R{\"o}ser}, {Russel}, {Rutz}, {Schilbach}, {Schultz},
  {Scolnic}, {Strolger}, {Szalay}, {Seitz}, {Small}, {Smith}, {Soderblom},
  {Taylor}, {Thomson}, {Taylor}, {Thakar}, {Thiel}, {Thilker}, {Unger},
  {Urata}, {Valenti}, {Wagner}, {Walder}, {Walter}, {Watters}, {Werner},
  {Wood-Vasey}, \& {Wyse}}]{chambers2016}
{Chambers}, K.~C., {Magnier}, E.~A., {Metcalfe}, N., {et~al.} 2016, arXiv
  e-prints, arXiv:1612.05560.
\newblock \doarXiv{1612.05560}

\bibitem[{Chan {et~al.}(2018)Chan, Messenger, Heng, \& Hendry}]{chan2018binary}
Chan, M.~L., Messenger, C., Heng, I.~S., \& Hendry, M. 2018, \prd, 97, 123014

\bibitem[{{Chan} {et~al.}(2018){Chan}, {Messenger}, {Heng}, \&
  {Hendry}}]{chan2018}
{Chan}, M.~L., {Messenger}, C., {Heng}, I.~S., \& {Hendry}, M. 2018, \prd, 97,
  123014, \dodoi{10.1103/PhysRevD.97.123014}

\bibitem[{{Chornock} {et~al.}(2017){Chornock}, {Berger}, {Kasen},
  {Cowperthwaite}, {Nicholl}, {Villar}, {Alexand er}, {Blanchard}, {Eftekhari},
  {Fong}, {Margutti}, {Williams}, {Annis}, {Brout}, {Brown}, {Chen}, {Drout},
  {Farr}, {Foley}, {Frieman}, {Fryer}, {Herner}, {Holz}, {Kessler}, {Matheson},
  {Metzger}, {Quataert}, {Rest}, {Sako}, {Scolnic}, {Smith}, \&
  {Soares-Santos}}]{chornock2017}
{Chornock}, R., {Berger}, E., {Kasen}, D., {et~al.} 2017, \apjl, 848, L19,
  \dodoi{10.3847/2041-8213/aa905c}

\bibitem[{{Colpi} \& {Sesana}(2017)}]{colpi2017}
{Colpi}, M., \& {Sesana}, A. 2017, {Gravitational Wave Sources in the Era of
  Multi-Band Gravitational Wave Astronomy}, 43--140,
  \dodoi{10.1142/9789813141766_0002}

\bibitem[{{Connaughton} {et~al.}(2015){Connaughton}, {Briggs}, {Goldstein},
  {Meegan}, {Paciesas}, {Preece}, {Wilson-Hodge}, {Gibby}, {Greiner}, {Gruber},
  {Jenke}, {Kippen}, {Pelassa}, {Xiong}, {Yu}, {Bhat}, {Burgess}, {Byrne},
  {Fitzpatrick}, {Foley}, {Giles}, {Guiriec}, {van der Horst}, {von Kienlin},
  {McBreen}, {McGlynn}, {Tierney}, \& {Zhang}}]{connaughton2015}
{Connaughton}, V., {Briggs}, M.~S., {Goldstein}, A., {et~al.} 2015, \apjs, 216,
  32, \dodoi{10.1088/0067-0049/216/2/32}

\bibitem[{{Coughlin} {et~al.}(2017){Coughlin}, {Dietrich}, {Kawaguchi},
  {Smartt}, {Stubbs}, \& {Ujevic}}]{coughlin2017}
{Coughlin}, M., {Dietrich}, T., {Kawaguchi}, K., {et~al.} 2017, \apj, 849, 12,
  \dodoi{10.3847/1538-4357/aa9114}

\bibitem[{{Coughlin} {et~al.}(2020{\natexlab{a}}){Coughlin}, {Dietrich},
  {Antier}, {Bulla}, {Foucart}, {Hotokezaka}, {Raaijmakers}, {Hinderer}, \&
  {Nissanke}}]{coughlin2020a}
{Coughlin}, M.~W., {Dietrich}, T., {Antier}, S., {et~al.} 2020{\natexlab{a}},
  \mnras, 492, 863, \dodoi{10.1093/mnras/stz3457}

\bibitem[{{Coughlin} {et~al.}(2020{\natexlab{b}}){Coughlin}, {Dietrich},
  {Antier}, {Almualla}, {Anand}, {Bulla}, {Foucart}, {Guessoum}, {Hotokezaka},
  {Kumar}, {Raaijmakers}, \& {Nissanke}}]{coughlin2020b}
---. 2020{\natexlab{b}}, \mnras, 497, 1181, \dodoi{10.1093/mnras/staa1925}

\bibitem[{{Coulter} {et~al.}(2017){Coulter}, {Foley}, {Kilpatrick}, {Drout},
  {Piro}, {Shappee}, {Siebert}, {Simon}, {Ulloa}, {Kasen}, {Madore},
  {Murguia-Berthier}, {Pan}, {Prochaska}, {Ramirez-Ruiz}, {Rest}, \&
  {Rojas-Bravo}}]{coulter2017}
{Coulter}, D.~A., {Foley}, R.~J., {Kilpatrick}, C.~D., {et~al.} 2017, Science,
  358, 1556, \dodoi{10.1126/science.aap9811}

\bibitem[{{Covino} {et~al.}(2017){Covino}, {Wiersema}, {Fan}, {Toma},
  {Higgins}, {Melandri}, {D'Avanzo}, {Mundell}, {Palazzi}, {Tanvir},
  {Bernardini}, {Branchesi}, {Brocato}, {Campana}, {di Serego Alighieri},
  {G{\"o}tz}, {Fynbo}, {Gao}, {Gomboc}, {Gompertz}, {Greiner}, {Hjorth}, {Jin},
  {Kaper}, {Klose}, {Kobayashi}, {Kopac}, {Kouveliotou}, {Levan}, {Mao},
  {Malesani}, {Pian}, {Rossi}, {Salvaterra}, {Starling}, {Steele},
  {Tagliaferri}, {Troja}, {van der Horst}, \& {Wijers}}]{covino2017}
{Covino}, S., {Wiersema}, K., {Fan}, Y.~Z., {et~al.} 2017, Nature Astronomy, 1,
  791, \dodoi{10.1038/s41550-017-0285-z}

\bibitem[{{Cowperthwaite} \& {Berger}(2015)}]{cowperthwaite2015}
{Cowperthwaite}, P.~S., \& {Berger}, E. 2015, \apj, 814, 25,
  \dodoi{10.1088/0004-637X/814/1/25}

\bibitem[{{Cowperthwaite} {et~al.}(2019){Cowperthwaite}, {Villar}, {Scolnic},
  \& {Berger}}]{cowperthwaite2019}
{Cowperthwaite}, P.~S., {Villar}, V.~A., {Scolnic}, D.~M., \& {Berger}, E.
  2019, \apj, 874, 88, \dodoi{10.3847/1538-4357/ab07b6}

\bibitem[{{Cowperthwaite} {et~al.}(2017){Cowperthwaite}, {Berger}, {Villar},
  {Metzger}, {Nicholl}, {Chornock}, {Blanchard}, {Fong}, {Margutti},
  {Soares-Santos}, {Alexander}, {Allam}, {Annis}, {Brout}, {Brown}, {Butler},
  {Chen}, {Diehl}, {Doctor}, {Drout}, {Eftekhari}, {Farr}, {Finley}, {Foley},
  {Frieman}, {Fryer}, {Garc{\'\i}a-Bellido}, {Gill}, {Guillochon}, {Herner},
  {Holz}, {Kasen}, {Kessler}, {Marriner}, {Matheson}, {Neilsen}, {Quataert},
  {Palmese}, {Rest}, {Sako}, {Scolnic}, {Smith}, {Tucker}, {Williams},
  {Balbinot}, {Carlin}, {Cook}, {Durret}, {Li}, {Lopes}, {Louren{\c{c}}o},
  {Marshall}, {Medina}, {Muir}, {Mu{\~n}oz}, {Sauseda}, {Schlegel}, {Secco},
  {Vivas}, {Wester}, {Zenteno}, {Zhang}, {Abbott}, {Banerji}, {Bechtol},
  {Benoit-L{\'e}vy}, {Bertin}, {Buckley-Geer}, {Burke}, {Capozzi}, {Carnero
  Rosell}, {Carrasco Kind}, {Castander}, {Crocce}, {Cunha}, {D'Andrea}, {da
  Costa}, {Davis}, {DePoy}, {Desai}, {Dietrich}, {Drlica-Wagner}, {Eifler},
  {Evrard}, {Fernand ez}, {Flaugher}, {Fosalba}, {Gaztanaga}, {Gerdes},
  {Giannantonio}, {Goldstein}, {Gruen}, {Gruendl}, {Gutierrez}, {Honscheid},
  {Jain}, {James}, {Jeltema}, {Johnson}, {Johnson}, {Kent}, {Krause}, {Kron},
  {Kuehn}, {Nuropatkin}, {Lahav}, {Lima}, {Lin}, {Maia}, {March}, {Martini},
  {McMahon}, {Menanteau}, {Miller}, {Miquel}, {Mohr}, {Neilsen}, {Nichol},
  {Ogando}, {Plazas}, {Roe}, {Romer}, {Roodman}, {Rykoff}, {Sanchez},
  {Scarpine}, {Schindler}, {Schubnell}, {Sevilla-Noarbe}, {Smith}, {Smith},
  {Sobreira}, {Suchyta}, {Swanson}, {Tarle}, {Thomas}, {Thomas}, {Troxel},
  {Vikram}, {Walker}, {Wechsler}, {Weller}, {Yanny}, \&
  {Zuntz}}]{cowperthwaite2017}
{Cowperthwaite}, P.~S., {Berger}, E., {Villar}, V.~A., {et~al.} 2017, \apjl,
  848, L17, \dodoi{10.3847/2041-8213/aa8fc7}

\bibitem[{{Creighton} \& {Anderson}(2011)}]{creighton2011}
{Creighton}, J., \& {Anderson}, W. 2011, {Gravitational-Wave Physics and
  Astronomy: An Introduction to Theory, Experiment and Data Analysis.}

\bibitem[{{Cutler} \& {Thorne}(2002)}]{cutler2002}
{Cutler}, C., \& {Thorne}, K.~S. 2002, arXiv e-prints, gr.
\newblock \doarXiv{gr-qc/0204090}

\bibitem[{{Dai}(2019)}]{dai2019}
{Dai}, Z.~G. 2019, \apjl, 873, L13, \dodoi{10.3847/2041-8213/ab0b45}

\bibitem[{Damour {et~al.}(2012)Damour, Nagar, \&
  Villain}]{damour2012measurability}
Damour, T., Nagar, A., \& Villain, L. 2012, \prd, 85, 123007

\bibitem[{{Darbha} \& {Kasen}(2020)}]{darbha2020}
{Darbha}, S., \& {Kasen}, D. 2020, \apj, 897, 150,
  \dodoi{10.3847/1538-4357/ab9a34}

\bibitem[{{Darbha} {et~al.}(2021){Darbha}, {Kasen}, {Foucart}, \&
  {Price}}]{darbha2021}
{Darbha}, S., {Kasen}, D., {Foucart}, F., \& {Price}, D.~J. 2021, arXiv
  e-prints, arXiv:2103.03378.
\newblock \doarXiv{2103.03378}

\bibitem[{{D'Avanzo} {et~al.}(2009){D'Avanzo}, {Malesani}, {Covino},
  {Piranomonte}, {Grazian}, {Fugazza}, {Margutti}, {D'Elia}, {Antonelli},
  {Campana}, {Chincarini}, {Della Valle}, {Fiore}, {Goldoni}, {Mao}, {Perna},
  {Salvaterra}, {Stella}, {Stratta}, \& {Tagliaferri}}]{davanzo2009}
{D'Avanzo}, P., {Malesani}, D., {Covino}, S., {et~al.} 2009, \aap, 498, 711,
  \dodoi{10.1051/0004-6361/200811294}

\bibitem[{{D'Avanzo} {et~al.}(2018){D'Avanzo}, {Campana}, {Salafia}, {Ghirland
  a}, {Ghisellini}, {Melandri}, {Bernardini}, {Branchesi}, {Chassande-Mottin},
  {Covino}, {D'Elia}, {Nava}, {Salvaterra}, {Tagliaferri}, \&
  {Vergani}}]{davanzo2018}
{D'Avanzo}, P., {Campana}, S., {Salafia}, O.~S., {et~al.} 2018, \aap, 613, L1,
  \dodoi{10.1051/0004-6361/201832664}

\bibitem[{{Deng}(2020)}]{Deng2020}
{Deng}, C.-M. 2020, \mnras, 497, 643, \dodoi{10.1093/mnras/staa1998}

\bibitem[{{D{\'\i}az} {et~al.}(2017){D{\'\i}az}, {Macri}, {Garcia Lambas},
  {Mendes de Oliveira}, {Nilo Castell{\'o}n}, {Ribeiro}, {S{\'a}nchez},
  {Schoenell}, {Abramo}, {Akras}, {Alcaniz}, {Artola}, {Beroiz}, {Bonoli},
  {Cabral}, {Camuccio}, {Castillo}, {Chavushyan}, {Coelho}, {Colazo},
  {Costa-Duarte}, {Cuevas Larenas}, {DePoy}, {Dom{\'\i}nguez Romero},
  {Dultzin}, {Fern{\'a}ndez}, {Garc{\'\i}a}, {Girardini}, {Gon{\c{c}}alves},
  {Gon{\c{c}}alves}, {Gurovich}, {Jim{\'e}nez-Teja}, {Kanaan}, {Lares}, {Lopes
  de Oliveira}, {L{\'o}pez-Cruz}, {Marshall}, {Melia}, {Molino}, {Padilla},
  {Pe{\~n}uela}, {Placco}, {Qui{\~n}ones}, {Ram{\'\i}rez Rivera}, {Renzi},
  {Riguccini}, {R{\'\i}os-L{\'o}pez}, {Rodriguez}, {Sampedro}, {Schneiter},
  {Sodr{\'e}}, {Starck}, {Torres-Flores}, {Tornatore}, \&
  {Zadro{\.z}ny}}]{diaz2017}
{D{\'\i}az}, M.~C., {Macri}, L.~M., {Garcia Lambas}, D., {et~al.} 2017, \apjl,
  848, L29, \dodoi{10.3847/2041-8213/aa9060}

\bibitem[{{Dobie} {et~al.}(2018){Dobie}, {Kaplan}, {Murphy}, {Lenc}, {Mooley},
  {Lynch}, {Corsi}, {Frail}, {Kasliwal}, \& {Hallinan}}]{dobie2018}
{Dobie}, D., {Kaplan}, D.~L., {Murphy}, T., {et~al.} 2018, \apjl, 858, L15,
  \dodoi{10.3847/2041-8213/aac105}

\bibitem[{{Drout} {et~al.}(2014){Drout}, {Chornock}, {Soderberg}, {Sand ers},
  {McKinnon}, {Rest}, {Foley}, {Milisavljevic}, {Margutti}, {Berger},
  {Calkins}, {Fong}, {Gezari}, {Huber}, {Kankare}, {Kirshner}, {Leibler},
  {Lunnan}, {Mattila}, {Marion}, {Narayan}, {Riess}, {Roth}, {Scolnic},
  {Smartt}, {Tonry}, {Burgett}, {Chambers}, {Hodapp}, {Jedicke}, {Kaiser},
  {Magnier}, {Metcalfe}, {Morgan}, {Price}, \& {Waters}}]{drout2014}
{Drout}, M.~R., {Chornock}, R., {Soderberg}, A.~M., {et~al.} 2014, \apj, 794,
  23, \dodoi{10.1088/0004-637X/794/1/23}

\bibitem[{{Drout} {et~al.}(2017){Drout}, {Piro}, {Shappee}, {Kilpatrick},
  {Simon}, {Contreras}, {Coulter}, {Foley}, {Siebert}, {Morrell}, {Boutsia},
  {Di Mille}, {Holoien}, {Kasen}, {Kollmeier}, {Madore}, {Monson},
  {Murguia-Berthier}, {Pan}, {Prochaska}, {Ramirez-Ruiz}, {Rest}, {Adams},
  {Alatalo}, {Ba{\~n}ados}, {Baughman}, {Beers}, {Bernstein}, {Bitsakis},
  {Campillay}, {Hansen}, {Higgs}, {Ji}, {Maravelias}, {Marshall}, {Moni Bidin},
  {Prieto}, {Rasmussen}, {Rojas-Bravo}, {Strom}, {Ulloa},
  {Vargas-Gonz{\'a}lez}, {Wan}, \& {Whitten}}]{drout2017}
{Drout}, M.~R., {Piro}, A.~L., {Shappee}, B.~J., {et~al.} 2017, Science, 358,
  1570, \dodoi{10.1126/science.aaq0049}

\bibitem[{{Drozda} {et~al.}(2020){Drozda}, {Belczynski}, {O'Shaughnessy},
  {Bulik}, \& {Fryer}}]{Drozda2020}
{Drozda}, P., {Belczynski}, K., {O'Shaughnessy}, R., {Bulik}, T., \& {Fryer},
  C.~L. 2020, arXiv e-prints, arXiv:2009.06655.
\newblock \doarXiv{2009.06655}

\bibitem[{{Eichler} {et~al.}(1989){Eichler}, {Livio}, {Piran}, \&
  {Schramm}}]{eichler1989}
{Eichler}, D., {Livio}, M., {Piran}, T., \& {Schramm}, D.~N. 1989, \nat, 340,
  126, \dodoi{10.1038/340126a0}

\bibitem[{{Er} {et~al.}(2020){Er}, {Bai}, {Chen}, {Cui}, {et~al.}}]{er2020}
{Er}, X., {Bai}, J., {Chen}, B., {Cui}, X., {et~al.} 2020, In Preparation

\bibitem[{{Evans} {et~al.}(2017){Evans}, {Cenko}, {Kennea}, {Emery}, {Kuin},
  {Korobkin}, {Wollaeger}, {Fryer}, {Madsen}, {Harrison}, {Xu}, {Nakar},
  {Hotokezaka}, {Lien}, {Campana}, {Oates}, {Troja}, {Breeveld}, {Marshall},
  {Barthelmy}, {Beardmore}, {Burrows}, {Cusumano}, {D'A{\`\i}}, {D'Avanzo},
  {D'Elia}, {de Pasquale}, {Even}, {Fontes}, {Forster}, {Garcia}, {Giommi},
  {Grefenstette}, {Gronwall}, {Hartmann}, {Heida}, {Hungerford}, {Kasliwal},
  {Krimm}, {Levan}, {Malesani}, {Melandri}, {Miyasaka}, {Nousek}, {O'Brien},
  {Osborne}, {Pagani}, {Page}, {Palmer}, {Perri}, {Pike}, {Racusin}, {Rosswog},
  {Siegel}, {Sakamoto}, {Sbarufatti}, {Tagliaferri}, {Tanvir}, \&
  {Tohuvavohu}}]{evans2017}
{Evans}, P.~A., {Cenko}, S.~B., {Kennea}, J.~A., {et~al.} 2017, Science, 358,
  1565, \dodoi{10.1126/science.aap9580}

\bibitem[{{Fan} {et~al.}(2013){Fan}, {Yu}, {Xu}, {Jin}, {Wu}, {Wei}, \&
  {Zhang}}]{fan2013}
{Fan}, Y.-Z., {Yu}, Y.-W., {Xu}, D., {et~al.} 2013, \apjl, 779, L25,
  \dodoi{10.1088/2041-8205/779/2/L25}

\bibitem[{{Fern{\'a}ndez} {et~al.}(2015){Fern{\'a}ndez}, {Kasen}, {Metzger}, \&
  {Quataert}}]{fernandez2015}
{Fern{\'a}ndez}, R., {Kasen}, D., {Metzger}, B.~D., \& {Quataert}, E. 2015,
  \mnras, 446, 750, \dodoi{10.1093/mnras/stu2112}

\bibitem[{{Fern{\'a}ndez} \& {Metzger}(2013)}]{fernandez2013}
{Fern{\'a}ndez}, R., \& {Metzger}, B.~D. 2013, \mnras, 435, 502,
  \dodoi{10.1093/mnras/stt1312}

\bibitem[{{Fern{\'a}ndez} \& {Metzger}(2016)}]{fernandez2016}
---. 2016, Annual Review of Nuclear and Particle Science, 66, 23,
  \dodoi{10.1146/annurev-nucl-102115-044819}

\bibitem[{{Flanagan} \& {Hinderer}(2008)}]{flanagan2008}
{Flanagan}, {\'E}.~{\'E}., \& {Hinderer}, T. 2008, \prd, 77, 021502,
  \dodoi{10.1103/PhysRevD.77.021502}

\bibitem[{{Fong} {et~al.}(2015){Fong}, {Berger}, {Margutti}, \&
  {Zauderer}}]{fong2015}
{Fong}, W., {Berger}, E., {Margutti}, R., \& {Zauderer}, B.~A. 2015, \apj, 815,
  102, \dodoi{10.1088/0004-637X/815/2/102}

\bibitem[{{Foucart}(2012)}]{foucart2012}
{Foucart}, F. 2012, \prd, 86, 124007, \dodoi{10.1103/PhysRevD.86.124007}

\bibitem[{{Foucart} {et~al.}(2019){Foucart}, {Duez}, {Kidder}, {Nissanke},
  {Pfeiffer}, \& {Scheel}}]{foucart2019}
{Foucart}, F., {Duez}, M.~D., {Kidder}, L.~E., {et~al.} 2019, \prd, 99, 103025,
  \dodoi{10.1103/PhysRevD.99.103025}

\bibitem[{{Foucart} {et~al.}(2018){Foucart}, {Hinderer}, \&
  {Nissanke}}]{foucart2018}
{Foucart}, F., {Hinderer}, T., \& {Nissanke}, S. 2018, \prd, 98, 081501,
  \dodoi{10.1103/PhysRevD.98.081501}

\bibitem[{{Foucart} {et~al.}(2014){Foucart}, {Deaton}, {Duez}, {O'Connor},
  {Ott}, {Haas}, {Kidder}, {Pfeiffer}, {Scheel}, \& {Szilagyi}}]{foucart2014}
{Foucart}, F., {Deaton}, M.~B., {Duez}, M.~D., {et~al.} 2014, \prd, 90, 024026,
  \dodoi{10.1103/PhysRevD.90.024026}

\bibitem[{{Foucart} {et~al.}(2017){Foucart}, {Desai}, {Brege}, {Duez}, {Kasen},
  {Hemberger}, {Kidder}, {Pfeiffer}, \& {Scheel}}]{foucart2017}
{Foucart}, F., {Desai}, D., {Brege}, W., {et~al.} 2017, Classical and Quantum
  Gravity, 34, 044002, \dodoi{10.1088/1361-6382/aa573b}

\bibitem[{{Fryer} {et~al.}(2012){Fryer}, {Belczynski}, {Wiktorowicz},
  {Dominik}, {Kalogera}, \& {Holz}}]{fryer2012}
{Fryer}, C.~L., {Belczynski}, K., {Wiktorowicz}, G., {et~al.} 2012, \apj, 749,
  91, \dodoi{10.1088/0004-637X/749/1/91}

\bibitem[{{Fryer} \& {Kalogera}(2001)}]{fryer2001}
{Fryer}, C.~L., \& {Kalogera}, V. 2001, \apj, 554, 548, \dodoi{10.1086/321359}

\bibitem[{{Fujibayashi} {et~al.}(2017){Fujibayashi}, {Sekiguchi}, {Kiuchi}, \&
  {Shibata}}]{fujibayashi2017}
{Fujibayashi}, S., {Sekiguchi}, Y., {Kiuchi}, K., \& {Shibata}, M. 2017, \apj,
  846, 114, \dodoi{10.3847/1538-4357/aa8039}

\bibitem[{{Fujibayashi} {et~al.}(2020){Fujibayashi}, {Shibata}, {Wanajo},
  {Kiuchi}, {Kyutoku}, \& {Sekiguchi}}]{fujibayashi2020}
{Fujibayashi}, S., {Shibata}, M., {Wanajo}, S., {et~al.} 2020, \prd, 101,
  083029, \dodoi{10.1103/PhysRevD.101.083029}

\bibitem[{{Fuller} \& {Ma}(2019)}]{fuller2019}
{Fuller}, J., \& {Ma}, L. 2019, \apjl, 881, L1,
  \dodoi{10.3847/2041-8213/ab339b}

\bibitem[{{Gao} {et~al.}(2020){Gao}, {Ai}, {Cao}, {Zhang}, {Zhu}, {Li},
  {Zhang}, \& {Bauswein}}]{gao2020}
{Gao}, H., {Ai}, S.-K., {Cao}, Z.-J., {et~al.} 2020, Frontiers of Physics, 15,
  24603, \dodoi{10.1007/s11467-019-0945-9}

\bibitem[{{Gao} {et~al.}(2017{\natexlab{a}}){Gao}, {Cao}, {Ai}, \&
  {Zhang}}]{gao2017b}
{Gao}, H., {Cao}, Z., {Ai}, S., \& {Zhang}, B. 2017{\natexlab{a}}, \apjl, 851,
  L45, \dodoi{10.3847/2041-8213/aaa0c6}

\bibitem[{{Gao} {et~al.}(2015){Gao}, {Ding}, {Wu}, {Dai}, \& {Zhang}}]{gao2015}
{Gao}, H., {Ding}, X., {Wu}, X.-F., {Dai}, Z.-G., \& {Zhang}, B. 2015, \apj,
  807, 163, \dodoi{10.1088/0004-637X/807/2/163}

\bibitem[{{Gao} {et~al.}(2013{\natexlab{a}}){Gao}, {Ding}, {Wu}, {Zhang}, \&
  {Dai}}]{gao2013b}
{Gao}, H., {Ding}, X., {Wu}, X.-F., {Zhang}, B., \& {Dai}, Z.-G.
  2013{\natexlab{a}}, \apj, 771, 86, \dodoi{10.1088/0004-637X/771/2/86}

\bibitem[{{Gao} {et~al.}(2013{\natexlab{b}}){Gao}, {Lei}, {Zou}, {Wu}, \&
  {Zhang}}]{gao2013}
{Gao}, H., {Lei}, W.-H., {Zou}, Y.-C., {Wu}, X.-F., \& {Zhang}, B.
  2013{\natexlab{b}}, \nar, 57, 141, \dodoi{10.1016/j.newar.2013.10.001}

\bibitem[{{Gao} {et~al.}(2016){Gao}, {Zhang}, \& {L{\"u}}}]{gao2016}
{Gao}, H., {Zhang}, B., \& {L{\"u}}, H.-J. 2016, \prd, 93, 044065,
  \dodoi{10.1103/PhysRevD.93.044065}

\bibitem[{{Gao} {et~al.}(2017{\natexlab{b}}){Gao}, {Zhang}, {L{\"u}}, \&
  {Li}}]{gao2017}
{Gao}, H., {Zhang}, B., {L{\"u}}, H.-J., \& {Li}, Y. 2017{\natexlab{b}}, \apj,
  837, 50, \dodoi{10.3847/1538-4357/aa5be3}

\bibitem[{{Gehrels} {et~al.}(2016){Gehrels}, {Cannizzo}, {Kanner}, {Kasliwal},
  {Nissanke}, \& {Singer}}]{gehrels2016}
{Gehrels}, N., {Cannizzo}, J.~K., {Kanner}, J., {et~al.} 2016, \apj, 820, 136,
  \dodoi{10.3847/0004-637X/820/2/136}

\bibitem[{{Ghirlanda} {et~al.}(2019){Ghirlanda}, {Salafia}, {Paragi},
  {Giroletti}, {Yang}, {Marcote}, {Blanchard}, {Agudo}, {An}, {Bernardini},
  {Beswick}, {Branchesi}, {Campana}, {Casadio}, {Chassand e-Mottin}, {Colpi},
  {Covino}, {D'Avanzo}, {D'Elia}, {Frey}, {Gawronski}, {Ghisellini}, {Gurvits},
  {Jonker}, {van Langevelde}, {Melandri}, {Moldon}, {Nava}, {Perego},
  {Perez-Torres}, {Reynolds}, {Salvaterra}, {Tagliaferri}, {Venturi},
  {Vergani}, \& {Zhang}}]{ghirlanda2019}
{Ghirlanda}, G., {Salafia}, O.~S., {Paragi}, Z., {et~al.} 2019, Science, 363,
  968, \dodoi{10.1126/science.aau8815}

\bibitem[{{Giacobbo} \& {Mapelli}(2018)}]{giacobbo2018}
{Giacobbo}, N., \& {Mapelli}, M. 2018, \mnras, 480, 2011,
  \dodoi{10.1093/mnras/sty1999}

\bibitem[{{Gill} \& {Granot}(2018)}]{gill2018}
{Gill}, R., \& {Granot}, J. 2018, \mnras, 478, 4128,
  \dodoi{10.1093/mnras/sty1214}

\bibitem[{{Gill} {et~al.}(2019){Gill}, {Nathanail}, \& {Rezzolla}}]{gill2019}
{Gill}, R., {Nathanail}, A., \& {Rezzolla}, L. 2019, \apj, 876, 139,
  \dodoi{10.3847/1538-4357/ab16da}

\bibitem[{{Goldstein} {et~al.}(2017){Goldstein}, {Veres}, {Burns}, {Briggs},
  {Hamburg}, {Kocevski}, {Wilson-Hodge}, {Preece}, {Poolakkil}, {Roberts},
  {Hui}, {Connaughton}, {Racusin}, {von Kienlin}, {Dal Canton}, {Christensen},
  {Littenberg}, {Siellez}, {Blackburn}, {Broida}, {Bissaldi}, {Cleveland},
  {Gibby}, {Giles}, {Kippen}, {McBreen}, {McEnery}, {Meegan}, {Paciesas}, \&
  {Stanbro}}]{goldstein2017}
{Goldstein}, A., {Veres}, P., {Burns}, E., {et~al.} 2017, \apjl, 848, L14,
  \dodoi{10.3847/2041-8213/aa8f41}

\bibitem[{{Goldstein} {et~al.}(2019){Goldstein}, {Hamburg}, {Wood}, {Hui},
  {Cleveland}, {Kocevski}, {Littenberg}, {Burns}, {Dal Canton}, {Veres},
  {Mailyan}, {Malacaria}, {Briggs}, \& {Wilson-Hodge}}]{goldstein2019}
{Goldstein}, A., {Hamburg}, R., {Wood}, J., {et~al.} 2019, arXiv e-prints,
  arXiv:1903.12597.
\newblock \doarXiv{1903.12597}

\bibitem[{{Gompertz} {et~al.}(2014){Gompertz}, {O'Brien}, \&
  {Wynn}}]{compertz2014}
{Gompertz}, B.~P., {O'Brien}, P.~T., \& {Wynn}, G.~A. 2014, \mnras, 438, 240,
  \dodoi{10.1093/mnras/stt2165}

\bibitem[{{Gompertz} {et~al.}(2013){Gompertz}, {O'Brien}, {Wynn}, \&
  {Rowlinson}}]{compertz2013}
{Gompertz}, B.~P., {O'Brien}, P.~T., {Wynn}, G.~A., \& {Rowlinson}, A. 2013,
  \mnras, 431, 1745, \dodoi{10.1093/mnras/stt293}

\bibitem[{{Gompertz} {et~al.}(2018){Gompertz}, {Levan}, {Tanvir}, {Hjorth},
  {Covino}, {Evans}, {Fruchter}, {Gonz{\'a}lez-Fern{\'a}ndez}, {Jin}, {Lyman},
  {Oates}, {O'Brien}, \& {Wiersema}}]{compertz2018}
{Gompertz}, B.~P., {Levan}, A.~J., {Tanvir}, N.~R., {et~al.} 2018, \apj, 860,
  62, \dodoi{10.3847/1538-4357/aac206}

\bibitem[{{Gompertz} {et~al.}(2020){Gompertz}, {Cutter}, {Steeghs}, {Galloway},
  {Lyman}, {Ulaczyk}, {Dyer}, {Ackley}, {Dhillon}, {O'Brien}, {Ramsay},
  {Poshyachinda}, {Kotak}, {Nuttall}, {Breton}, {Pall{\'e}}, {Pollacco},
  {Thrane}, {Aukkaravittayapun}, {Awiphan}, {Brown}, {Burhanudin}, {Chote},
  {Chrimes}, {Daw}, {Duffy}, {Eyles-Ferris}, {Heikkil{\"a}}, {Irawati},
  {Kennedy}, {Killestein}, {Levan}, {Littlefair}, {Makrygianni}, {Marsh}, {Mata
  S{\'a}nchez}, {Mattila}, {Maund}, {McCormac}, {Mkrtichian}, {Mong},
  {Mullaney}, {M{\"u}ller}, {Obradovic}, {Rol}, {Sawangwit}, {Stanway},
  {Starling}, {Str{\o}m}, {Tooke}, {West}, \& {Wiersema}}]{compertz2020}
{Gompertz}, B.~P., {Cutter}, R., {Steeghs}, D., {et~al.} 2020, \mnras, 497,
  726, \dodoi{10.1093/mnras/staa1845}

\bibitem[{{Gong} {et~al.}(2019){Gong}, {Liu}, {Cao}, {Chen}, {Fan}, {Li}, {Li},
  {Li}, {Zhang}, \& {Zhan}}]{gong2019}
{Gong}, Y., {Liu}, X., {Cao}, Y., {et~al.} 2019, \apj, 883, 203,
  \dodoi{10.3847/1538-4357/ab391e}

\bibitem[{{Graham} {et~al.}(2019){Graham}, {Kulkarni}, {Bellm}, {Adams},
  {Barbarino}, {Blagorodnova}, {Bodewits}, {Bolin}, {Brady}, {Cenko}, {Chang},
  {Coughlin}, {De}, {Eadie}, {Farnham}, {Feindt}, {Franckowiak}, {Fremling},
  {Gezari}, {Ghosh}, {Goldstein}, {Golkhou}, {Goobar}, {Ho}, {Huppenkothen},
  {Ivezi{\'c}}, {Jones}, {Juric}, {Kaplan}, {Kasliwal}, {Kelley}, {Kupfer},
  {Lee}, {Lin}, {Lunnan}, {Mahabal}, {Miller}, {Ngeow}, {Nugent}, {Ofek},
  {Prince}, {Rauch}, {van Roestel}, {Schulze}, {Singer}, {Sollerman}, {Taddia},
  {Yan}, {Ye}, {Yu}, {Barlow}, {Bauer}, {Beck}, {Belicki}, {Biswas}, {Brinnel},
  {Brooke}, {Bue}, {Bulla}, {Burruss}, {Connolly}, {Cromer}, {Cunningham},
  {Dekany}, {Delacroix}, {Desai}, {Duev}, {Feeney}, {Flynn}, {Frederick},
  {Gal-Yam}, {Giomi}, {Groom}, {Hacopians}, {Hale}, {Helou}, {Henning},
  {Hover}, {Hillenbrand}, {Howell}, {Hung}, {Imel}, {Ip}, {Jackson}, {Kaspi},
  {Kaye}, {Kowalski}, {Kramer}, {Kuhn}, {Landry}, {Laher}, {Mao}, {Masci},
  {Monkewitz}, {Murphy}, {Nordin}, {Patterson}, {Penprase}, {Porter},
  {Rebbapragada}, {Reiley}, {Riddle}, {Rigault}, {Rodriguez}, {Rusholme}, {van
  Santen}, {Shupe}, {Smith}, {Soumagnac}, {Stein}, {Surace}, {Szkody}, {Terek},
  {Van Sistine}, {van Velzen}, {Vestrand}, {Walters}, {Ward}, {Zhang}, \&
  {Zolkower}}]{graham2019}
{Graham}, M.~J., {Kulkarni}, S.~R., {Bellm}, E.~C., {et~al.} 2019, \pasp, 131,
  078001, \dodoi{10.1088/1538-3873/ab006c}

\bibitem[{{Graham} {et~al.}(2020){Graham}, {Ford}, {McKernan}, {Ross}, {Stern},
  {Burdge}, {Coughlin}, {Djorgovski}, {Drake}, {Duev}, {Kasliwal}, {Mahabal},
  {van Velzen}, {Belecki}, {Bellm}, {Burruss}, {Cenko}, {Cunningham}, {Helou},
  {Kulkarni}, {Masci}, {Prince}, {Reiley}, {Rodriguez}, {Rusholme}, {Smith}, \&
  {Soumagnac}}]{graham2020P}
{Graham}, M.~J., {Ford}, K.~E.~S., {McKernan}, B., {et~al.} 2020, \prl, 124,
  251102, \dodoi{10.1103/PhysRevLett.124.251102}

\bibitem[{{Grossman} {et~al.}(2014){Grossman}, {Korobkin}, {Rosswog}, \&
  {Piran}}]{grossman2014}
{Grossman}, D., {Korobkin}, O., {Rosswog}, S., \& {Piran}, T. 2014, \mnras,
  439, 757, \dodoi{10.1093/mnras/stt2503}

\bibitem[{{Haggard} {et~al.}(2017){Haggard}, {Nynka}, {Ruan}, {Kalogera},
  {Cenko}, {Evans}, \& {Kennea}}]{haggard2017}
{Haggard}, D., {Nynka}, M., {Ruan}, J.~J., {et~al.} 2017, \apjl, 848, L25,
  \dodoi{10.3847/2041-8213/aa8ede}

\bibitem[{{Hall} \& {Evans}(2019)}]{hall2019}
{Hall}, E.~D., \& {Evans}, M. 2019, Classical and Quantum Gravity, 36, 225002,
  \dodoi{10.1088/1361-6382/ab41d6}

\bibitem[{{Hallinan} {et~al.}(2017){Hallinan}, {Corsi}, {Mooley}, {Hotokezaka},
  {Nakar}, {Kasliwal}, {Kaplan}, {Frail}, {Myers}, {Murphy}, {De}, {Dobie},
  {Allison}, {Bannister}, {Bhalerao}, {Chandra}, {Clarke}, {Giacintucci}, {Ho},
  {Horesh}, {Kassim}, {Kulkarni}, {Lenc}, {Lockman}, {Lynch}, {Nichols},
  {Nissanke}, {Palliyaguru}, {Peters}, {Piran}, {Rana}, {Sadler}, \&
  {Singer}}]{hallinan2017}
{Hallinan}, G., {Corsi}, A., {Mooley}, K.~P., {et~al.} 2017, Science, 358,
  1579, \dodoi{10.1126/science.aap9855}

\bibitem[{Harry {et~al.}(2010)Harry, Collaboration,
  {et~al.}}]{harry2010advanced}
Harry, G.~M., Collaboration, L.~S., {et~al.} 2010, Classical and Quantum
  Gravity, 27, 084006

\bibitem[{{Hounsell} {et~al.}(2018){Hounsell}, {Scolnic}, {Foley}, {Kessler},
  {Miranda}, {Avelino}, {Bohlin}, {Filippenko}, {Frieman}, {Jha}, {Kelly},
  {Kirshner}, {Mandel}, {Rest}, {Riess}, {Rodney}, \&
  {Strolger}}]{hounsell2018}
{Hounsell}, R., {Scolnic}, D., {Foley}, R.~J., {et~al.} 2018, \apj, 867, 23,
  \dodoi{10.3847/1538-4357/aac08b}

\bibitem[{{Hu} {et~al.}(2017){Hu}, {Wu}, {Andreoni}, {Ashley}, {Cooke}, {Cui},
  {Du}, {Dai}, {Gu}, {Hu}, {Lu}, {Li}, {Li}, {Liang}, {Liu}, {Ma}, {Shang},
  {Sun}, {Suntzeff}, {Tao}, {Udden}, {Wang}, {Wang}, {Wen}, {Xiao}, {Su},
  {Yang}, {Yang}, {Yuan}, {Zhou}, {Zhang}, {Zhou}, \& {Zhu}}]{hu2017}
{Hu}, L., {Wu}, X., {Andreoni}, I., {et~al.} 2017, Science Bulletin, 62, 1433,
  \dodoi{10.1016/j.scib.2017.10.006}

\bibitem[{{Jin} {et~al.}(2020){Jin}, {Covino}, {Liao}, {Li}, {D'Avanzo}, {Fan},
  \& {Wei}}]{jin2020}
{Jin}, Z.-P., {Covino}, S., {Liao}, N.-H., {et~al.} 2020, Nature Astronomy, 4,
  77, \dodoi{10.1038/s41550-019-0892-y}

\bibitem[{{Jin} {et~al.}(2015){Jin}, {Li}, {Cano}, {Covino}, {Fan}, \&
  {Wei}}]{jin2015}
{Jin}, Z.-P., {Li}, X., {Cano}, Z., {et~al.} 2015, \apjl, 811, L22,
  \dodoi{10.1088/2041-8205/811/2/L22}

\bibitem[{{Jin} {et~al.}(2016){Jin}, {Hotokezaka}, {Li}, {Tanaka}, {D'Avanzo},
  {Fan}, {Covino}, {Wei}, \& {Piran}}]{jin2016}
{Jin}, Z.-P., {Hotokezaka}, K., {Li}, X., {et~al.} 2016, Nature Communications,
  7, 12898, \dodoi{10.1038/ncomms12898}

\bibitem[{{Just} {et~al.}(2015){Just}, {Bauswein}, {Ardevol Pulpillo},
  {Goriely}, \& {Janka}}]{just2015}
{Just}, O., {Bauswein}, A., {Ardevol Pulpillo}, R., {Goriely}, S., \& {Janka},
  H.~T. 2015, \mnras, 448, 541, \dodoi{10.1093/mnras/stv009}

\bibitem[{{Kaiser} {et~al.}(2010){Kaiser}, {Burgett}, {Chambers}, {Denneau},
  {Heasley}, {Jedicke}, {Magnier}, {Morgan}, {Onaka}, \& {Tonry}}]{kaiser2010}
{Kaiser}, N., {Burgett}, W., {Chambers}, K., {et~al.} 2010, in Society of
  Photo-Optical Instrumentation Engineers (SPIE) Conference Series, Vol. 7733,
  Ground-based and Airborne Telescopes III, 77330E, \dodoi{10.1117/12.859188}

\bibitem[{{Kapadia} {et~al.}(2020){Kapadia}, {Singh}, {Shaikh}, {Chatterjee},
  \& {Ajith}}]{2020ApJ...898L..39K}
{Kapadia}, S.~J., {Singh}, M.~K., {Shaikh}, M.~A., {Chatterjee}, D., \&
  {Ajith}, P. 2020, \apjl, 898, L39, \dodoi{10.3847/2041-8213/aba42d}

\bibitem[{{Kasen} {et~al.}(2013){Kasen}, {Badnell}, \& {Barnes}}]{kasen2013}
{Kasen}, D., {Badnell}, N.~R., \& {Barnes}, J. 2013, \apj, 774, 25,
  \dodoi{10.1088/0004-637X/774/1/25}

\bibitem[{{Kasen} {et~al.}(2015){Kasen}, {Fern{\'a}ndez}, \&
  {Metzger}}]{kasen2015}
{Kasen}, D., {Fern{\'a}ndez}, R., \& {Metzger}, B.~D. 2015, \mnras, 450, 1777,
  \dodoi{10.1093/mnras/stv721}

\bibitem[{{Kasen} {et~al.}(2017){Kasen}, {Metzger}, {Barnes}, {Quataert}, \&
  {Ramirez-Ruiz}}]{kasen2017}
{Kasen}, D., {Metzger}, B., {Barnes}, J., {Quataert}, E., \& {Ramirez-Ruiz}, E.
  2017, \nat, 551, 80, \dodoi{10.1038/nature24453}

\bibitem[{{Kasliwal} {et~al.}(2017){Kasliwal}, {Nakar}, {Singer}, {Kaplan},
  {Cook}, {Van Sistine}, {Lau}, {Fremling}, {Gottlieb}, {Jencson}, {Adams},
  {Feindt}, {Hotokezaka}, {Ghosh}, {Perley}, {Yu}, {Piran}, {Allison},
  {Anupama}, {Balasubramanian}, {Bannister}, {Bally}, {Barnes}, {Barway},
  {Bellm}, {Bhalerao}, {Bhattacharya}, {Blagorodnova}, {Bloom}, {Brady},
  {Cannella}, {Chatterjee}, {Cenko}, {Cobb}, {Copperwheat}, {Corsi}, {De},
  {Dobie}, {Emery}, {Evans}, {Fox}, {Frail}, {Frohmaier}, {Goobar}, {Hallinan},
  {Harrison}, {Helou}, {Hinderer}, {Ho}, {Horesh}, {Ip}, {Itoh}, {Kasen},
  {Kim}, {Kuin}, {Kupfer}, {Lynch}, {Madsen}, {Mazzali}, {Miller}, {Mooley},
  {Murphy}, {Ngeow}, {Nichols}, {Nissanke}, {Nugent}, {Ofek}, {Qi}, {Quimby},
  {Rosswog}, {Rusu}, {Sadler}, {Schmidt}, {Sollerman}, {Steele}, {Williamson},
  {Xu}, {Yan}, {Yatsu}, {Zhang}, \& {Zhao}}]{kasliwal2017}
{Kasliwal}, M.~M., {Nakar}, E., {Singer}, L.~P., {et~al.} 2017, Science, 358,
  1559, \dodoi{10.1126/science.aap9455}

\bibitem[{{Kasliwal} {et~al.}(2020){Kasliwal}, {Anand}, {Ahumada}, {Stein},
  {Sagues Carracedo}, {Andreoni}, {Coughlin}, {Singer}, {Kool}, {De}, {Kumar},
  {AlMualla}, {Yao}, {Bulla}, {Dobie}, {Reusch}, {Perley}, {Cenko}, {Bhalerao},
  {Kaplan}, {Sollerman}, {Goobar}, {Copperwheat}, {Bellm}, {Anupama}, {Corsi},
  {Nissanke}, {Agudo}, {Bagdasaryan}, {Barway}, {Belicki}, {Bloom}, {Bolin},
  {Buckley}, {Burdge}, {Burruss}, {Caballero-Garc{\i}a}, {Cannella},
  {Castro-Tirado}, {Cook}, {Cooke}, {Cunningham}, {Dahiwale}, {Deshmukh},
  {Dichiara}, {Duev}, {Dutta}, {Feeney}, {Franckowiak}, {Frederick},
  {Fremling}, {Gal-Yam}, {Gatkine}, {Ghosh}, {Goldstein}, {Golkhou}, {Graham},
  {Graham}, {Hankins}, {Helou}, {Hu}, {Ip}, {Jaodand}, {Karambelkar}, {Kong},
  {Kowalski}, {Khandagale}, {Kulkarni}, {Kumar}, {Laher}, {Li}, {Mahabal},
  {Masci}, {Miller}, {Mogotsi}, {Mohite}, {Mooley}, {Mroz}, {Newman}, {Ngeow},
  {Oates}, {Patil}, {Pandey}, {Pavana}, {Pian}, {Riddle}, {Sanchez-Ram{\i}rez},
  {Sharma}, {Singh}, {Smith}, {Soumagnac}, {Taggart}, {Tan}, {Tzanidakis},
  {Troja}, {Valeev}, {Walters}, {Waratkar}, {Webb}, {Yu}, {Zhang}, {Zhou}, \&
  {Zolkower}}]{kasliwal2020}
{Kasliwal}, M.~M., {Anand}, S., {Ahumada}, T., {et~al.} 2020, arXiv e-prints,
  arXiv:2006.11306.
\newblock \doarXiv{2006.11306}

\bibitem[{{Kawaguchi} {et~al.}(2015){Kawaguchi}, {Kyutoku}, {Nakano}, {Okawa},
  {Shibata}, \& {Taniguchi}}]{kawaguchi2015}
{Kawaguchi}, K., {Kyutoku}, K., {Nakano}, H., {et~al.} 2015, \prd, 92, 024014,
  \dodoi{10.1103/PhysRevD.92.024014}

\bibitem[{{Kawaguchi} {et~al.}(2016){Kawaguchi}, {Kyutoku}, {Shibata}, \&
  {Tanaka}}]{kawaguchi2016}
{Kawaguchi}, K., {Kyutoku}, K., {Shibata}, M., \& {Tanaka}, M. 2016, \apj, 825,
  52, \dodoi{10.3847/0004-637X/825/1/52}

\bibitem[{{Kawaguchi} {et~al.}(2018){Kawaguchi}, {Shibata}, \&
  {Tanaka}}]{kawaguchi2018}
{Kawaguchi}, K., {Shibata}, M., \& {Tanaka}, M. 2018, \apjl, 865, L21,
  \dodoi{10.3847/2041-8213/aade02}

\bibitem[{{Kawaguchi} {et~al.}(2020){Kawaguchi}, {Shibata}, \&
  {Tanaka}}]{kawaguchi2020}
---. 2020, \apj, 889, 171, \dodoi{10.3847/1538-4357/ab61f6}

\bibitem[{{Kilpatrick} {et~al.}(2017){Kilpatrick}, {Foley}, {Kasen},
  {Murguia-Berthier}, {Ramirez-Ruiz}, {Coulter}, {Drout}, {Piro}, {Shappee},
  {Boutsia}, {Contreras}, {Di Mille}, {Madore}, {Morrell}, {Pan}, {Prochaska},
  {Rest}, {Rojas-Bravo}, {Siebert}, {Simon}, \& {Ulloa}}]{kilpatrick2017}
{Kilpatrick}, C.~D., {Foley}, R.~J., {Kasen}, D., {et~al.} 2017, Science, 358,
  1583, \dodoi{10.1126/science.aaq0073}

\bibitem[{{Kiziltan} {et~al.}(2013){Kiziltan}, {Kottas}, {De Yoreo}, \&
  {Thorsett}}]{kiziltan2013}
{Kiziltan}, B., {Kottas}, A., {De Yoreo}, M., \& {Thorsett}, S.~E. 2013, \apj,
  778, 66, \dodoi{10.1088/0004-637X/778/1/66}

\bibitem[{{K{\'o}bori} {et~al.}(2020){K{\'o}bori}, {Bagoly}, \&
  {Bal{\'a}zs}}]{kobori2020}
{K{\'o}bori}, J., {Bagoly}, Z., \& {Bal{\'a}zs}, L.~G. 2020, \mnras, 494, 4343,
  \dodoi{10.1093/mnras/staa1034}

\bibitem[{{Korobkin} {et~al.}(2012){Korobkin}, {Rosswog}, {Arcones}, \&
  {Winteler}}]{korobkin2012}
{Korobkin}, O., {Rosswog}, S., {Arcones}, A., \& {Winteler}, C. 2012, \mnras,
  426, 1940, \dodoi{10.1111/j.1365-2966.2012.21859.x}

\bibitem[{{Korobkin} {et~al.}(2020){Korobkin}, {Wollaeger}, {Fryer},
  {Hungerford}, {Rosswog}, {Fontes}, {Mumpower}, {Chase}, {Even}, {Miller},
  {Misch}, \& {Lippuner}}]{korobkin2020}
{Korobkin}, O., {Wollaeger}, R., {Fryer}, C., {et~al.} 2020, arXiv e-prints,
  arXiv:2004.00102.
\newblock \doarXiv{2004.00102}

\bibitem[{{Kr{\"u}ger} \& {Foucart}(2020)}]{kruger2020}
{Kr{\"u}ger}, C.~J., \& {Foucart}, F. 2020, \prd, 101, 103002,
  \dodoi{10.1103/PhysRevD.101.103002}

\bibitem[{{Kulkarni}(2005)}]{kulkarni2005}
{Kulkarni}, S.~R. 2005, arXiv e-prints, astro.
\newblock \doarXiv{astro-ph/0510256}

\bibitem[{{Kyutoku} {et~al.}(2015){Kyutoku}, {Ioka}, {Okawa}, {Shibata}, \&
  {Taniguchi}}]{kyutoku2015}
{Kyutoku}, K., {Ioka}, K., {Okawa}, H., {Shibata}, M., \& {Taniguchi}, K. 2015,
  \prd, 92, 044028, \dodoi{10.1103/PhysRevD.92.044028}

\bibitem[{{Kyutoku} {et~al.}(2013){Kyutoku}, {Ioka}, \&
  {Shibata}}]{kyutoku2013}
{Kyutoku}, K., {Ioka}, K., \& {Shibata}, M. 2013, \prd, 88, 041503,
  \dodoi{10.1103/PhysRevD.88.041503}

\bibitem[{{Kyutoku} {et~al.}(2018){Kyutoku}, {Kiuchi}, {Sekiguchi}, {Shibata},
  \& {Taniguchi}}]{kyutoku2018}
{Kyutoku}, K., {Kiuchi}, K., {Sekiguchi}, Y., {Shibata}, M., \& {Taniguchi}, K.
  2018, \prd, 97, 023009, \dodoi{10.1103/PhysRevD.97.023009}

\bibitem[{{Kyutoku} {et~al.}(2011){Kyutoku}, {Okawa}, {Shibata}, \&
  {Taniguchi}}]{kyutoku2011}
{Kyutoku}, K., {Okawa}, H., {Shibata}, M., \& {Taniguchi}, K. 2011, \prd, 84,
  064018, \dodoi{10.1103/PhysRevD.84.064018}

\bibitem[{{Lamb} \& {Kobayashi}(2017)}]{lamb2017}
{Lamb}, G.~P., \& {Kobayashi}, S. 2017, \mnras, 472, 4953,
  \dodoi{10.1093/mnras/stx2345}

\bibitem[{{Lattimer}(2012)}]{lattimer2012}
{Lattimer}, J.~M. 2012, Annual Review of Nuclear and Particle Science, 62, 485,
  \dodoi{10.1146/annurev-nucl-102711-095018}

\bibitem[{{Lattimer} \& {Schramm}(1974)}]{lattimer1974}
{Lattimer}, J.~M., \& {Schramm}, D.~N. 1974, \apjl, 192, L145,
  \dodoi{10.1086/181612}

\bibitem[{{Lattimer} \& {Schramm}(1976)}]{lattimer1976}
---. 1976, \apj, 210, 549, \dodoi{10.1086/154860}

\bibitem[{{Lazzati} {et~al.}(2018){Lazzati}, {Perna}, {Morsony},
  {Lopez-Camara}, {Cantiello}, {Ciolfi}, {Giacomazzo}, \&
  {Workman}}]{lazzati2018}
{Lazzati}, D., {Perna}, R., {Morsony}, B.~J., {et~al.} 2018, \prl, 120, 241103,
  \dodoi{10.1103/PhysRevLett.120.241103}

\bibitem[{{Lei} {et~al.}(2021){Lei}, {Li}, {Wu}, {Jiang}, \& {Chen}}]{lei2021}
{Lei}, L., {Li}, J., {Wu}, J., {Jiang}, S., \& {Chen}, B. 2021, Astronomical
  Research \& Technology, 18, L18,
  \dodoi{10.14005/j.cnki.issn1672-7673.20200713.001}

\bibitem[{{Li} \& {Paczy{\'n}ski}(1998)}]{li1998}
{Li}, L.-X., \& {Paczy{\'n}ski}, B. 1998, \apjl, 507, L59,
  \dodoi{10.1086/311680}

\bibitem[{{Li} {et~al.}(2018){Li}, {Liu}, {Yu}, \& {Zhang}}]{li2018}
{Li}, S.-Z., {Liu}, L.-D., {Yu}, Y.-W., \& {Zhang}, B. 2018, \apjl, 861, L12,
  \dodoi{10.3847/2041-8213/aace61}

\bibitem[{{LIGO Scientific Collaboration}(2018)}]{lalsuite}
{LIGO Scientific Collaboration}. 2018, {LIGO} {A}lgorithm {L}ibrary -
  {LALS}uite, free software (GPL), \dodoi{10.7935/GT1W-FZ16}

\bibitem[{{Lippuner} {et~al.}(2017){Lippuner}, {Fern{\'a}ndez}, {Roberts},
  {Foucart}, {Kasen}, {Metzger}, \& {Ott}}]{lippuner2017}
{Lippuner}, J., {Fern{\'a}ndez}, R., {Roberts}, L.~F., {et~al.} 2017, \mnras,
  472, 904, \dodoi{10.1093/mnras/stx1987}

\bibitem[{{Lipunov} {et~al.}(2017){Lipunov}, {Gorbovskoy}, {Kornilov}, {.
  Tyurina}, {Balanutsa}, {Kuznetsov}, {Vlasenko}, {Kuvshinov}, {Gorbunov},
  {Buckley}, {Krylov}, {Podesta}, {Lopez}, {Podesta}, {Levato}, {Saffe},
  {Mallamachi}, {Potter}, {Budnev}, {Gress}, {Ishmuhametova}, {Vladimirov},
  {Zimnukhov}, {Yurkov}, {Sergienko}, {Gabovich}, {Rebolo}, {Serra-Ricart},
  {Israelyan}, {Chazov}, {Wang}, {Tlatov}, \& {Panchenko}}]{lipunov2017}
{Lipunov}, V.~M., {Gorbovskoy}, E., {Kornilov}, V.~G., {et~al.} 2017, \apjl,
  850, L1, \dodoi{10.3847/2041-8213/aa92c0}

\bibitem[{{Liu} {et~al.}(2020){Liu}, {Gao}, \& {Zhang}}]{liu2020}
{Liu}, L.-D., {Gao}, H., \& {Zhang}, B. 2020, \apj, 890, 102,
  \dodoi{10.3847/1538-4357/ab6b24}

\bibitem[{{LSST Science Collaboration} {et~al.}(2009){LSST Science
  Collaboration}, {Abell}, {Allison}, {Anderson}, {Andrew}, {Angel}, {Armus},
  {Arnett}, {Asztalos}, {Axelrod}, {Bailey}, {Ballantyne}, {Bankert},
  {Barkhouse}, {Barr}, {Barrientos}, {Barth}, {Bartlett}, {Becker}, {Becla},
  {Beers}, {Bernstein}, {Biswas}, {Blanton}, {Bloom}, {Bochanski}, {Boeshaar},
  {Borne}, {Bradac}, {Brandt}, {Bridge}, {Brown}, {Brunner}, {Bullock},
  {Burgasser}, {Burge}, {Burke}, {Cargile}, {Chand rasekharan}, {Chartas},
  {Chesley}, {Chu}, {Cinabro}, {Claire}, {Claver}, {Clowe}, {Connolly}, {Cook},
  {Cooke}, {Cooray}, {Covey}, {Culliton}, {de Jong}, {de Vries}, {Debattista},
  {Delgado}, {Dell'Antonio}, {Dhital}, {Di Stefano}, {Dickinson}, {Dilday},
  {Djorgovski}, {Dobler}, {Donalek}, {Dubois-Felsmann}, {Durech},
  {Eliasdottir}, {Eracleous}, {Eyer}, {Falco}, {Fan}, {Fassnacht}, {Ferguson},
  {Fernandez}, {Fields}, {Finkbeiner}, {Figueroa}, {Fox}, {Francke}, {Frank},
  {Frieman}, {Fromenteau}, {Furqan}, {Galaz}, {Gal-Yam}, {Garnavich},
  {Gawiser}, {Geary}, {Gee}, {Gibson}, {Gilmore}, {Grace}, {Green}, {Gressler},
  {Grillmair}, {Habib}, {Haggerty}, {Hamuy}, {Harris}, {Hawley}, {Heavens},
  {Hebb}, {Henry}, {Hileman}, {Hilton}, {Hoadley}, {Holberg}, {Holman},
  {Howell}, {Infante}, {Ivezic}, {Jacoby}, {Jain}, {R}, {Jedicke}, {Jee},
  {Garrett Jernigan}, {Jha}, {Johnston}, {Jones}, {Juric}, {Kaasalainen},
  {Styliani}, {Kafka}, {Kahn}, {Kaib}, {Kalirai}, {Kantor}, {Kasliwal},
  {Keeton}, {Kessler}, {Knezevic}, {Kowalski}, {Krabbendam}, {Krughoff},
  {Kulkarni}, {Kuhlman}, {Lacy}, {Lepine}, {Liang}, {Lien}, {Lira}, {Long},
  {Lorenz}, {Lotz}, {Lupton}, {Lutz}, {Macri}, {Mahabal}, {Mandelbaum},
  {Marshall}, {May}, {McGehee}, {Meadows}, {Meert}, {Milani}, {Miller},
  {Miller}, {Mills}, {Minniti}, {Monet}, {Mukadam}, {Nakar}, {Neill}, {Newman},
  {Nikolaev}, {Nordby}, {O'Connor}, {Oguri}, {Oliver}, {Olivier}, {Olsen},
  {Olsen}, {Olszewski}, {Oluseyi}, {Padilla}, {Parker}, {Pepper}, {Peterson},
  {Petry}, {Pinto}, {Pizagno}, {Popescu}, {Prsa}, {Radcka}, {Raddick},
  {Rasmussen}, {Rau}, {Rho}, {Rhoads}, {Richards}, {Ridgway}, {Robertson},
  {Roskar}, {Saha}, {Sarajedini}, {Scannapieco}, {Schalk}, {Schindler},
  {Schmidt}, {Schmidt}, {Schneider}, {Schumacher}, {Scranton}, {Sebag},
  {Seppala}, {Shemmer}, {Simon}, {Sivertz}, {Smith}, {Allyn Smith}, {Smith},
  {Spitz}, {Stanford}, {Stassun}, {Strader}, {Strauss}, {Stubbs}, {Sweeney},
  {Szalay}, {Szkody}, {Takada}, {Thorman}, {Trilling}, {Trimble}, {Tyson}, {Van
  Berg}, {Vand en Berk}, {VanderPlas}, {Verde}, {Vrsnak}, {Walkowicz}, {Wand
  elt}, {Wang}, {Wang}, {Warner}, {Wechsler}, {West}, {Wiecha}, {Williams},
  {Willman}, {Wittman}, {Wolff}, {Wood-Vasey}, {Wozniak}, {Young}, {Zentner},
  \& {Zhan}}]{lsst2009}
{LSST Science Collaboration}, {Abell}, P.~A., {Allison}, J., {et~al.} 2009,
  arXiv e-prints, arXiv:0912.0201.
\newblock \doarXiv{0912.0201}

\bibitem[{{L{\"u}} {et~al.}(2015){L{\"u}}, {Zhang}, {Lei}, {Li}, \&
  {Lasky}}]{lv2015}
{L{\"u}}, H.-J., {Zhang}, B., {Lei}, W.-H., {Li}, Y., \& {Lasky}, P.~D. 2015,
  \apj, 805, 89, \dodoi{10.1088/0004-637X/805/2/89}

\bibitem[{{Lyman} {et~al.}(2018){Lyman}, {Lamb}, {Levan}, {Mandel}, {Tanvir},
  {Kobayashi}, {Gompertz}, {Hjorth}, {Fruchter}, {Kangas}, {Steeghs}, {Steele},
  {Cano}, {Copperwheat}, {Evans}, {Fynbo}, {Gall}, {Im}, {Izzo}, {Jakobsson},
  {Milvang-Jensen}, {O'Brien}, {Osborne}, {Palazzi}, {Perley}, {Pian},
  {Rosswog}, {Rowlinson}, {Schulze}, {Stanway}, {Sutton}, {Th{\"o}ne}, {de
  Ugarte Postigo}, {Watson}, {Wiersema}, \& {Wijers}}]{lyman2018}
{Lyman}, J.~D., {Lamb}, G.~P., {Levan}, A.~J., {et~al.} 2018, Nature Astronomy,
  2, 751, \dodoi{10.1038/s41550-018-0511-3}

\bibitem[{{Ma} {et~al.}(2018){Ma}, {Lei}, {Gao}, {Xie}, {Chen}, {Zhang}, \&
  {Wang}}]{ma2018}
{Ma}, S.-B., {Lei}, W.-H., {Gao}, H., {et~al.} 2018, \apjl, 852, L5,
  \dodoi{10.3847/2041-8213/aaa0cd}

\bibitem[{Maggiore {et~al.}(2020)Maggiore, Van Den~Broeck, Bartolo, Belgacem,
  Bertacca, Bizouard, Branchesi, Clesse, Foffa, Garc{\'\i}a-Bellido,
  {et~al.}}]{maggiore2020science}
Maggiore, M., Van Den~Broeck, C., Bartolo, N., {et~al.} 2020, Journal of
  Cosmology and Astroparticle Physics, 2020, 050

\bibitem[{{Maggiore} {et~al.}(2020){Maggiore}, {Van Den Broeck}, {Bartolo},
  {Belgacem}, {Bertacca}, {Bizouard}, {Branchesi}, {Clesse}, {Foffa},
  {Garc{\'\i}a-Bellido}, {Grimm}, {Harms}, {Hinderer}, {Matarrese}, {Palomba},
  {Peloso}, {Ricciardone}, \& {Sakellariadou}}]{maggiore2020}
{Maggiore}, M., {Van Den Broeck}, C., {Bartolo}, N., {et~al.} 2020, \jcap,
  2020, 050, \dodoi{10.1088/1475-7516/2020/03/050}

\bibitem[{{Mapelli} \& {Giacobbo}(2018)}]{mapelli2018}
{Mapelli}, M., \& {Giacobbo}, N. 2018, \mnras, 479, 4391,
  \dodoi{10.1093/mnras/sty1613}

\bibitem[{{Margutti} {et~al.}(2017){Margutti}, {Berger}, {Fong}, {Guidorzi},
  {Alexander}, {Metzger}, {Blanchard}, {Cowperthwaite}, {Chornock},
  {Eftekhari}, {Nicholl}, {Villar}, {Williams}, {Annis}, {Brown}, {Chen},
  {Doctor}, {Frieman}, {Holz}, {Sako}, \& {Soares-Santos}}]{margutti2017}
{Margutti}, R., {Berger}, E., {Fong}, W., {et~al.} 2017, \apjl, 848, L20,
  \dodoi{10.3847/2041-8213/aa9057}

\bibitem[{{Margutti} {et~al.}(2018{\natexlab{a}}){Margutti}, {Alexander},
  {Xie}, {Sironi}, {Metzger}, {Kathirgamaraju}, {Fong}, {Blanchard}, {Berger},
  {MacFadyen}, {Giannios}, {Guidorzi}, {Hajela}, {Chornock}, {Cowperthwaite},
  {Eftekhari}, {Nicholl}, {Villar}, {Williams}, \&
  {Zrake}}]{margutti2018GW170817}
{Margutti}, R., {Alexander}, K.~D., {Xie}, X., {et~al.} 2018{\natexlab{a}},
  \apjl, 856, L18, \dodoi{10.3847/2041-8213/aab2ad}

\bibitem[{{Margutti} {et~al.}(2018{\natexlab{b}}){Margutti}, {Cowperthwaite},
  {Doctor}, {Mortensen}, {Pankow}, {Salafia}, {Villar}, {Alexander}, {Annis},
  {Andreoni}, {Baldeschi}, {Balmaverde}, {Berger}, {Bernardini}, {Berry},
  {Bianco}, {Blanchard}, {Brocato}, {Carnerero}, {Cartier}, {Cenko},
  {Chornock}, {Chomiuk}, {Copperwheat}, {Coughlin}, {Coppejans}, {Corsi},
  {D'Ammando}, {Datrier}, {D'Avanzo}, {Dimitriadis}, {Drout}, {Foley}, {Fong},
  {Fox}, {Ghirlanda}, {Goldstein}, {Grindlay}, {Guidorzi}, {Haiman}, {Hendry},
  {Holz}, {Hung}, {Inserra}, {Jones}, {Kalogera}, {Kilpatrick}, {Lamb},
  {Laskar}, {Levan}, {Mason}, {Maguire}, {Melandri}, {Milisavljevic}, {Miller},
  {Narayan}, {Nielsen}, {Nicholl}, {Nissanke}, {Nugent}, {Pan}, {Pasham},
  {Paterson}, {Piranomonte}, {Racusin}, {Rest}, {Righi}, {Sand}, {Seaman},
  {Scolnic}, {Siellez}, {Singer}, {Szkody}, {Smith}, {Steeghs}, {Sullivan},
  {Tanvir}, {Terreran}, {Trimble}, {Valenti}, {LSST Transient}, \& {Variable
  Stars Collaboration}}]{margutti2018ToO}
{Margutti}, R., {Cowperthwaite}, P., {Doctor}, Z., {et~al.} 2018{\natexlab{b}},
  arXiv e-prints, arXiv:1812.04051.
\newblock \doarXiv{1812.04051}

\bibitem[{{Martin} {et~al.}(2015){Martin}, {Perego}, {Arcones}, {Thielemann},
  {Korobkin}, \& {Rosswog}}]{martin2015}
{Martin}, D., {Perego}, A., {Arcones}, A., {et~al.} 2015, \apj, 813, 2,
  \dodoi{10.1088/0004-637X/813/1/2}

\bibitem[{{Masci} {et~al.}(2019){Masci}, {Laher}, {Rusholme}, {Shupe}, {Groom},
  {Surace}, {Jackson}, {Monkewitz}, {Beck}, {Flynn}, {Terek}, {Landry},
  {Hacopians}, {Desai}, {Howell}, {Brooke}, {Imel}, {Wachter}, {Ye}, {Lin},
  {Cenko}, {Cunningham}, {Rebbapragada}, {Bue}, {Miller}, {Mahabal}, {Bellm},
  {Patterson}, {Juri{\'c}}, {Golkhou}, {Ofek}, {Walters}, {Graham}, {Kasliwal},
  {Dekany}, {Kupfer}, {Burdge}, {Cannella}, {Barlow}, {Van Sistine}, {Giomi},
  {Fremling}, {Blagorodnova}, {Levitan}, {Riddle}, {Smith}, {Helou}, {Prince},
  \& {Kulkarni}}]{masci2019}
{Masci}, F.~J., {Laher}, R.~R., {Rusholme}, B., {et~al.} 2019, \pasp, 131,
  018003, \dodoi{10.1088/1538-3873/aae8ac}

\bibitem[{{McBrien} {et~al.}(2020){McBrien}, {Smartt}, {Huber}, {Rest},
  {Chambers}, {Barbieri}, {Bulla}, {Jha}, {Gromadzki}, {Srivastav}, {Smith},
  {Young}, {McLaughlin}, {Inserra}, {Nicholl}, {Fraser}, {Maguire}, {Chen},
  {Wevers}, {Anderson}, {M{\"u}ller-Bravo}, {Olivares E.}, {Kankare},
  {Gal-Yam}, \& {Waters}}]{mcbrien2020}
{McBrien}, O.~R., {Smartt}, S.~J., {Huber}, M.~E., {et~al.} 2020, arXiv
  e-prints, arXiv:2006.10442.
\newblock \doarXiv{2006.10442}

\bibitem[{{McClintock} {et~al.}(2011){McClintock}, {Narayan}, {Davis}, {Gou},
  {Kulkarni}, {Orosz}, {Penna}, {Remillard}, \& {Steiner}}]{mcclintock2011}
{McClintock}, J.~E., {Narayan}, R., {Davis}, S.~W., {et~al.} 2011, Classical
  and Quantum Gravity, 28, 114009, \dodoi{10.1088/0264-9381/28/11/114009}

\bibitem[{{McCully} {et~al.}(2017){McCully}, {Hiramatsu}, {Howell},
  {Hosseinzadeh}, {Arcavi}, {Kasen}, {Barnes}, {Shara}, {Williams},
  {V{\"a}is{\"a}nen}, {Potter}, {Romero-Colmenero}, {Crawford}, {Buckley},
  {Cooke}, {Andreoni}, {Pritchard}, {Mao}, {Gromadzki}, \&
  {Burke}}]{mccully2017}
{McCully}, C., {Hiramatsu}, D., {Howell}, D.~A., {et~al.} 2017, \apjl, 848,
  L32, \dodoi{10.3847/2041-8213/aa9111}

\bibitem[{{McKernan} {et~al.}(2020){McKernan}, {Ford}, \&
  {O'Shaughnessy}}]{mckernan2020}
{McKernan}, B., {Ford}, K.~E.~S., \& {O'Shaughnessy}, R. 2020, \mnras, 498,
  4088, \dodoi{10.1093/mnras/staa2681}

\bibitem[{{Meszaros} \& {Rees}(1993)}]{meszaros1993}
{Meszaros}, P., \& {Rees}, M.~J. 1993, \apj, 405, 278, \dodoi{10.1086/172360}

\bibitem[{{M{\'e}sz{\'a}ros} \& {Rees}(1997)}]{meszaros1997}
{M{\'e}sz{\'a}ros}, P., \& {Rees}, M.~J. 1997, \apj, 476, 232,
  \dodoi{10.1086/303625}

\bibitem[{{Metzger}(2017)}]{metzger2017}
{Metzger}, B.~D. 2017, Living Reviews in Relativity, 20, 3,
  \dodoi{10.1007/s41114-017-0006-z}

\bibitem[{{Metzger}(2019)}]{metzger2019}
---. 2019, Living Reviews in Relativity, 23, 1,
  \dodoi{10.1007/s41114-019-0024-0}

\bibitem[{{Metzger} \& {Berger}(2012)}]{metzger2012}
{Metzger}, B.~D., \& {Berger}, E. 2012, \apj, 746, 48,
  \dodoi{10.1088/0004-637X/746/1/48}

\bibitem[{{Metzger} \& {Fern{\'a}ndez}(2014)}]{metzger2014b}
{Metzger}, B.~D., \& {Fern{\'a}ndez}, R. 2014, \mnras, 441, 3444,
  \dodoi{10.1093/mnras/stu802}

\bibitem[{{Metzger} \& {Piro}(2014)}]{metzger2014a}
{Metzger}, B.~D., \& {Piro}, A.~L. 2014, \mnras, 439, 3916,
  \dodoi{10.1093/mnras/stu247}

\bibitem[{{Metzger} {et~al.}(2010){Metzger}, {Mart{\'\i}nez-Pinedo}, {Darbha},
  {Quataert}, {Arcones}, {Kasen}, {Thomas}, {Nugent}, {Panov}, \&
  {Zinner}}]{metzger2010}
{Metzger}, B.~D., {Mart{\'\i}nez-Pinedo}, G., {Darbha}, S., {et~al.} 2010,
  \mnras, 406, 2650, \dodoi{10.1111/j.1365-2966.2010.16864.x}

\bibitem[{Michimura {et~al.}(2014)Michimura, Komori, Enomoto, Nagano, \&
  Somiya}]{michimura2014example}
Michimura, Y., Komori, K., Enomoto, Y., Nagano, K., \& Somiya, K. 2014, Phys.
  Rev. D, 90, 062006

\bibitem[{Michimura {et~al.}(2020)Michimura, Komori, Enomoto, Nagano,
  Nishizawa, Hirose, Leonardi, Capocasa, Aritomi, Zhao,
  {et~al.}}]{michimura2020prospects}
Michimura, Y., Komori, K., Enomoto, Y., {et~al.} 2020, \prd, 102, 022008

\bibitem[{Miller {et~al.}(2015)Miller, Barsotti, Vitale, Fritschel, Evans, \&
  Sigg}]{miller2015prospects}
Miller, J., Barsotti, L., Vitale, S., {et~al.} 2015, \prd, 91, 062005

\bibitem[{{Miller} {et~al.}(2011){Miller}, {Miller}, \&
  {Reynolds}}]{miller2011}
{Miller}, J.~M., {Miller}, M.~C., \& {Reynolds}, C.~S. 2011, \apjl, 731, L5,
  \dodoi{10.1088/2041-8205/731/1/L5}

\bibitem[{{Miller} \& {Miller}(2015)}]{miller2015}
{Miller}, M.~C., \& {Miller}, J.~M. 2015, \physrep, 548, 1,
  \dodoi{10.1016/j.physrep.2014.09.003}

\bibitem[{{Mooley} {et~al.}(2018){Mooley}, {Deller}, {Gottlieb}, {Nakar},
  {Hallinan}, {Bourke}, {Frail}, {Horesh}, {Corsi}, \&
  {Hotokezaka}}]{mooley2018}
{Mooley}, K.~P., {Deller}, A.~T., {Gottlieb}, O., {et~al.} 2018, \nat, 561,
  355, \dodoi{10.1038/s41586-018-0486-3}

\bibitem[{{M{\"u}ller} \& {Serot}(1996)}]{muller1996}
{M{\"u}ller}, H., \& {Serot}, B.~D. 1996, \nphysa, 606, 508,
  \dodoi{10.1016/0375-9474(96)00187-X}

\bibitem[{{Narayan} {et~al.}(1992){Narayan}, {Paczynski}, \&
  {Piran}}]{narayan1992}
{Narayan}, R., {Paczynski}, B., \& {Piran}, T. 1992, \apjl, 395, L83,
  \dodoi{10.1086/186493}

\bibitem[{{Nicholl} {et~al.}(2017){Nicholl}, {Berger}, {Kasen}, {Metzger},
  {Elias}, {Brice{\~n}o}, {Alexander}, {Blanchard}, {Chornock},
  {Cowperthwaite}, {Eftekhari}, {Fong}, {Margutti}, {Villar}, {Williams},
  {Brown}, {Annis}, {Bahramian}, {Brout}, {Brown}, {Chen}, {Clemens},
  {Dennihy}, {Dunlap}, {Holz}, {Marchesini}, {Massaro}, {Moskowitz},
  {Pelisoli}, {Rest}, {Ricci}, {Sako}, {Soares-Santos}, \&
  {Strader}}]{nicholl2017}
{Nicholl}, M., {Berger}, E., {Kasen}, D., {et~al.} 2017, \apjl, 848, L18,
  \dodoi{10.3847/2041-8213/aa9029}

\bibitem[{{Nitz} {et~al.}(2020){Nitz}, {Sch{\"a}fer}, \& {Dal
  Canton}}]{nitz2020}
{Nitz}, A.~H., {Sch{\"a}fer}, M., \& {Dal Canton}, T. 2020, arXiv e-prints,
  arXiv:2009.04439.
\newblock \doarXiv{2009.04439}

\bibitem[{{O'Connor} {et~al.}(2020){O'Connor}, {Beniamini}, \&
  {Kouveliotou}}]{oconnor2020}
{O'Connor}, B., {Beniamini}, P., \& {Kouveliotou}, C. 2020, \mnras, 495, 4782,
  \dodoi{10.1093/mnras/staa1433}

\bibitem[{{O'Shaughnessy} {et~al.}(2008){O'Shaughnessy}, {Kim}, {Kalogera}, \&
  {Belczynski}}]{oshaughnessy2008}
{O'Shaughnessy}, R., {Kim}, C., {Kalogera}, V., \& {Belczynski}, K. 2008, \apj,
  672, 479, \dodoi{10.1086/523620}

\bibitem[{{{\"O}zel} {et~al.}(2010){{\"O}zel}, {Psaltis}, {Narayan}, \&
  {McClintock}}]{ozel2010}
{{\"O}zel}, F., {Psaltis}, D., {Narayan}, R., \& {McClintock}, J.~E. 2010,
  \apj, 725, 1918, \dodoi{10.1088/0004-637X/725/2/1918}

\bibitem[{{Paczynski}(1991)}]{paczynski1991}
{Paczynski}, B. 1991, \actaa, 41, 257

\bibitem[{{Paczynski} \& {Rhoads}(1993)}]{paczynski1993}
{Paczynski}, B., \& {Rhoads}, J.~E. 1993, \apjl, 418, L5,
  \dodoi{10.1086/187102}

\bibitem[{{Page} {et~al.}(2020){Page}, {Evans}, {Tohuvavohu}, {Kennea},
  {Klingler}, {Cenko}, {Oates}, {Ambrosi}, {Barthelmy}, {Beardmore},
  {Bernardini}, {Breeveld}, {Brown}, {Burrows}, {Campana}, {Caputo},
  {Cusumano}, {D'Ai}, {D'Avanzo}, {D'Elia}, {De Pasquale}, {Emery}, {Giommi},
  {Gronwall}, {Hartmann}, {Krimm}, {Kuin}, {Malesani}, {Marshall}, {Meland ri},
  {Nousek}, {O'Brien}, {Osborne}, {Pagani}, {Page}, {Palmer}, {Perri},
  {Racusin}, {Sakamoto}, {Sbarufatti}, {Schlieder}, {Siegel}, {Tagliaferri}, \&
  {Troja}}]{page2020}
{Page}, K.~L., {Evans}, P.~A., {Tohuvavohu}, A., {et~al.} 2020, arXiv e-prints,
  arXiv:2009.13804.
\newblock \doarXiv{2009.13804}

\bibitem[{{Pannarale}(2013)}]{pannarale2013}
{Pannarale}, F. 2013, \prd, 88, 104025, \dodoi{10.1103/PhysRevD.88.104025}

\bibitem[{{Perego} {et~al.}(2017){Perego}, {Radice}, \&
  {Bernuzzi}}]{perego2017}
{Perego}, A., {Radice}, D., \& {Bernuzzi}, S. 2017, \apjl, 850, L37,
  \dodoi{10.3847/2041-8213/aa9ab9}

\bibitem[{{Perego} {et~al.}(2014){Perego}, {Rosswog}, {Cabez{\'o}n},
  {Korobkin}, {K{\"a}ppeli}, {Arcones}, \& {Liebend{\"o}rfer}}]{perego2014}
{Perego}, A., {Rosswog}, S., {Cabez{\'o}n}, R.~M., {et~al.} 2014, \mnras, 443,
  3134, \dodoi{10.1093/mnras/stu1352}

\bibitem[{{Perna} {et~al.}(2021){Perna}, {Lazzati}, \& {Cantiello}}]{perna2021}
{Perna}, R., {Lazzati}, D., \& {Cantiello}, M. 2021, \apjl, 906, L7,
  \dodoi{10.3847/2041-8213/abd319}

\bibitem[{{Pian} {et~al.}(2017){Pian}, {D'Avanzo}, {Benetti}, {Branchesi},
  {Brocato}, {Campana}, {Cappellaro}, {Covino}, {D'Elia}, {Fynbo}, {Getman},
  {Ghirland a}, {Ghisellini}, {Grado}, {Greco}, {Hjorth}, {Kouveliotou},
  {Levan}, {Limatola}, {Malesani}, {Mazzali}, {Melandri}, {M{\o}ller},
  {Nicastro}, {Palazzi}, {Piranomonte}, {Rossi}, {Salafia}, {Selsing},
  {Stratta}, {Tanaka}, {Tanvir}, {Tomasella}, {Watson}, {Yang}, {Amati},
  {Antonelli}, {Ascenzi}, {Bernardini}, {Bo{\"e}r}, {Bufano}, {Bulgarelli},
  {Capaccioli}, {Casella}, {Castro-Tirado}, {Chassande-Mottin}, {Ciolfi},
  {Copperwheat}, {Dadina}, {De Cesare}, {di Paola}, {Fan}, {Gendre},
  {Giuffrida}, {Giunta}, {Hunt}, {Israel}, {Jin}, {Kasliwal}, {Klose}, {Lisi},
  {Longo}, {Maiorano}, {Mapelli}, {Masetti}, {Nava}, {Patricelli}, {Perley},
  {Pescalli}, {Piran}, {Possenti}, {Pulone}, {Razzano}, {Salvaterra},
  {Schipani}, {Spera}, {Stamerra}, {Stella}, {Tagliaferri}, {Testa}, {Troja},
  {Turatto}, {Vergani}, \& {Vergani}}]{pian2017}
{Pian}, E., {D'Avanzo}, P., {Benetti}, S., {et~al.} 2017, \nat, 551, 67,
  \dodoi{10.1038/nature24298}

\bibitem[{{Piro} {et~al.}(2019){Piro}, {Troja}, {Zhang}, {Ryan}, {van Eerten},
  {Ricci}, {Wieringa}, {Tiengo}, {Butler}, {Cenko}, {Fox}, {Khandrika},
  {Novara}, {Rossi}, \& {Sakamoto}}]{piro2019}
{Piro}, L., {Troja}, E., {Zhang}, B., {et~al.} 2019, \mnras, 483, 1912,
  \dodoi{10.1093/mnras/sty3047}

\bibitem[{{Planck Collaboration} {et~al.}(2016){Planck Collaboration}, {Ade},
  {Aghanim}, {Arnaud}, {Ashdown}, {Aumont}, {Baccigalupi}, {Banday},
  {Barreiro}, {Bartlett}, {Bartolo}, {Battaner}, {Battye}, {Benabed},
  {Beno{\^\i}t}, {Benoit-L{\'e}vy}, {Bernard}, {Bersanelli}, {Bielewicz},
  {Bock}, {Bonaldi}, {Bonavera}, {Bond}, {Borrill}, {Bouchet}, {Boulanger},
  {Bucher}, {Burigana}, {Butler}, {Calabrese}, {Cardoso}, {Catalano},
  {Challinor}, {Chamballu}, {Chary}, {Chiang}, {Chluba}, {Christensen},
  {Church}, {Clements}, {Colombi}, {Colombo}, {Combet}, {Coulais}, {Crill},
  {Curto}, {Cuttaia}, {Danese}, {Davies}, {Davis}, {de Bernardis}, {de Rosa},
  {de Zotti}, {Delabrouille}, {D{\'e}sert}, {Di Valentino}, {Dickinson},
  {Diego}, {Dolag}, {Dole}, {Donzelli}, {Dor{\'e}}, {Douspis}, {Ducout},
  {Dunkley}, {Dupac}, {Efstathiou}, {Elsner}, {En{\ss}lin}, {Eriksen},
  {Farhang}, {Fergusson}, {Finelli}, {Forni}, {Frailis}, {Fraisse},
  {Franceschi}, {Frejsel}, {Galeotta}, {Galli}, {Ganga}, {Gauthier}, {Gerbino},
  {Ghosh}, {Giard}, {Giraud-H{\'e}raud}, {Giusarma}, {Gjerl{\o}w},
  {Gonz{\'a}lez-Nuevo}, {G{\'o}rski}, {Gratton}, {Gregorio}, {Gruppuso},
  {Gudmundsson}, {Hamann}, {Hansen}, {Hanson}, {Harrison}, {Helou},
  {Henrot-Versill{\'e}}, {Hern{\'a}ndez-Monteagudo}, {Herranz}, {Hildebrand t},
  {Hivon}, {Hobson}, {Holmes}, {Hornstrup}, {Hovest}, {Huang}, {Huffenberger},
  {Hurier}, {Jaffe}, {Jaffe}, {Jones}, {Juvela}, {Keih{\"a}nen}, {Keskitalo},
  {Kisner}, {Kneissl}, {Knoche}, {Knox}, {Kunz}, {Kurki-Suonio}, {Lagache},
  {L{\"a}hteenm{\"a}ki}, {Lamarre}, {Lasenby}, {Lattanzi}, {Lawrence}, {Leahy},
  {Leonardi}, {Lesgourgues}, {Levrier}, {Lewis}, {Liguori}, {Lilje},
  {Linden-V{\o}rnle}, {L{\'o}pez-Caniego}, {Lubin}, {Mac{\'\i}as-P{\'e}rez},
  {Maggio}, {Maino}, {Mandolesi}, {Mangilli}, {Marchini}, {Maris}, {Martin},
  {Martinelli}, {Mart{\'\i}nez-Gonz{\'a}lez}, {Masi}, {Matarrese}, {McGehee},
  {Meinhold}, {Melchiorri}, {Melin}, {Mendes}, {Mennella}, {Migliaccio},
  {Millea}, {Mitra}, {Miville-Desch{\^e}nes}, {Moneti}, {Montier}, {Morgante},
  {Mortlock}, {Moss}, {Munshi}, {Murphy}, {Naselsky}, {Nati}, {Natoli},
  {Netterfield}, {N{\o}rgaard-Nielsen}, {Noviello}, {Novikov}, {Novikov},
  {Oxborrow}, {Paci}, {Pagano}, {Pajot}, {Paladini}, {Paoletti}, {Partridge},
  {Pasian}, {Patanchon}, {Pearson}, {Perdereau}, {Perotto}, {Perrotta},
  {Pettorino}, {Piacentini}, {Piat}, {Pierpaoli}, {Pietrobon}, {Plaszczynski},
  {Pointecouteau}, {Polenta}, {Popa}, {Pratt}, {Pr{\'e}zeau}, {Prunet},
  {Puget}, {Rachen}, {Reach}, {Rebolo}, {Reinecke}, {Remazeilles}, {Renault},
  {Renzi}, {Ristorcelli}, {Rocha}, {Rosset}, {Rossetti}, {Roudier},
  {Rouill{\'e} d'Orfeuil}, {Rowan-Robinson}, {Rubi{\~n}o-Mart{\'\i}n},
  {Rusholme}, {Said}, {Salvatelli}, {Salvati}, {Sandri}, {Santos},
  {Savelainen}, {Savini}, {Scott}, {Seiffert}, {Serra}, {Shellard}, {Spencer},
  {Spinelli}, {Stolyarov}, {Stompor}, {Sudiwala}, {Sunyaev}, {Sutton},
  {Suur-Uski}, {Sygnet}, {Tauber}, {Terenzi}, {Toffolatti}, {Tomasi},
  {Tristram}, {Trombetti}, {Tucci}, {Tuovinen}, {T{\"u}rler}, {Umana},
  {Valenziano}, {Valiviita}, {Van Tent}, {Vielva}, {Villa}, {Wade}, {Wandelt},
  {Wehus}, {White}, {White}, {Wilkinson}, {Yvon}, {Zacchei}, \&
  {Zonca}}]{planck2016}
{Planck Collaboration}, {Ade}, P.~A.~R., {Aghanim}, N., {et~al.} 2016, \aap,
  594, A13, \dodoi{10.1051/0004-6361/201525830}

\bibitem[{{Prentice} {et~al.}(2018){Prentice}, {Maguire}, {Smartt}, {Magee},
  {Schady}, {Sim}, {Chen}, {Clark}, {Colin}, {Fulton}, {McBrien}, {O'Neill},
  {Smith}, {Ashall}, {Chambers}, {Denneau}, {Flewelling}, {Heinze}, {Holoien},
  {Huber}, {Kochanek}, {Mazzali}, {Prieto}, {Rest}, {Shappee}, {Stalder},
  {Stanek}, {Stritzinger}, {Thompson}, \& {Tonry}}]{prentice2018}
{Prentice}, S.~J., {Maguire}, K., {Smartt}, S.~J., {et~al.} 2018, \apjl, 865,
  L3, \dodoi{10.3847/2041-8213/aadd90}

\bibitem[{Punturo {et~al.}(2010{\natexlab{a}})Punturo, Abernathy, Acernese,
  Allen, Andersson, Arun, Barone, Barr, Barsuglia, Beker,
  {et~al.}}]{punturo2010einstein}
Punturo, M., Abernathy, M., Acernese, F., {et~al.} 2010{\natexlab{a}},
  Classical and Quantum Gravity, 27, 194002

\bibitem[{Punturo {et~al.}(2010{\natexlab{b}})Punturo, Abernathy, Acernese,
  Allen, Andersson, Arun, Barone, Barr, Barsuglia, Beker,
  {et~al.}}]{punturo2010third}
---. 2010{\natexlab{b}}, Classical and Quantum Gravity, 27, 084007

\bibitem[{{Qin} {et~al.}(2018){Qin}, {Fragos}, {Meynet}, {Andrews},
  {S{\o}rensen}, \& {Song}}]{qin2018}
{Qin}, Y., {Fragos}, T., {Meynet}, G., {et~al.} 2018, \aap, 616, A28,
  \dodoi{10.1051/0004-6361/201832839}

\bibitem[{{Qin} {et~al.}(2019){Qin}, {Marchant}, {Fragos}, {Meynet}, \&
  {Kalogera}}]{qin2019}
{Qin}, Y., {Marchant}, P., {Fragos}, T., {Meynet}, G., \& {Kalogera}, V. 2019,
  \apjl, 870, L18, \dodoi{10.3847/2041-8213/aaf97b}

\bibitem[{{Rastello} {et~al.}(2020){Rastello}, {Mapelli}, {Di Carlo},
  {Giacobbo}, {Santoliquido}, {Spera}, {Ballone}, \& {Iorio}}]{rastello2020}
{Rastello}, S., {Mapelli}, M., {Di Carlo}, U.~N., {et~al.} 2020, \mnras, 497,
  1563, \dodoi{10.1093/mnras/staa2018}

\bibitem[{{Rees} \& {Meszaros}(1992)}]{rees1992}
{Rees}, M.~J., \& {Meszaros}, P. 1992, \mnras, 258, 41,
  \dodoi{10.1093/mnras/258.1.41P}

\bibitem[{Reitze {et~al.}(2019)Reitze, Adhikari, Ballmer, Barish, Barsotti,
  Billingsley, Brown, Chen, Coyne, Eisenstein, {et~al.}}]{reitze2019cosmic}
Reitze, D., Adhikari, R.~X., Ballmer, S., {et~al.} 2019, arXiv preprint
  arXiv:1907.04833

\bibitem[{{Ren} {et~al.}(2019){Ren}, {Lin}, {Zhang}, {Li}, {Liu}, {Lu}, {Wang},
  \& {Liang}}]{ren2019}
{Ren}, J., {Lin}, D.-B., {Zhang}, L.-L., {et~al.} 2019, \apj, 885, 60,
  \dodoi{10.3847/1538-4357/ab4188}

\bibitem[{{Ren} {et~al.}(2020){Ren}, {Lin}, {Zhang}, {Wang}, {Li}, {Wang}, \&
  {Liang}}]{ren2020}
{Ren}, J., {Lin}, D.-B., {Zhang}, L.-l., {et~al.} 2020, arXiv e-prints,
  arXiv:2009.04735.
\newblock \doarXiv{2009.04735}

\bibitem[{{Rest} {et~al.}(2018){Rest}, {Garnavich}, {Khatami}, {Kasen},
  {Tucker}, {Shaya}, {Olling}, {Mushotzky}, {Zenteno}, {Margheim},
  {Strampelli}, {James}, {Smith}, {F{\"o}rster}, \& {Villar}}]{rest2018}
{Rest}, A., {Garnavich}, P.~M., {Khatami}, D., {et~al.} 2018, Nature Astronomy,
  2, 307, \dodoi{10.1038/s41550-018-0423-2}

\bibitem[{{Rhoads}(1999)}]{rhoads1999}
{Rhoads}, J.~E. 1999, \apj, 525, 737, \dodoi{10.1086/307907}

\bibitem[{{Roberts} {et~al.}(2011){Roberts}, {Kasen}, {Lee}, \&
  {Ramirez-Ruiz}}]{roberts2011}
{Roberts}, L.~F., {Kasen}, D., {Lee}, W.~H., \& {Ramirez-Ruiz}, E. 2011, \apjl,
  736, L21, \dodoi{10.1088/2041-8205/736/1/L21}

\bibitem[{{Rossi} {et~al.}(2020){Rossi}, {Stratta}, {Maiorano}, {Spighi},
  {Masetti}, {Palazzi}, {Gardini}, {Melandri}, {Nicastro}, {Pian}, {Branchesi},
  {Dadina}, {Testa}, {Brocato}, {Benetti}, {Ciolfi}, {Covino}, {D'Elia},
  {Grado}, {Izzo}, {Perego}, {Piranomonte}, {Salvaterra}, {Selsing},
  {Tomasella}, {Yang}, {Vergani}, {Amati}, \& {Stephen}}]{rossi2020}
{Rossi}, A., {Stratta}, G., {Maiorano}, E., {et~al.} 2020, \mnras, 493, 3379,
  \dodoi{10.1093/mnras/staa479}

\bibitem[{{Rossi} {et~al.}(2002){Rossi}, {Lazzati}, \& {Rees}}]{rossi2002}
{Rossi}, E., {Lazzati}, D., \& {Rees}, M.~J. 2002, \mnras, 332, 945,
  \dodoi{10.1046/j.1365-8711.2002.05363.x}

\bibitem[{{Rosswog}(2007)}]{rosswog2007}
{Rosswog}, S. 2007, \mnras, 376, L48, \dodoi{10.1111/j.1745-3933.2007.00284.x}

\bibitem[{{Rowlinson} {et~al.}(2014){Rowlinson}, {Gompertz}, {Dainotti},
  {O'Brien}, {Wijers}, \& {van der Horst}}]{rowlinson2014}
{Rowlinson}, A., {Gompertz}, B.~P., {Dainotti}, M., {et~al.} 2014, \mnras, 443,
  1779, \dodoi{10.1093/mnras/stu1277}

\bibitem[{Ruiz {et~al.}(2008)Ruiz, Alcubierre, N{\'u}{\~n}ez, \&
  Takahashi}]{ruiz2008multiple}
Ruiz, M., Alcubierre, M., N{\'u}{\~n}ez, D., \& Takahashi, R. 2008, General
  Relativity and Gravitation, 40, 1705

\bibitem[{Sachdev {et~al.}(2020)Sachdev, Magee, Hanna, Cannon, Singer,
  Mukherjee, Caudill, Chan, Creighton, Ewing, {et~al.}}]{sachdev2020early}
Sachdev, S., Magee, R., Hanna, C., {et~al.} 2020, arXiv preprint
  arXiv:2008.04288

\bibitem[{{Sagu{\'e}s Carracedo} {et~al.}(2020){Sagu{\'e}s Carracedo}, {Bulla},
  {Feindt}, \& {Goobar}}]{saguescarracedo2020}
{Sagu{\'e}s Carracedo}, A., {Bulla}, M., {Feindt}, U., \& {Goobar}, A. 2020,
  arXiv e-prints, arXiv:2004.06137.
\newblock \doarXiv{2004.06137}

\bibitem[{{Salafia} {et~al.}(2019){Salafia}, {Ghirlanda}, {Ascenzi}, \&
  {Ghisellini}}]{salafia2019}
{Salafia}, O.~S., {Ghirlanda}, G., {Ascenzi}, S., \& {Ghisellini}, G. 2019,
  \aap, 628, A18, \dodoi{10.1051/0004-6361/201935831}

\bibitem[{{Santana} {et~al.}(2014){Santana}, {Barniol Duran}, \&
  {Kumar}}]{santana2014}
{Santana}, R., {Barniol Duran}, R., \& {Kumar}, P. 2014, \apj, 785, 29,
  \dodoi{10.1088/0004-637X/785/1/29}

\bibitem[{{Santoliquido} {et~al.}(2020){Santoliquido}, {Mapelli}, {Bouffanais},
  {Giacobbo}, {Di Carlo}, {Rastello}, {Artale}, \&
  {Ballone}}]{santoliquido2020}
{Santoliquido}, F., {Mapelli}, M., {Bouffanais}, Y., {et~al.} 2020, \apj, 898,
  152, \dodoi{10.3847/1538-4357/ab9b78}

\bibitem[{{Sari} {et~al.}(1999){Sari}, {Piran}, \& {Halpern}}]{sari1999}
{Sari}, R., {Piran}, T., \& {Halpern}, J.~P. 1999, \apjl, 519, L17,
  \dodoi{10.1086/312109}

\bibitem[{{Sari} {et~al.}(1998){Sari}, {Piran}, \& {Narayan}}]{sari1998}
{Sari}, R., {Piran}, T., \& {Narayan}, R. 1998, \apjl, 497, L17,
  \dodoi{10.1086/311269}

\bibitem[{{Savchenko} {et~al.}(2017){Savchenko}, {Ferrigno}, {Kuulkers},
  {Bazzano}, {Bozzo}, {Brandt}, {Chenevez}, {Courvoisier}, {Diehl}, {Domingo},
  {Hanlon}, {Jourdain}, {von Kienlin}, {Laurent}, {Lebrun}, {Lutovinov},
  {Martin-Carrillo}, {Mereghetti}, {Natalucci}, {Rodi}, {Roques}, {Sunyaev}, \&
  {Ubertini}}]{savchenko2017}
{Savchenko}, V., {Ferrigno}, C., {Kuulkers}, E., {et~al.} 2017, \apjl, 848,
  L15, \dodoi{10.3847/2041-8213/aa8f94}

\bibitem[{{Scolnic} {et~al.}(2018){Scolnic}, {Kessler}, {Brout},
  {Cowperthwaite}, {Soares-Santos}, {Annis}, {Herner}, {Chen}, {Sako},
  {Doctor}, {Butler}, {Palmese}, {Diehl}, {Frieman}, {Holz}, {Berger},
  {Chornock}, {Villar}, {Nicholl}, {Biswas}, {Hounsell}, {Foley}, {Metzger},
  {Rest}, {Garc{\'\i}a-Bellido}, {M{\"o}ller}, {Nugent}, {Abbott}, {Abdalla},
  {Allam}, {Bechtol}, {Benoit-L{\'e}vy}, {Bertin}, {Brooks}, {Buckley-Geer},
  {Carnero Rosell}, {Carrasco Kind}, {Carretero}, {Castander}, {Cunha},
  {D'Andrea}, {da Costa}, {Davis}, {Doel}, {Drlica-Wagner}, {Eifler},
  {Flaugher}, {Fosalba}, {Gaztanaga}, {Gerdes}, {Gruen}, {Gruendl}, {Gschwend},
  {Gutierrez}, {Hartley}, {Honscheid}, {James}, {Johnson}, {Johnson}, {Krause},
  {Kuehn}, {Kuhlmann}, {Lahav}, {Li}, {Lima}, {Maia}, {March}, {Marshall},
  {Menanteau}, {Miquel}, {Neilsen}, {Plazas}, {Sanchez}, {Scarpine},
  {Schubnell}, {Sevilla-Noarbe}, {Smith}, {Smith}, {Sobreira}, {Suchyta},
  {Swanson}, {Tarle}, {Thomas}, {Tucker}, {Walker}, \& {DES
  Collaboration}}]{scolnic2018}
{Scolnic}, D., {Kessler}, R., {Brout}, D., {et~al.} 2018, \apjl, 852, L3,
  \dodoi{10.3847/2041-8213/aa9d82}

\bibitem[{{Setzer} {et~al.}(2019){Setzer}, {Biswas}, {Peiris}, {Rosswog},
  {Korobkin}, {Wollaeger}, \& {LSST Dark Energy Science
  Collaboration}}]{setzer2019}
{Setzer}, C.~N., {Biswas}, R., {Peiris}, H.~V., {et~al.} 2019, \mnras, 485,
  4260, \dodoi{10.1093/mnras/stz506}

\bibitem[{{Shappee} {et~al.}(2017){Shappee}, {Simon}, {Drout}, {Piro},
  {Morrell}, {Prieto}, {Kasen}, {Holoien}, {Kollmeier}, {Kelson}, {Coulter},
  {Foley}, {Kilpatrick}, {Siebert}, {Madore}, {Murguia-Berthier}, {Pan},
  {Prochaska}, {Ramirez-Ruiz}, {Rest}, {Adams}, {Alatalo}, {Ba{\~n}ados},
  {Baughman}, {Bernstein}, {Bitsakis}, {Boutsia}, {Bravo}, {Di Mille}, {Higgs},
  {Ji}, {Maravelias}, {Marshall}, {Placco}, {Prieto}, \& {Wan}}]{shappee2017}
{Shappee}, B.~J., {Simon}, J.~D., {Drout}, M.~R., {et~al.} 2017, Science, 358,
  1574, \dodoi{10.1126/science.aaq0186}

\bibitem[{{Shibata} \& {Taniguchi}(2011)}]{shibata2011}
{Shibata}, M., \& {Taniguchi}, K. 2011, Living Reviews in Relativity, 14, 6,
  \dodoi{10.12942/lrr-2011-6}

\bibitem[{{Siegel} \& {Metzger}(2017)}]{siegel2017}
{Siegel}, D.~M., \& {Metzger}, B.~D. 2017, \prl, 119, 231102,
  \dodoi{10.1103/PhysRevLett.119.231102}

\bibitem[{{Singh} {et~al.}(2020){Singh}, {Kapadia}, {Arif Shaikh},
  {Chatterjee}, \& {Ajith}}]{2020arXiv201012407S}
{Singh}, M.~K., {Kapadia}, S.~J., {Arif Shaikh}, M., {Chatterjee}, D., \&
  {Ajith}, P. 2020, arXiv e-prints, arXiv:2010.12407.
\newblock \doarXiv{2010.12407}

\bibitem[{{Smartt} {et~al.}(2017){Smartt}, {Chen}, {Jerkstrand}, {Coughlin},
  {Kankare}, {Sim}, {Fraser}, {Inserra}, {Maguire}, {Chambers}, {Huber},
  {Kr{\"u}hler}, {Leloudas}, {Magee}, {Shingles}, {Smith}, {Young}, {Tonry},
  {Kotak}, {Gal-Yam}, {Lyman}, {Homan}, {Agliozzo}, {Anderson}, {Angus},
  {Ashall}, {Barbarino}, {Bauer}, {Berton}, {Botticella}, {Bulla}, {Bulger},
  {Cannizzaro}, {Cano}, {Cartier}, {Cikota}, {Clark}, {De Cia}, {Della Valle},
  {Denneau}, {Dennefeld}, {Dessart}, {Dimitriadis}, {Elias-Rosa}, {Firth},
  {Flewelling}, {Fl{\"o}rs}, {Franckowiak}, {Frohmaier}, {Galbany},
  {Gonz{\'a}lez-Gait{\'a}n}, {Greiner}, {Gromadzki}, {Guelbenzu},
  {Guti{\'e}rrez}, {Hamanowicz}, {Hanlon}, {Harmanen}, {Heintz}, {Heinze},
  {Hernandez}, {Hodgkin}, {Hook}, {Izzo}, {James}, {Jonker}, {Kerzendorf},
  {Klose}, {Kostrzewa-Rutkowska}, {Kowalski}, {Kromer}, {Kuncarayakti},
  {Lawrence}, {Lowe}, {Magnier}, {Manulis}, {Martin-Carrillo}, {Mattila},
  {McBrien}, {M{\"u}ller}, {Nordin}, {O'Neill}, {Onori}, {Palmerio},
  {Pastorello}, {Patat}, {Pignata}, {Podsiadlowski}, {Pumo}, {Prentice}, {Rau},
  {Razza}, {Rest}, {Reynolds}, {Roy}, {Ruiter}, {Rybicki}, {Salmon}, {Schady},
  {Schultz}, {Schweyer}, {Seitenzahl}, {Smith}, {Sollerman}, {Stalder},
  {Stubbs}, {Sullivan}, {Szegedi}, {Taddia}, {Taubenberger}, {Terreran}, {van
  Soelen}, {Vos}, {Wainscoat}, {Walton}, {Waters}, {Weiland}, {Willman},
  {Wiseman}, {Wright}, {Wyrzykowski}, \& {Yaron}}]{smartt2017N}
{Smartt}, S.~J., {Chen}, T.~W., {Jerkstrand}, A., {et~al.} 2017, \nat, 551, 75,
  \dodoi{10.1038/nature24303}

\bibitem[{{Soares-Santos} {et~al.}(2017){Soares-Santos}, {Holz}, {Annis},
  {Chornock}, {Herner}, {Berger}, {Brout}, {Chen}, {Kessler}, {Sako}, {Allam},
  {Tucker}, {Butler}, {Palmese}, {Doctor}, {Diehl}, {Frieman}, {Yanny}, {Lin},
  {Scolnic}, {Cowperthwaite}, {Neilsen}, {Marriner}, {Kuropatkin}, {Hartley},
  {Paz-Chinch{\'o}n}, {Alexander}, {Balbinot}, {Blanchard}, {Brown}, {Carlin},
  {Conselice}, {Cook}, {Drlica-Wagner}, {Drout}, {Durret}, {Eftekhari}, {Farr},
  {Finley}, {Foley}, {Fong}, {Fryer}, {Garc{\'\i}a-Bellido}, {Gill}, {Gruendl},
  {Hanna}, {Kasen}, {Li}, {Lopes}, {Louren{\c{c}}o}, {Margutti}, {Marshall},
  {Matheson}, {Medina}, {Metzger}, {Mu{\~n}oz}, {Muir}, {Nicholl}, {Quataert},
  {Rest}, {Sauseda}, {Schlegel}, {Secco}, {Sobreira}, {Stebbins}, {Villar},
  {Vivas}, {Walker}, {Wester}, {Williams}, {Zenteno}, {Zhang}, {Abbott},
  {Abdalla}, {Banerji}, {Bechtol}, {Benoit-L{\'e}vy}, {Bertin}, {Brooks},
  {Buckley-Geer}, {Burke}, {Carnero Rosell}, {Carrasco Kind}, {Carretero},
  {Castander}, {Crocce}, {Cunha}, {D'Andrea}, {da Costa}, {Davis}, {Desai},
  {Dietrich}, {Doel}, {Eifler}, {Fernand ez}, {Flaugher}, {Fosalba},
  {Gaztanaga}, {Gerdes}, {Giannantonio}, {Goldstein}, {Gruen}, {Gschwend},
  {Gutierrez}, {Honscheid}, {Jain}, {James}, {Jeltema}, {Johnson}, {Johnson},
  {Kent}, {Krause}, {Kron}, {Kuehn}, {Kuhlmann}, {Lahav}, {Lima}, {Maia},
  {March}, {McMahon}, {Menanteau}, {Miquel}, {Mohr}, {Nichol}, {Nord}, {Ogand
  o}, {Petravick}, {Plazas}, {Romer}, {Roodman}, {Rykoff}, {Sanchez},
  {Scarpine}, {Schubnell}, {Sevilla-Noarbe}, {Smith}, {Smith}, {Suchyta},
  {Swanson}, {Tarle}, {Thomas}, {Thomas}, {Troxel}, {Vikram}, {Wechsler},
  {Weller}, {Dark Energy Survey}, \& {Dark Energy Camera GW-EM
  Collaboration}}]{soaressantos2017}
{Soares-Santos}, M., {Holz}, D.~E., {Annis}, J., {et~al.} 2017, \apjl, 848,
  L16, \dodoi{10.3847/2041-8213/aa9059}

\bibitem[{{Song} {et~al.}(2019){Song}, {Ai}, {Wang}, {Xing}, {Gao}, \&
  {Zhang}}]{song2019}
{Song}, H.-R., {Ai}, S.-K., {Wang}, M.-H., {et~al.} 2019, \apjl, 881, L40,
  \dodoi{10.3847/2041-8213/ab3921}

\bibitem[{{Spergel} {et~al.}(2015){Spergel}, {Gehrels}, {Baltay}, {Bennett},
  {Breckinridge}, {Donahue}, {Dressler}, {Gaudi}, {Greene}, {Guyon}, {Hirata},
  {Kalirai}, {Kasdin}, {Macintosh}, {Moos}, {Perlmutter}, {Postman},
  {Rauscher}, {Rhodes}, {Wang}, {Weinberg}, {Benford}, {Hudson}, {Jeong},
  {Mellier}, {Traub}, {Yamada}, {Capak}, {Colbert}, {Masters}, {Penny},
  {Savransky}, {Stern}, {Zimmerman}, {Barry}, {Bartusek}, {Carpenter}, {Cheng},
  {Content}, {Dekens}, {Demers}, {Grady}, {Jackson}, {Kuan}, {Kruk}, {Melton},
  {Nemati}, {Parvin}, {Poberezhskiy}, {Peddie}, {Ruffa}, {Wallace}, {Whipple},
  {Wollack}, \& {Zhao}}]{spergel2015}
{Spergel}, D., {Gehrels}, N., {Baltay}, C., {et~al.} 2015, arXiv e-prints,
  arXiv:1503.03757.
\newblock \doarXiv{1503.03757}

\bibitem[{{Steiner} {et~al.}(2012){Steiner}, {McClintock}, \&
  {Reid}}]{steiner2012}
{Steiner}, J.~F., {McClintock}, J.~E., \& {Reid}, M.~J. 2012, \apjl, 745, L7,
  \dodoi{10.1088/2041-8205/745/1/L7}

\bibitem[{{Sun} {et~al.}(2019){Sun}, {Li}, {Zhang}, {Zhang}, {Bauer}, {Xue}, \&
  {Yuan}}]{sun2019}
{Sun}, H., {Li}, Y., {Zhang}, B.-B., {et~al.} 2019, \apj, 886, 129,
  \dodoi{10.3847/1538-4357/ab4bc7}

\bibitem[{{Sun} {et~al.}(2017){Sun}, {Zhang}, \& {Gao}}]{sun2017}
{Sun}, H., {Zhang}, B., \& {Gao}, H. 2017, \apj, 835, 7,
  \dodoi{10.3847/1538-4357/835/1/7}

\bibitem[{{Sun} {et~al.}(2015){Sun}, {Zhang}, \& {Li}}]{sun2015}
{Sun}, H., {Zhang}, B., \& {Li}, Z. 2015, \apj, 812, 33,
  \dodoi{10.1088/0004-637X/812/1/33}

\bibitem[{{Symbalisty} \& {Schramm}(1982)}]{symbalisty1982}
{Symbalisty}, E., \& {Schramm}, D.~N. 1982, \aplett, 22, 143

\bibitem[{{Tan} \& {Yu}(2020)}]{tan2020}
{Tan}, W.-W., \& {Yu}, Y.-W. 2020, arXiv e-prints, arXiv:2006.02060.
\newblock \doarXiv{2006.02060}

\bibitem[{{Tanaka} \& {Hotokezaka}(2013)}]{tanaka2013}
{Tanaka}, M., \& {Hotokezaka}, K. 2013, \apj, 775, 113,
  \dodoi{10.1088/0004-637X/775/2/113}

\bibitem[{{Tanaka} {et~al.}(2020){Tanaka}, {Kato}, {Gaigalas}, \&
  {Kawaguchi}}]{tanaka2020}
{Tanaka}, M., {Kato}, D., {Gaigalas}, G., \& {Kawaguchi}, K. 2020, \mnras, 496,
  1369, \dodoi{10.1093/mnras/staa1576}

\bibitem[{{Tanaka} {et~al.}(2017){Tanaka}, {Utsumi}, {Mazzali}, {Tominaga},
  {Yoshida}, {Sekiguchi}, {Morokuma}, {Motohara}, {Ohta}, {Kawabata}, {Abe},
  {Aoki}, {Asakura}, {Baar}, {Barway}, {Bond}, {Doi}, {Fujiyoshi}, {Furusawa},
  {Honda}, {Itoh}, {Kawabata}, {Kawai}, {Kim}, {Lee}, {Miyazaki}, {Morihana},
  {Nagashima}, {Nagayama}, {Nakaoka}, {Nakata}, {Ohsawa}, {Ohshima}, {Okita},
  {Saito}, {Sumi}, {Tajitsu}, {Takahashi}, {Takayama}, {Tamura}, {Tanaka},
  {Terai}, {Tristram}, {Yasuda}, \& {Zenko}}]{tanaka2017}
{Tanaka}, M., {Utsumi}, Y., {Mazzali}, P.~A., {et~al.} 2017, \pasj, 69, 102,
  \dodoi{10.1093/pasj/psx121}

\bibitem[{{Tanaka} {et~al.}(2018){Tanaka}, {Kato}, {Gaigalas}, {Rynkun},
  {Rad{\v{z}}i{\={u}}t{\.{e}}}, {Wanajo}, {Sekiguchi}, {Nakamura}, {Tanuma},
  {Murakami}, \& {Sakaue}}]{tanaka2018}
{Tanaka}, M., {Kato}, D., {Gaigalas}, G., {et~al.} 2018, \apj, 852, 109,
  \dodoi{10.3847/1538-4357/aaa0cb}

\bibitem[{{Tanvir} {et~al.}(2013){Tanvir}, {Levan}, {Fruchter}, {Hjorth},
  {Hounsell}, {Wiersema}, \& {Tunnicliffe}}]{tanvir2013}
{Tanvir}, N.~R., {Levan}, A.~J., {Fruchter}, A.~S., {et~al.} 2013, \nat, 500,
  547, \dodoi{10.1038/nature12505}

\bibitem[{{Tanvir} {et~al.}(2017){Tanvir}, {Levan},
  {Gonz{\'a}lez-Fern{\'a}ndez}, {Korobkin}, {Mandel}, {Rosswog}, {Hjorth},
  {D'Avanzo}, {Fruchter}, {Fryer}, {Kangas}, {Milvang-Jensen}, {Rosetti},
  {Steeghs}, {Wollaeger}, {Cano}, {Copperwheat}, {Covino}, {D'Elia}, {de Ugarte
  Postigo}, {Evans}, {Even}, {Fairhurst}, {Figuera Jaimes}, {Fontes}, {Fujii},
  {Fynbo}, {Gompertz}, {Greiner}, {Hodosan}, {Irwin}, {Jakobsson},
  {J{\o}rgensen}, {Kann}, {Lyman}, {Malesani}, {McMahon}, {Melandri},
  {O'Brien}, {Osborne}, {Palazzi}, {Perley}, {Pian}, {Piranomonte}, {Rabus},
  {Rol}, {Rowlinson}, {Schulze}, {Sutton}, {Th{\"o}ne}, {Ulaczyk}, {Watson},
  {Wiersema}, \& {Wijers}}]{tanvir2017}
{Tanvir}, N.~R., {Levan}, A.~J., {Gonz{\'a}lez-Fern{\'a}ndez}, C., {et~al.}
  2017, \apjl, 848, L27, \dodoi{10.3847/2041-8213/aa90b6}

\bibitem[{{Tchekhovskoy} {et~al.}(2010){Tchekhovskoy}, {Narayan}, \&
  {McKinney}}]{tchekhovskoy2010}
{Tchekhovskoy}, A., {Narayan}, R., \& {McKinney}, J.~C. 2010, \apj, 711, 50,
  \dodoi{10.1088/0004-637X/711/1/50}

\bibitem[{{Troja} {et~al.}(2017){Troja}, {Piro}, {van Eerten}, {Wollaeger},
  {Im}, {Fox}, {Butler}, {Cenko}, {Sakamoto}, {Fryer}, {Ricci}, {Lien}, {Ryan},
  {Korobkin}, {Lee}, {Burgess}, {Lee}, {Watson}, {Choi}, {Covino}, {D'Avanzo},
  {Fontes}, {Gonz{\'a}lez}, {Khandrika}, {Kim}, {Kim}, {Lee}, {Lee}, {Kutyrev},
  {Lim}, {S{\'a}nchez-Ram{\'\i}rez}, {Veilleux}, {Wieringa}, \&
  {Yoon}}]{troja2017}
{Troja}, E., {Piro}, L., {van Eerten}, H., {et~al.} 2017, \nat, 551, 71,
  \dodoi{10.1038/nature24290}

\bibitem[{{Troja} {et~al.}(2018){Troja}, {Piro}, {Ryan}, {van Eerten}, {Ricci},
  {Wieringa}, {Lotti}, {Sakamoto}, \& {Cenko}}]{troja2018}
{Troja}, E., {Piro}, L., {Ryan}, G., {et~al.} 2018, \mnras, 478, L18,
  \dodoi{10.1093/mnrasl/sly061}

\bibitem[{{Troja} {et~al.}(2020){Troja}, {van Eerten}, {Zhang}, {Ryan}, {Piro},
  {Ricci}, {O'Connor}, {Wieringa}, {Cenko}, \& {Sakamoto}}]{troja2020}
{Troja}, E., {van Eerten}, H., {Zhang}, B., {et~al.} 2020, arXiv e-prints,
  arXiv:2006.01150.
\newblock \doarXiv{2006.01150}

\bibitem[{{Typel} {et~al.}(2010){Typel}, {R{\"o}pke}, {Kl{\"a}hn}, {Blaschke},
  \& {Wolter}}]{typel2010}
{Typel}, S., {R{\"o}pke}, G., {Kl{\"a}hn}, T., {Blaschke}, D., \& {Wolter},
  H.~H. 2010, \prc, 81, 015803, \dodoi{10.1103/PhysRevC.81.015803}

\bibitem[{Unnikrishnan(2013)}]{unnikrishnan2013indigo}
Unnikrishnan, C. 2013, International Journal of Modern Physics D, 22, 1341010

\bibitem[{{Utsumi} {et~al.}(2017){Utsumi}, {Tanaka}, {Tominaga}, {Yoshida},
  {Barway}, {Nagayama}, {Zenko}, {Aoki}, {Fujiyoshi}, {Furusawa}, {Kawabata},
  {Koshida}, {Lee}, {Morokuma}, {Motohara}, {Nakata}, {Ohsawa}, {Ohta},
  {Okita}, {Tajitsu}, {Tanaka}, {Terai}, {Yasuda}, {Abe}, {Asakura}, {Bond},
  {Miyazaki}, {Sumi}, {Tristram}, {Honda}, {Itoh}, {Itoh}, {Kawabata},
  {Morihana}, {Nagashima}, {Nakaoka}, {Ohshima}, {Takahashi}, {Takayama},
  {Aoki}, {Baar}, {Doi}, {Finet}, {Kanda}, {Kawai}, {Kim}, {Kuroda}, {Liu},
  {Matsubayashi}, {Murata}, {Nagai}, {Saito}, {Saito}, {Sako}, {Sekiguchi},
  {Tamura}, {Tanaka}, {Uemura}, \& {Yamaguchi}}]{utsumi2017}
{Utsumi}, Y., {Tanaka}, M., {Tominaga}, N., {et~al.} 2017, \pasj, 69, 101,
  \dodoi{10.1093/pasj/psx118}

\bibitem[{{Valenti} {et~al.}(2017){Valenti}, {Sand}, {Yang}, {Cappellaro},
  {Tartaglia}, {Corsi}, {Jha}, {Reichart}, {Haislip}, \&
  {Kouprianov}}]{valenti2017}
{Valenti}, S., {Sand}, D.~J., {Yang}, S., {et~al.} 2017, \apjl, 848, L24,
  \dodoi{10.3847/2041-8213/aa8edf}

\bibitem[{{van Eerten} {et~al.}(2010){van Eerten}, {Zhang}, \&
  {MacFadyen}}]{vanerten2010}
{van Eerten}, H., {Zhang}, W., \& {MacFadyen}, A. 2010, \apj, 722, 235,
  \dodoi{10.1088/0004-637X/722/1/235}

\bibitem[{{Villar} {et~al.}(2017){Villar}, {Guillochon}, {Berger}, {Metzger},
  {Cowperthwaite}, {Nicholl}, {Alexand er}, {Blanchard}, {Chornock},
  {Eftekhari}, {Fong}, {Margutti}, \& {Williams}}]{villar2017}
{Villar}, V.~A., {Guillochon}, J., {Berger}, E., {et~al.} 2017, \apjl, 851,
  L21, \dodoi{10.3847/2041-8213/aa9c84}

\bibitem[{{Virgili} {et~al.}(2011){Virgili}, {Zhang}, {O'Brien}, \&
  {Troja}}]{virgili2011}
{Virgili}, F.~J., {Zhang}, B., {O'Brien}, P., \& {Troja}, E. 2011, \apj, 727,
  109, \dodoi{10.1088/0004-637X/727/2/109}

\bibitem[{Vitale \& Evans(2017)}]{vitale2017parameter}
Vitale, S., \& Evans, M. 2017, \prd, 95, 064052

\bibitem[{Vitale \& Whittle(2018)}]{vitale2018characterization}
Vitale, S., \& Whittle, C. 2018, \prd, 98, 024029

\bibitem[{{Wanajo}(2018)}]{wanajo2018}
{Wanajo}, S. 2018, \apj, 868, 65, \dodoi{10.3847/1538-4357/aae0f2}

\bibitem[{{Wanajo} {et~al.}(2014){Wanajo}, {Sekiguchi}, {Nishimura}, {Kiuchi},
  {Kyutoku}, \& {Shibata}}]{wanajo2014}
{Wanajo}, S., {Sekiguchi}, Y., {Nishimura}, N., {et~al.} 2014, \apjl, 789, L39,
  \dodoi{10.1088/2041-8205/789/2/L39}

\bibitem[{{Wanderman} \& {Piran}(2015)}]{wanderman2015}
{Wanderman}, D., \& {Piran}, T. 2015, \mnras, 448, 3026,
  \dodoi{10.1093/mnras/stv123}

\bibitem[{{Wang} {et~al.}(2015){Wang}, {Zhang}, {Liang}, {Gao}, {Li}, {Deng},
  {Qin}, {Tang}, {Kann}, {Ryde}, \& {Kumar}}]{wang2015}
{Wang}, X.-G., {Zhang}, B., {Liang}, E.-W., {et~al.} 2015, \apjs, 219, 9,
  \dodoi{10.1088/0067-0049/219/1/9}

\bibitem[{{Wollaeger} {et~al.}(2018){Wollaeger}, {Korobkin}, {Fontes},
  {Rosswog}, {Even}, {Fryer}, {Sollerman}, {Hungerford}, {van Rossum}, \&
  {Wollaber}}]{wollaeger2018}
{Wollaeger}, R.~T., {Korobkin}, O., {Fontes}, C.~J., {et~al.} 2018, \mnras,
  478, 3298, \dodoi{10.1093/mnras/sty1018}

\bibitem[{{Wu} {et~al.}(2019){Wu}, {Barnes}, {Mart{\'\i}nez-Pinedo}, \&
  {Metzger}}]{wu2019}
{Wu}, M.-R., {Barnes}, J., {Mart{\'\i}nez-Pinedo}, G., \& {Metzger}, B.~D.
  2019, \prl, 122, 062701, \dodoi{10.1103/PhysRevLett.122.062701}

\bibitem[{{Wu} {et~al.}(2016){Wu}, {Fern{\'a}ndez}, {Mart{\'\i}nez-Pinedo}, \&
  {Metzger}}]{wu2016}
{Wu}, M.-R., {Fern{\'a}ndez}, R., {Mart{\'\i}nez-Pinedo}, G., \& {Metzger},
  B.~D. 2016, \mnras, 463, 2323, \dodoi{10.1093/mnras/stw2156}

\bibitem[{Wu {et~al.}(2020)Wu, Cao, \& Zhu}]{wu2020measuring}
Wu, S., Cao, Z., \& Zhu, Z.-H. 2020, \mnras, 495, 466

\bibitem[{{Wyatt} {et~al.}(2020){Wyatt}, {Tohuvavohu}, {Arcavi}, {Lundquist},
  {Howell}, \& {Sand}}]{wyatt2020}
{Wyatt}, S.~D., {Tohuvavohu}, A., {Arcavi}, I., {et~al.} 2020, \apj, 894, 127,
  \dodoi{10.3847/1538-4357/ab855e}

\bibitem[{{Xiao} {et~al.}(2019){Xiao}, {Zhang}, \& {Dai}}]{xiao2019}
{Xiao}, D., {Zhang}, B.-B., \& {Dai}, Z.-G. 2019, \apjl, 879, L7,
  \dodoi{10.3847/2041-8213/ab2980}

\bibitem[{{Xie} {et~al.}(2018){Xie}, {Zrake}, \& {MacFadyen}}]{xie2018}
{Xie}, X., {Zrake}, J., \& {MacFadyen}, A. 2018, \apj, 863, 58,
  \dodoi{10.3847/1538-4357/aacf9c}

\bibitem[{{Xue} {et~al.}(2019){Xue}, {Zheng}, {Li}, {Brandt}, {Zhang}, {Luo},
  {Zhang}, {Bauer}, {Sun}, {Lehmer}, {Wu}, {Yang}, {Kong}, {Li}, {Sun}, {Wang},
  \& {Vito}}]{xue2019}
{Xue}, Y.~Q., {Zheng}, X.~C., {Li}, Y., {et~al.} 2019, \nat, 568, 198,
  \dodoi{10.1038/s41586-019-1079-5}

\bibitem[{{Yang} {et~al.}(2015){Yang}, {Jin}, {Li}, {Covino}, {Zheng},
  {Hotokezaka}, {Fan}, {Piran}, \& {Wei}}]{yang2015}
{Yang}, B., {Jin}, Z.-P., {Li}, X., {et~al.} 2015, Nature Communications, 6,
  7323, \dodoi{10.1038/ncomms8323}

\bibitem[{{Yang} {et~al.}(2020){Yang}, {Zhong}, {Zhang}, {Wu}, {Zhang}, {Yang},
  {Cao}, {Gao}, {Zou}, {Wang}, {L{\"u}}, {Cang}, \& {Dai}}]{Yang2020}
{Yang}, Y.-S., {Zhong}, S.-Q., {Zhang}, B.-B., {et~al.} 2020, \apj, 899, 60,
  \dodoi{10.3847/1538-4357/ab9ff5}

\bibitem[{{Yu} {et~al.}(2021){Yu}, {Song}, {Ai}, {Gao}, {Wang}, {Wang}, {Lu},
  {Fang}, \& {Zhao}}]{yu2021}
{Yu}, J., {Song}, H., {Ai}, S., {et~al.} 2021, arXiv e-prints,
  arXiv:2104.12374.
\newblock \doarXiv{2104.12374}

\bibitem[{{Yu} {et~al.}(2018){Yu}, {Liu}, \& {Dai}}]{yu2018}
{Yu}, Y.-W., {Liu}, L.-D., \& {Dai}, Z.-G. 2018, \apj, 861, 114,
  \dodoi{10.3847/1538-4357/aac6e5}

\bibitem[{{Yu} {et~al.}(2013){Yu}, {Zhang}, \& {Gao}}]{yu2013}
{Yu}, Y.-W., {Zhang}, B., \& {Gao}, H. 2013, \apjl, 776, L40,
  \dodoi{10.1088/2041-8205/776/2/L40}

\bibitem[{{Y{\"u}ksel} {et~al.}(2008){Y{\"u}ksel}, {Kistler}, {Beacom}, \&
  {Hopkins}}]{yuksel2008}
{Y{\"u}ksel}, H., {Kistler}, M.~D., {Beacom}, J.~F., \& {Hopkins}, A.~M. 2008,
  \apjl, 683, L5, \dodoi{10.1086/591449}

\bibitem[{{Zappa} {et~al.}(2019){Zappa}, {Bernuzzi}, {Pannarale}, {Mapelli}, \&
  {Giacobbo}}]{zappa2019}
{Zappa}, F., {Bernuzzi}, S., {Pannarale}, F., {Mapelli}, M., \& {Giacobbo}, N.
  2019, \prl, 123, 041102, \dodoi{10.1103/PhysRevLett.123.041102}

\bibitem[{{Zhan}(2011)}]{zhan2011}
{Zhan}, H. 2011, SciSn, 41, 1441

\bibitem[{{Zhang}(2013)}]{zhang2013}
{Zhang}, B. 2013, \apjl, 763, L22, \dodoi{10.1088/2041-8205/763/1/L22}

\bibitem[{{Zhang}(2018)}]{zhang2018}
---. 2018, {The Physics of Gamma-Ray Bursts}, \dodoi{10.1017/9781139226530}

\bibitem[{{Zhang}(2019)}]{zhang2019}
---. 2019, \apjl, 873, L9, \dodoi{10.3847/2041-8213/ab0ae8}

\bibitem[{{Zhang} \& {M{\'e}sz{\'a}ros}(2002)}]{zhang2002}
{Zhang}, B., \& {M{\'e}sz{\'a}ros}, P. 2002, \apj, 571, 876,
  \dodoi{10.1086/339981}

\bibitem[{{Zhang} {et~al.}(2018){Zhang}, {Zhang}, {Sun}, {Lei}, {Gao}, {Li},
  {Shao}, {Zhao}, {Hu}, {L{\"u}}, {Wu}, {Fan}, {Wang}, {Castro-Tirado},
  {Zhang}, {Yu}, {Cao}, \& {Liang}}]{zhangbb2018}
{Zhang}, B.~B., {Zhang}, B., {Sun}, H., {et~al.} 2018, Nature Communications,
  9, 447, \dodoi{10.1038/s41467-018-02847-3}

\bibitem[{{Zhao} \& {Wen}(2018)}]{zhao2018}
{Zhao}, W., \& {Wen}, L. 2018, \prd, 97, 064031,
  \dodoi{10.1103/PhysRevD.97.064031}

\bibitem[{{Zhu} {et~al.}(2021{\natexlab{a}}){Zhu}, {Wang}, {Zhang}, {Yang},
  {Yu}, \& {Gao}}]{zhu2021b}
{Zhu}, J.-P., {Wang}, K., {Zhang}, B., {et~al.} 2021{\natexlab{a}}, \apjl, 911,
  L19, \dodoi{10.3847/2041-8213/abf2c3}

\bibitem[{{Zhu} {et~al.}(2020){Zhu}, {Yang}, {Liu}, {Huang}, {Zhang}, {Li},
  {Yu}, \& {Gao}}]{zhu2020}
{Zhu}, J.-P., {Yang}, Y.-P., {Liu}, L.-D., {et~al.} 2020, \apj, 897, 20,
  \dodoi{10.3847/1538-4357/ab93bf}

\bibitem[{{Zhu} {et~al.}(2021{\natexlab{b}}){Zhu}, {Zhang}, {Yu}, \&
  {Gao}}]{zhu2021a}
{Zhu}, J.-P., {Zhang}, B., {Yu}, Y.-W., \& {Gao}, H. 2021{\natexlab{b}}, \apjl,
  906, L11, \dodoi{10.3847/2041-8213/abd412}

\end{thebibliography}

\appendix

\section{Empirical formulae of Redshift Distribution\label{Sec:EmpiricalFormula}}

The dimensionless redshift distribution factor $f(z)$ describes the redshift-dependent merger event rate density, i.e.,
\begin{equation}
    \dot{\rho}(z) = \dot{\rho}_0f(z).
\end{equation}
The merger event rate density of BH--NS binary system $\dot{\rho}(z)$ at redshift $z$ can be connected to the cosmological star formation rate density by accounting for the merger delay time from the formation due of the BH--NS binary system and the orbital decay time scale through gravitational wave radiation. Therefore, the merger event rate density $\dot{\rho}(z)$ can be expressed as a convolution of the cosmological star formation rate density $\dot{\rho}_*(z)$ and the probability density function of delay time $P(\tau)$, i.e.,
\begin{equation}
\begin{split}
\label{Eq. Convolution}
    \dot{\rho}(z) \propto \int_{\tau_{\min}}^{\infty}\dot{\rho}_*[z'(\tau)]P(\tau)d\tau\propto \int_{z}^{\infty}\dot{\rho}_*(z')P[\tau(z')]\frac{dt(z')}{dz'}dz',
\end{split}
\end{equation}
where $\tau = t(z) - t(z')$ is the delay time, $\tau_{\min}$ is the minimum delay time, $t(z')$ is the time when the binaries are formed, and $t(z)$ is the delay time for binaries to merge, which is related to the source redshift as
\begin{equation}
    t(z) = \frac{1}{H_0} \int_z^\infty\frac{dz}{(1 + z)\sqrt{\Omega_\Lambda + \Omega_{\rm m}(1 + z)^ 3}}.
\end{equation}

\cite{sun2015} collected three major types of merger delay time models, including the Gaussian delay model \citep{virgili2011}, log-normal delay model \citep{wanderman2015}, and power-law delay model \citep{virgili2011}. They simulated the binary systems following the distribution of star formation rate density $\dot{\rho}_*(z)$ in the redshift space, and randomly generated the merger delay timescales of all the systems based on different delay models. Then, they derived the lookback time of each system and converted the lookback time to redshift. By these steps, \cite{sun2015} suggested some empirical formulae of the dimensionless redshift distribution $f(z)$ for the three merger delay models. However, due to the different independent variables of $\dot{\rho}_*(z)$ and $P(\tau)$, one needs to use unified variables to calculate event rate density of BH--NS mergers. Here, we use the method of convolution with redshift $z$ by Equation (\ref{Eq. Convolution}) to obtain the redshift distributions, which are normalized to unity at the local universe ($z = 0$) for the three merger delay models collected by \cite{sun2015}. We adopt the analytical model of \cite{yuksel2008} as the star formation rate density which reads
\begin{equation}
    \dot{\rho}_{*}(z) \propto \left[ (1 + z)^{3.4\eta} + \left( \frac{1 + z}{5000} \right)^{-0.3\eta} + \left( \frac{1 + z}{9} \right) ^{-3.5\eta} \right]^{1 / \eta},
\end{equation}
where $\eta = -10$.

\begin{figure}
    \centering
    \includegraphics[width = 0.5\linewidth , trim = 70 30 100 60, clip]{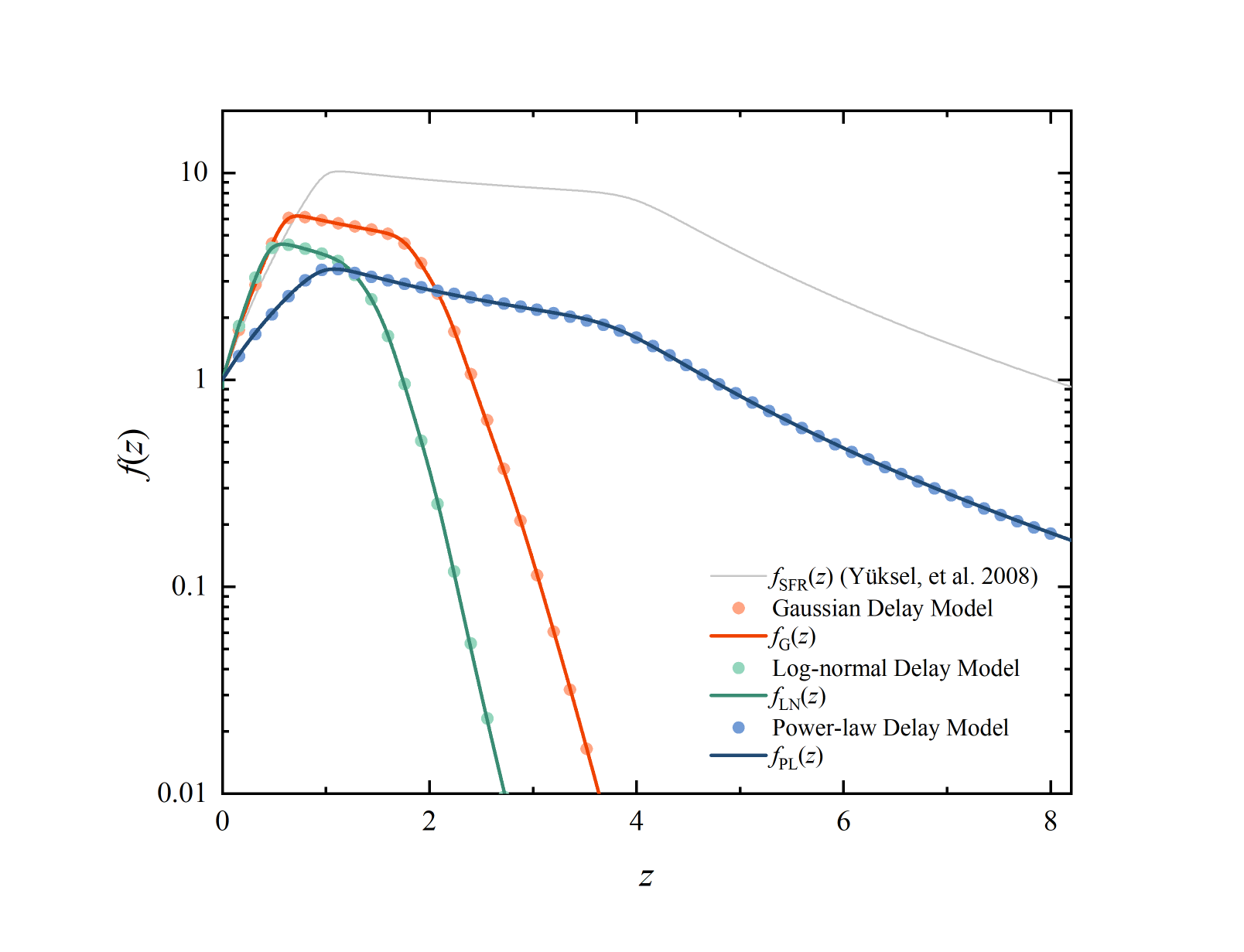}
    \caption{Redshift distribution derived from the method of convolution considering three merger delay models: Gaussian delay model (red), log-normal delay model (green), and power-law delay model (blue). The corresponding colored curves represent empirical fits for each merger delay model. The gray curve is the star formation rate modeled by \cite{yuksel2008}}
    \label{fig:RedshiftDistribution}
\end{figure}

 {In order to provide convenient analytical forms of $f(z)$ for future use, we present updated empirical formulae for the three merger delay models collected by \cite{sun2015}}. The updated empirical formulae of $f(z)$ shown in Figure \ref{fig:RedshiftDistribution} are given as follows. All of these three empirical formulae can well fit the simulated data from $z = 0$ to $z = 8$. For the Gaussian delay model \citep{virgili2011}
\begin{equation}
    P_{\rm G}(\tau) = \frac{\exp\left[-(\tau - t_{\rm d,G})^2/2\sigma_{\rm t,G}^2)\right]}{\sqrt{2\pi}\sigma_{\rm t,G}},
\end{equation}
where $t_{\rm d,G} = 2\,{\rm Gyr}$ and $\sigma_{\rm t,G} = 0.3$, one has
\begin{equation}
\begin{split}
    f_{\rm G}(z) = &\left[ (1 + z)^{3.879\eta} + \left( \frac{1 + z}{73.5} \right)^{-0.4901\eta} + \left( \frac{1 + z}{3.672} \right) ^{-5.691\eta} + \right.\\ 
    &\left.\left(\frac{1+z}{3.411}\right)^{-11.46\eta} + \left( \frac{1 + z}{3.546} \right) ^ {-16.38 \eta} + \left( \frac{1 + z}{3.716} \right) ^ {-20.66 \eta} \right]^{1 / \eta},
\end{split}
\end{equation}
where $\eta = -7.553$.

For the log-normal delay model \citep{wanderman2015}
\begin{equation}
    P_{\rm LN}(\tau) =\frac{\exp\left[-(\ln\tau - \ln t_{\rm d,LN})^2/2\sigma^2_{\rm t,LN}\right]}{\sqrt{2\pi}\sigma_{\rm t,LN}},
\end{equation}
where $t_{\rm d,LN} = 2.9\,{\rm Gyr}$ and $\sigma_{\rm t,LN} = 0.2$, the empirical dimensionless redshift distribution $f(z)$ is 
\begin{equation}
\begin{split}
\label{Eq: LogNomalDistribution}
    f_{\rm LN}(z) = &\left[ (1 + z)^{4.131\eta} + \left( \frac{1 + z}{22.37} \right)^{-0.5789\eta} + \left( \frac{1 + z}{2.978} \right) ^{-4.735\eta} + \left(\frac{1+z}{2.749}\right)^{-10.77\eta} + \right. \\
    &\left.\left( \frac{1 + z}{2.867} \right) ^{-17.51\eta} + \left( \frac{1 + z}{3.04} \right) ^ {-(0.08148 + z^{0.574}/0.08682)\eta} \right]^{1 / \eta},
\end{split}
\end{equation}
where $\eta = -5.51$.

For the power-law delay model \citep{wanderman2015}
\begin{equation}
    P_{\rm PL}(\tau) = \tau^{-\alpha_{\rm t}},
\end{equation}
where $\alpha_{\rm t} = 0.81$, one has
\begin{equation}
    f_{\rm PL}(z) = \left[ (1 + z)^{1.895\eta} + \left( \frac{1 + z}{5.722} \right)^{-3.759\eta} + \left( \frac{1 + z}{11.55} \right) ^{-0.7426\eta}. \right]^{1 / \eta},
\end{equation}
where $\eta = -8.161$.

\end{document}